\documentclass[preprint2]{aastex}
\usepackage{ifthen}
\usepackage{array}
\usepackage{multicol}
\usepackage{amssymb}

\newboolean{ispreprint}
\setboolean{ispreprint}{true}

\newcommand{\pab}{Pa$\beta$}
\newcommand{\lpab}{$\lambda$=1.2818$\mu$m}
\newcommand{\feii}{[\ion{Fe}{2}]}
\newcommand{\lfeii}{$\lambda$=1.2567$\mu$m}

\newcommand{\brg}{Br$\gamma$}
\newcommand{\lbrg}{$\lambda$=2.1655$\mu$m}
\newcommand{\six}{[\ion{S}{9}]}
\newcommand{\lsix}{$\lambda$=1.2525$\mu$m}
\newcommand{\htwo}{H$_2$}
\newcommand{\lhtwo}{$\lambda$=2.1212$\mu$m}
\newcommand{\hei}{\ion{He}{1}}

\newcommand{\etal}{\textit{et al.}}
\newcommand{\cnc}[1]{\multicolumn{1}{c}{#1}}
\newcommand{\kms}{km~s$^{-1}$}

\newlength{\pmsp}
\newlength{\digsp}

\newlength{\dotsp}

\newsavebox{\thistable}
\newsavebox{\thisbox}
\newlength{\thiswid}

\slugcomment{to appear in \textit{The Astronomical Journal}}
\shorttitle{Near-IR Spectroscopy of Seyfert 2 Galaxies}
\shortauthors{Knop \etal}

\begin{document}

\title{Spatially Resolved Near-Infrared Spectroscopy of Seyfert 2
Galaxies Mk~1066, NGC~2110, NGC~4388, and Mk~3\altaffilmark{1}}

\author{R. A. Knop\altaffilmark{2}, L. Armus\altaffilmark{3}, K. Matthews,
T. W. Murphy\altaffilmark{4}, \and B. T. Soifer}

\affil{Palomar Observatory, The California Institute of Technology,
Pasadena, CA 91125}

\altaffiltext{1}{Accepted to appear in \emph{The Astronomical Journal}}

\altaffiltext{2}{Current address: MS 50-232, Lawrence Berkeley National
Laboratory, 1 Cyclotron Rd., Berkeley, CA, 94720; rknop@lbl.gov}

\altaffiltext{3}{Current address: SIRTF Science Center, CIT, 314-6,
Pasadena, CA 91125}

\altaffiltext{4}{Current address: University of Washington, Dept. of
Physics, Box 351560, Seattle, WA 98195}


\begin{abstract}

We present near-infrared spectra with resolutions of
$\lambda/\Delta\lambda\sim1200$ in the emission lines \pab, \feii\
(\lfeii), \brg, and \htwo~$\nu$=1--0S(1) of the nuclei and circumnuclear
regions of the four Seyfert~2 galaxies Mk~1066, NGC~2110, NGC~4388, and
Mk~3.  All of these galaxies show strong near-infrared line emission
that is detected at radii several times the spatial resolution,
corresponding to projected physical scales of 0.07 to 0.7 kpc.  Velocity
gradients are detected in these nuclei, as are spatial variations in
line profiles and flux ratios.  We compare the spatial and velocity
distribution of the line emission to previously observed optical line
and radio emission.  The evidence indicates that the \feii\ emission is
associated with the Seyfert activity in the galaxies.  Our data are
consistent with X-ray heating being responsible for most of the \feii\
emission, although differences in \feii\ and \pab\ line profiles
associated with radio emission suggests that the \feii\ emission is
enhanced by fast shocks associated with radio outflows.  The \htwo\
emission is not as strongly associated with outflows or ionization cones
as is the emission in other lines, but rather appears to be primarily
associated with the disk of the galaxy.
\ifthenelse{\boolean{ispreprint}}{\vspace*{0.2in}}{}

\end{abstract}

\section{Introduction}

The near-infrared wavelength range offers valuable probes of the dynamics, composition, and
excitation mechanisms present in the circumnuclear gas of active galactic
nuclei (AGN).
Among the near-infrared emission lines which have been observed to be
strongest in Seyfert galaxies are the recombination lines of hydrogen,
\pab\ (\lpab) and \brg\ (\lbrg), collisionally excited lines of singly
ionized iron, e.g., \feii (\lfeii), and rotational-vibrational lines of
molecular hydrogen, e.g., \htwo~$\nu$=1--0S(1) (\lhtwo).  Additionally,
there are very high excitation coronal lines, e.g., \six\ (\lsix), which
have been seen in a number of Seyfert galaxies
\citep[e.g.,][]{moo94,kno964151}.  To date, most published infrared
spectra of Seyfert 2 galaxies use single large beams centered on the
nucleus \citep[e.g.,][]{goo94}.  Thus, very little spatial information
on the strong near-infrared lines is available.  However, some studies
have shown that near-infrared line emission is extended over a spatial
scale easily probed with ground-based spectroscopy.  For example,
near-infrared \feii\ and \htwo\ emission has been seen extended over
several arcseconds ($\lesssim$1~kpc) in the nucleus of the Seyfert 2
galaxy NGC~1068 \citep{bli94}, the Seyfert 1.5 galaxy NGC~7469
\citep{gen95}, and the Seyfert 1.5 galaxy NGC~4151 \citep{kno964151}.
In each case, there were overall spatial and kinematic similarities
between the infrared and optical line emission, but differences were
also apparent on spatial and velocity scales within the NLR.  Clearly,
there is important diagnostic information about the circumnuclear
regions of Seyfert 2 galaxies to be gained from spatially resolved
near-infrared spectroscopy.

This paper presents longslit near-infrared spectra of the Seyfert 2
galaxies Mk~1066, NGC~2110, NGC~4388, and Mk~3, in wavelength ranges
which include the lines \pab, \brg, \feii\ (\lfeii),
\htwo~$\nu$=1--0~S(1), and \six (\lsix).  These four objects were chosen
from a larger project to search for spatially extended infrared line
emission in $\sim$20 Seyfert 1 and Seyfert 2 galaxies.  Because
desirable targets in this larger project were those likely to show
extended emission, Seyfert galaxies visible from Palomar Observatory
($\delta>-20^\circ$) were selected with redshift cz$<$5400~\kms,
corresponding to a spatial scale of 350~pc/$''$.  (Throughout this
paper, for the determination of distances and hence projected scales, we
assume $\mathrm{H}_0=75$~km s$^{-1}$ Mpc$^{-1}$.)  Galaxies with known
extended optical ionized gas were preferentially observed, followed in
priority by galaxies with spatially extended radio emission on scales of
$\gtrsim1''$.  The final sample of galaxies observed includes 11
Seyfert~1.x (1, 1.5, and 1.9) galaxies and 12 Seyfert~2 galaxies.  This
paper reports the results from this program for the four Seyfert~2
galaxies that show the best well-detected extended circumnuclear
emission line regions in the near-infrared.

Throughout this paper, for the determination of distances and hence
projected scales, we assume $\mathrm{H}_0=75$~km s$^{-1}$ Mpc$^{-1}$.

\section{Observations and Data Reduction}

The near-infrared spectra discussed in this paper were obtained at the
Hale~200$''$ telescope with the Palomar Near-Infrared Spectrometer
\citep{lar96}.  Observations at resolution $\lambda/\Delta\lambda\sim1200$
were made in two spectral ranges for each galaxy.  One spectral range in the
J-band atmospheric window covers 0.07$\mu$m and includes the \six, \feii,
and \pab\ lines in the rest frame of each galaxy.  The other spectral range in
the K-band atmospheric window covers 0.13$\mu$m and includes the
\htwo~$\nu$=1--0~S(1) and \brg\ lines.  Table~\ref{sey2obslog} is a log of the
observations, including the resolution (in \kms), on-source integration
time, and position angle of the slit for each observation.  Since we are
interested in exploring the near infrared emission line flux ratios and
dynamics as a function of distance from the Seyfert nucleus, we chose slit
position angles to coincide with the position angle of known ionization
cones or other anisotropy previously mapped through narrow band optical
emission line imaging.  In the case of two of the galaxies (Mk~1066 and
NGC~4388), there are additional observations along a secondary slit through
the nucleus, perpendicular to the primary position angle.  


The spectrometer is a reimaging system which disperses the light from
the telescope and feeds the spectra to an externally mounted infrared
camera.  The spatial scale of the 256$\times$256 NICMOS HgCdTe array in
the camera is 0.165$''$~pixel$^{-1}$.  The width of the adjustable slit
was approximately 0.6$''$, yielding the resolutions listed in
Table~\ref{sey2obslog}.  In order to accurately center the nucleus in
the slit, a fold mirror inside the spectrometer, but in front of the
grating, allows infrared imaging with a $10''\times40''$ field of view.
Between subsequent observations of each galaxy, the nucleus was moved
back and forth along the slit by 20$''$, to allow for background
subtraction while maximizing the amount of on-source integration time.
Each individual exposure was 600s for the J-band (1.2$\mu$m) wavelength
range, and 300s for the K-band (2.2$\mu$m) wavelength range.  Along with
the observation of each galaxy, we observed a G-dwarf star at a similar
airmass (within 0.1 of that of the galaxy), and where possible in the
same area of the sky, in order to correct for atmospheric transmission.
A chopping secondary mirror on the telescope uniformly spread the light
of this star along the length of the slit.  In this way, the spectrum
of the atmospheric calibrator is also used as a flat-field.

\ifthenelse{\boolean{ispreprint}}{
\onecolumn
\begin{table}[htbp]
\renewcommand{\baselinestretch}{1}\small
\begin{center}
\begin{tabular}{l l r r r r l}
\hline
\hline
Galaxy & Wavelength & \multicolumn{1}{c}{On-Source}   & 
        \multicolumn{1}{c}{Slit} & \multicolumn{1}{c}{Res.}  & 
        \multicolumn{1}{c}{psf}  & Date of    \\
       &            & \multicolumn{1}{c}{Integration} & 
        \multicolumn{1}{c}{PA}   & \multicolumn{1}{c}{\kms}  & 
        \multicolumn{1}{c}{FWHM} & Observation \\
\hline
Mk~1066  & 1.25$\mu$m-1.31$\mu$m & 2400s & 135$^\circ$
         & 242$\pm$15       & $0.7''$     & 1995/09/09 \\
Mk~1066  &                       & 2400s &  45$^\circ$
         & 228$\pm$15       & $0.6''$     & 1996/09/10 \\
Mk~1066  & 2.10$\mu$m-2.23$\mu$m & 2400s & 135$^\circ$
         & 289$\pm$19       & $0.6''$     & 1995/09/08 \\
Mk~1066  &                       & 2400s &  45$^\circ$
         & 289$\pm$19       & $0.5''$     & 1995/09/10 \\[12pt]
NGC~2110 & 1.25$\mu$m-1.31$\mu$m & 2400s & 160$^\circ$
         & 229$\pm$16       & $\sim1.1''$ & 1996/01/05 \\
NGC~2110 & 2.09$\mu$m-2.22$\mu$m & 1800s & 160$^\circ$
         & 285$\pm$19       & $\sim1.1''$ & 1996/01/05 \\[12pt]
NGC~4388 & 1.25$\mu$m-1.31$\mu$m & 2400s & 30$^\circ$
         & 229$\pm$15       & $\sim0.8''$ & 1996/01/03 \\
NGC~4388 &                       & 2400s & 120$^\circ$
         & 244$\pm$15       & $0.8''$ & 1996/04/08 \\
NGC~4388 & 2.10$\mu$m-2.23$\mu$m & 2400s & 30$^\circ$
         & 261$\pm$19       & $\sim0.7''$ & 1996/01/03 \\
NGC~4388 &                       & 1800s & 120$^\circ$
         & 303$\pm$19       & $0.8''$ & 1996/04/08 \\[12pt]
Mk~3     & 1.26$\mu$m-1.32$\mu$m & 3000s & 113$^\circ$
         & 243$\pm$15       & $\sim1.0''$ & 1995/11/06 \\
Mk~3     & 2.14$\mu$m-2.27$\mu$m & 1800s & 113$^\circ$
         & 292$\pm$19       & $\sim1.0''$ & 1995/11/07 \\[12pt]
\hline
\hline
\end{tabular}
\end{center}
\caption[Seyfert 2 Observation Log.]{Observation log.  References for
position angles are: Mk~1066: \protect\citet{bow95}; NGC~2110:
\protect\citet{mul94}; NGC~4388: \protect\citet{cor88}; Mk~3:
\protect\citet{mul96}.  The resolution column indicates the FWHM of an
unresolved line measured from OH sky lines in one of the galaxy frames.  The
point spread function (psf) of the atmospheric seeing was determined for
Mk~1066 and for the April 1996 observations of NGC~4388 by observing a star,
guided in the same manner as the spectra, in between spectral observations.
For Mk~3, the seeing was estimated based on 2.2$\mu$m imaging of a quasar
somewhat later than the observations of Mk~3.  For NGC~2110 and the January
1996 observations of NGC~4388, unguided G-star and photometric standard star
images were used to estimate the seeing.}
\label{sey2obslog}
\end{table}
\twocolumn
}{
\placetable{sey2obslog}
}

The data were reduced following the procedure described in \citet{thesis}.
Bad pixels were removed by linear interpolation, and the spatial and
spectral curvature of the field of view was measured and removed using a
point source spectrum and the spectrally unresolved OH night sky lines.
Spectral data frames were divided by the appropriate G-star spectrum, in
order to remove atmospheric absorption lines and to correct for efficiency
variations across the field of view.  In order to correct for the blackbody
slope of the G-star continuum, and to remove absorption features intrinsic
to the star, each G-star spectrum was first divided by a solar spectrum
following the procedure of \citet{mai96}.

\section{Results and Discussion: Individual Objects}

\subsection{Mk~1066}
\label{sec:mk1066}

Mk~1066 is an inclined SB0+ galaxy with a systemic velocity of
3625~\kms\ \citep{bow95}.  This corresponds to a calculated distance of
48~Mpc, which yields a spatial scale of 234~pc~arcsec$^{-1}$.  Hubble
Space Telescope (HST) imaging of the nuclear regions shows a narrow
``jetlike'' feature in a narrow band image which includes [OIII] and
H$\beta$.  This jet is oriented at 315$^\circ$ and extends $\sim 1.4''$
northwest of the nucleus.  In H$\alpha$ and [NII], the jet is observed
on both sides of the nucleus \citep{bow95}.  The nuclear radio source is
linear, and extended over 2.6$''$.  The 5~GHz radio emission is oriented
at the same position angle as the optical line emission \citep{ulv89}.

We obtained infrared spectra of Mk~1066 along two position angles,
135/315$^\circ$ and 45/225$^\circ$.  The 135$^\circ$ angle of the slit
was chosen to align with the position angle of the optical ionization
cones \citep{bow95}.  These spectra show strong line emission,
concentrated on the nucleus.  Four emission lines are visible in
Figures~\ref{mk1066grey}, \ref{mk1066spec135}, and \ref{mk1066spec45}:
\pab\ and \feii\ in the J-band spectra, and \brg\ and
\htwo~$\nu$=1--0~S(1) in the K-band spectra.  No \six\ (\lsix) is
visible on the nucleus (to an upper limit on the equivalent width of
$2\times10^{-4}$~$\mu$m).  In addition, there is data along a 45$^\circ$
slit chosen to be perpendicular to the primary slit.  The infrared lines
are visible over a spatial extent of $\sim5''$ (1.2~kpc) along the
135$^\circ$ slit, in comparison to the FWHM of the nuclear continuum
which was $\sim1.8''$ at 1.25$\mu$m and $\sim1.1''$ at 2.15$\mu$m in our
spectra (both of which are greater than the 0.5--0.7$''$ FWHM of the
atmospheric seeing).  In contrast, the lines are less spatially extended
along the 45$^\circ$ slit, and are visible only over an extent of
$\lesssim3''$.  The \brg\ line is strongest on the nucleus, but the
\pab, \feii, and \htwo\ lines peak about $0.5''$ northwest of the
nucleus (see Figure~\ref{mk1066lv}).

Figure~\ref{mk1066grey} shows two greyscale plots of the reduced
1.25--1.31$\mu$m spectra along both position angles.  The horizontal axis is
observed wavelength, and the vertical axis is position along the slit in
arcseconds from the nucleus.  The \pab\ and \feii\ lines are spatially
resolved along both position angles in Figure~\ref{mk1066grey}.  The tilt in
the lines indicates that there is a spatial velocity gradient in the \feii\
and \pab\ lines along the 135$^\circ$ slit; this will be discussed in
section~\ref{sec:mk1066velocitystruc}.  Along the 45$^\circ$ slit, no
velocity gradient is immediately apparent.

\ifthenelse{\boolean{ispreprint}}{
\begin{figure}[htb]
\begin{center}
\epsscale{1.0}
\plotone{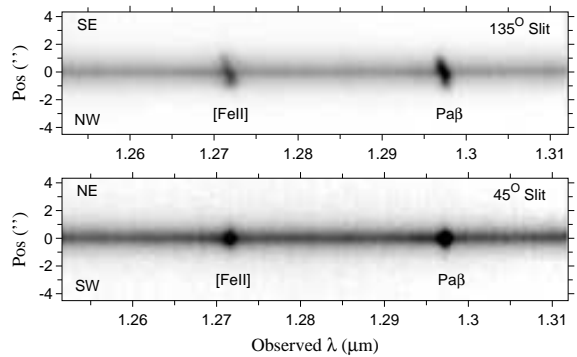}
\caption[Greyscale Spectra of Mk~1066]{Spectra of Mk~1066 in the
$1.25-1.31\mu$m range.  Position is distance from the nucleus in arcseconds
along the indicated position angle of the slit.}
\label{mk1066grey}
\end{center}
\end{figure}
}{}

Figures~\ref{mk1066spec135} and \ref{mk1066spec45} are one-dimensional
projections in various spatial bins extracted from the two-dimensional data.
Figure~\ref{mk1066spec135} shows the spectra along the 135$^\circ$ slit, and
Figure~\ref{mk1066spec45} along the 45$^\circ$ slit.  The size of each
spatial bin used in the extractions is 4 pixels, or 0.67$''$.  This matches
the point spread function (psf) at the time of the observations
(0.5$''$--0.7$''$).

The \pab\ profile near the nucleus shows enhanced emission blueward of
the line center.  These blueward features on \pab\ are visible not only
in Mk~1066, but also in several other Seyfert galaxies \citep{kno96supp}.
Were this \pab\ emission, it would be at a velocity of $\sim-700$~\kms\
relative to the centroid of the \pab\ emission, which is higher than the
velocities observed in other lines in Mk~1066. These blue features on
\pab\ are probably primarily due to a blend of two \hei\ lines at
1.2785$\mu$m and 1.2791$\mu$m, which have been resolved from \pab\ in
high resolution spectroscopy of the planetary nebula BD$+30^\circ3639$
\citep{goo94}.  There may also be a small contribution from \feii\
($\lambda=1.2788\mu$m).  For T$_\mathrm{e}\sim10^4$~K and
n$_\mathrm{e}\sim10^4$~cm$^{-3}$, the strength of the \feii\
($\lambda=1.2788\mu$m) line is expected to be $<0.1$ of the strength of
the \feii\ (\lfeii) line \citep{goo94}.  In Mk~1066, \feii\
($\lambda=1.2788\mu$m) could account for at most half of the emission on
the blue side of \pab.

\ifthenelse{\boolean{ispreprint}}{
\begin{figure}[htbp]
\renewcommand{\baselinestretch}{1}
\begin{center}
\epsscale{0.9}
\plotone{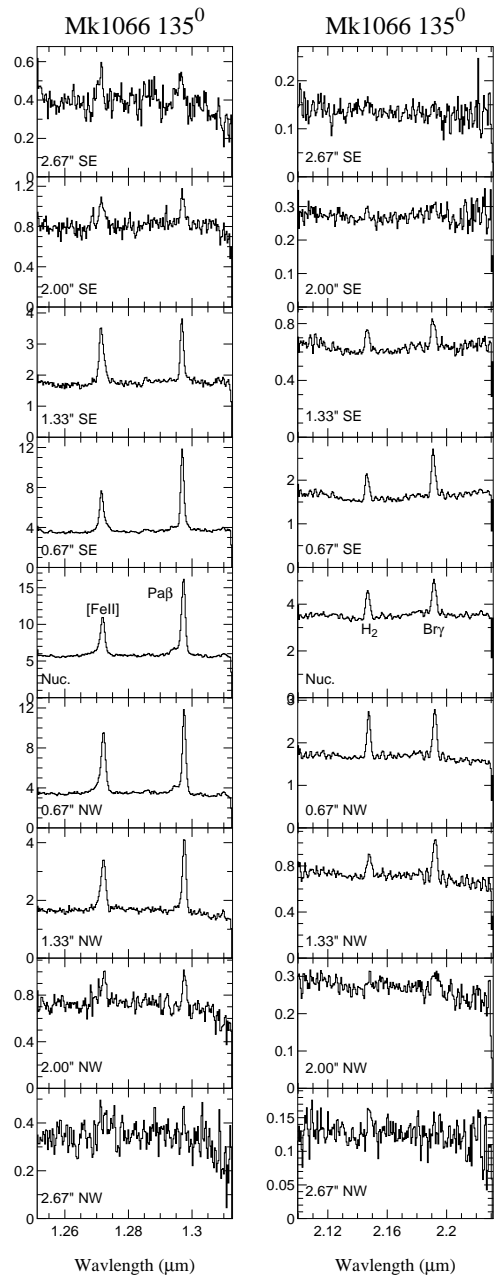}
\caption[Spectra of Mk~1066 at 135$^\circ$]{Spectra of Mk~1066 along the
135$^\circ$ slit.  Each spectrum represents the spectrum in a
rectangular beam which is $0.67''$ (4 pixels) along the slit by the
width of the slit ($\sim0.6''$).  Spatial bins are adjacent.  The
position of the center of each spatial bin, relative to the nucleus, is
indicated in the lower left of each panel.  Flux units (F$_\lambda$) are
arbitrary.}
\label{mk1066spec135}
\end{center}
\end{figure}
}{}

\ifthenelse{\boolean{ispreprint}}{
\begin{figure}[htb]
\begin{center}
\epsscale{0.9}
\plotone{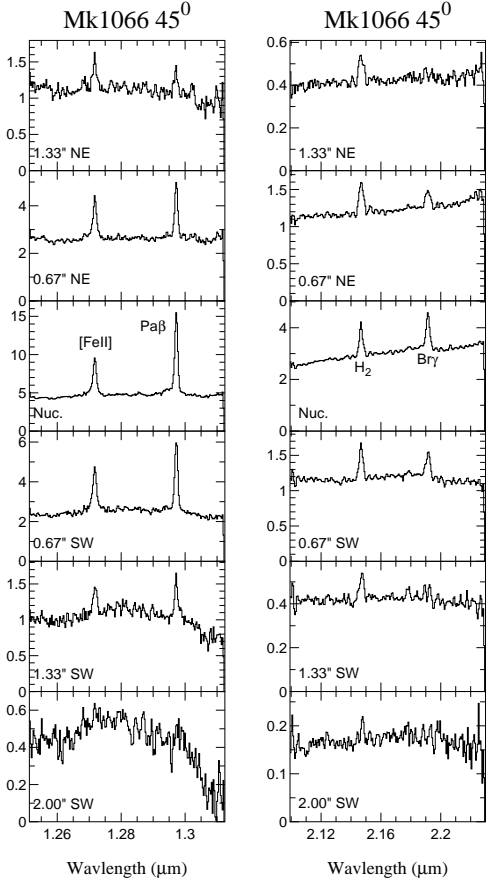}
\caption[Spectra of Mk~1066 at 45$^\circ$]{Spectra of Mk~1066 along the
45$^\circ$ slit.  Each spectrum represents the spectrum in a rectangular
beam which is $0.67''$ along the slit by the width of the slit
$\sim0.6''$.  Spatial bins are adjacent.  Flux units (F$_\lambda$) are
arbitrary.}\label{mk1066spec45}
\end{center}
\end{figure}
}{}

\subsubsection{Integrated Line Flux Ratios}
\label{sec:mk1066fluxrat}

Table~\ref{mk1066eqw} in Appendix~\ref{sec:tabappndx} lists the
equivalent widths and flux ratios of the lines observed in Mk~1066.
Because \feii\ and \pab\ were observed simultaneously and are close in
wavelength, we may obtain accurate \feii/\pab\ flux ratios which are
relatively insensitive to uncertainties in flux calibration or
differential reddening.  Similarly, we may obtain reliable \htwo/\brg\
flux ratios from our K-band spectra.  Figure~\ref{mk1066fluxrat} shows
these ratios as a function of position along the slit, using the total
integrated fluxes for each species in each spatial bin to derive the
line flux ratios.

\ifthenelse{\boolean{ispreprint}}{
\begin{figure}[htb]
\begin{center}
\epsscale{1.0}
\plotone{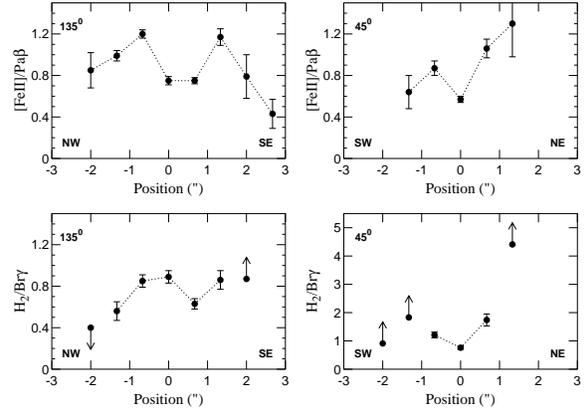}
\caption[Mk~1066 Flux Ratios]{Flux ratios for Mk~1066.  The two plots on the
left are for the 135$^\circ$ slit, the two plots on the right for the
45$^\circ$ slit.}
\label{mk1066fluxrat}
\end{center}
\end{figure}
}{}

A striking feature of Figures~\ref{mk1066spec45} and \ref{mk1066fluxrat} is
the rapid rise of the \htwo/\brg\ flux ratio with distance from the nucleus
along the 45$^\circ$ slit, perpendicular to the position angle of the
optical ionization cones.  The \htwo\ line is detected 2$''$ away from the
nucleus in both directions, while the \brg\ line is detected only within
0.67$''$ of the nucleus.  The \htwo/\brg\ flux ratio 1.33$''$ from the
nucleus is $>$3, in contrast to the nuclear value of $0.8$.  By comparison,
along the 135$^\circ$ slit, the \htwo/\brg\ flux ratio always remains
between about 0.4 and 0.9, with an indication of a gradient across the
nucleus.  The \feii/\pab\ flux ratio varies with position along the
45$^\circ$ slit, although not as rapidly as does the \htwo/\brg\ flux ratio.
There is a clear local minmum in the \feii/\pab\ ratio at the nucleus along
both slits.

\subsubsection{Velocity structure}
\label{sec:mk1066velocitystruc}
\label{sec:mk1066wings}

%
%
Figure~\ref{mk1066lv} is a set of position versus velocity plots along
the 135$^\circ$ slit for each of the four lines identified in
Figures~\ref{mk1066spec135} and \ref{mk1066spec45}.  In each case, a
smoothed continuum has been subtracted from the spectrum.  Velocity is
plotted relative to 3625~\kms, the systemic velocity of Mk~1066
\citep{bow95}, and position is distance in arcseconds on the sky from
the continuum peak.  A velocity gradient is visible in each of the
lines, in the sense that the lines are blueshifted southeast of the
nucleus and redshifted northwest of the nucleus.  The slope of the
velocity gradient does not appear to be the same for all of the lines.
Specifically, the \htwo\ line appears to have a steeper gradient
($\sim$300~\kms~arcsec$^{-1}$, or $\sim$1300~\kms~kpc$^{-1}$) in
comparison to the other three lines (whose gradients are
$\sim$200~\kms~arcsec$^{-1}$, or
$\sim$900\ifthenelse{\boolean{ispreprint}}{\linebreak}{}~\kms~kpc$^{-1}$).
In addition, \brg\ appears to have a slightly steeper gradient (by
$\sim$50~\kms~arcsec$^{-1}$) than \pab\ southeast of the nucleus.  The
gradients of the \feii\ and \pab\ lines are similar.

Another feature apparent in Figure~\ref{mk1066lv} is that all of the
lines, except for \brg, peak northwest of the nucleus.  This is discussed in
section~\ref{mk1066extinction}.

\ifthenelse{\boolean{ispreprint}}{
\begin{figure}[htb]
\begin{center}
\epsscale{1.0}
\plotone{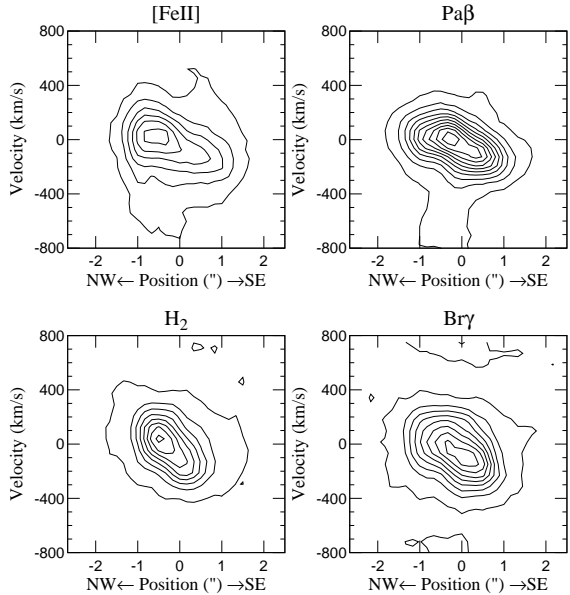}
\caption[Mk~1066 Longitude-Velocity Contour Plots]{Position versus velocity
plots for Mk~1066.  Position along a 135$^\circ$ slit.  Velocity is relative
to 3625~\kms, the systemic velocity of Mk~1066 \protect\citep{bow95}.}
\label{mk1066lv}
\end{center}
\end{figure}
}{}

Table~\ref{mk1066kinematics} in Appendix~\ref{sec:tabappndx} lists the
velocity centroids and full widths at half maxima of the complete line
profiles for the lines observed in Mk~1066 along both slits in each
spatial bin.  Many of the lines show a complicated asymmetric profile.
These lines were decomposed into the smallest number of Gaussian
components necessary to statistically describe the line profile.
Figure~\ref{mk1066vvspos} shows plots of the velocity of the strongest
narrow Gaussian component of each line as a function of position along
the slit.  Usually, this is equivalent to a plot of the mode (peak) of
the line profile as a function of position, because most of the line
profiles in Mk~1066 are dominated by a strong, central narrow component.
For reference, a line connecting the \brg\ velocity peaks is drawn on
each plot, and the H$\alpha$ velocity curve \citep{bow95} is shown on
the \pab\ plot.

\ifthenelse{\boolean{ispreprint}}{
\begin{figure}[htb]
\begin{center}
\epsscale{1.0}
\plotone{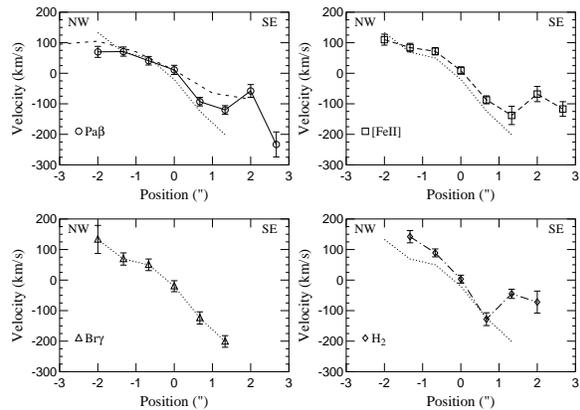}
\caption[Mk~1066 Position Versus Velocity]{Velocity of the primary narrow
component of each line versus position along the 135$^\circ$ slit for
Mk~1066.  The data for \brg\ are shown as a dotted line in each plot.  The
dashed line on the \pab\ plot is the H$\alpha$ data of
\protect\citet{bow95}.}
\label{mk1066vvspos}
\end{center}
\end{figure}
}{}

The same relative velocity gradients between the different species
mentioned for Figure~\ref{mk1066lv} are visible in
Figure~\ref{mk1066vvspos}.  It is apparent the difference between the
\htwo\ and \brg\ velocity slopes is greater than the difference between
the \htwo\ and \pab\ velocity slopes.  Two arcseconds southeast of the
nucleus, all of the velocity profiles except for that of \brg\ (which is
undetected) show a local velocity reversal of 50--75~\kms.

In addition to the primary narrow emission line components whose centroid
velocities are plotted in Figure~\ref{mk1066vvspos}, there are weaker
components which are displaced in velocity relative to the center of the
lines.  Southeast of the nucleus, where the peaks of the line profiles are
blueshifted relative to the systemic velocity, all of the lines show a
noticeable enhanced red wing (although it is very weak in \htwo).  On the
nucleus and to the northwest where the peaks of the line profiles are
redshifted, the \pab\ and \feii\ lines show an enhanced blue wing.  While
this blue wing is almost certainly a feature of the \feii\ line profile
itself, the blue wing of the \pab\ line suffers contamination from other
species as discussed above.

\ifthenelse{\boolean{ispreprint}}{
\begin{figure}[htb]
\begin{center}
\epsscale{1.0}
\plotone{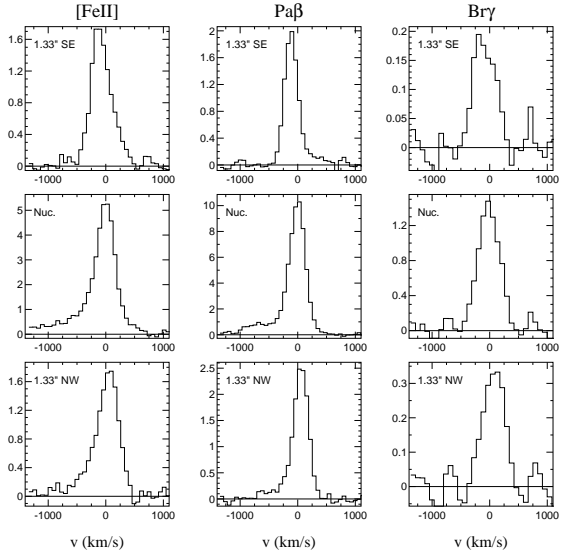}
\caption[Mk~1066 Line Profiles]{Selected line profiles of Mk~1066 along the
135$^\circ$ slit, chosen to illustrate shoulders seen in asymmetrical line
profiles.  One resolution element is approximately 3 bins on the plot.}
\label{mk1066closeups}
\end{center}
\end{figure}
}{}

For illustration, Figure~\ref{mk1066closeups} shows the profiles of the
\feii, \pab, and \brg\ lines on the nucleus and 1.33$''$ southeast and
northwest of the nucleus.  Southeast of the nucleus, the red wing is
stronger relative to the rest of the line in \feii\ and \brg\ than it is in
\pab.  Similarly, northwest of the nucleus, the blue wing is most prominent
in \feii.  It is not visible in the \brg\ line, and in \pab\ it may be
contaminated by the aforementioned \hei\ and \feii\ lines.

\subsubsection{Comparison with Optical Spectroscopy}
\label{sec:mk1066compop}

\citet{bow95} present longslit optical spectroscopy of Mk~1066 along a
134$^\circ$ slit.  The H$\alpha$, H$\beta$, [NII], [SII], and [OI] lines all
have a velocity gradient extending from +100~\kms\ 2-3$''$ northwest of the
nucleus to -90~\kms\ 2-3$''$ southeast of the nucleus.  This is similar to
the velocity structure seen in the infrared lines (see
Figure~\ref{mk1066vvspos}), although the peaks of the \pab\ and particularly
the \brg\ lines are slightly blueshifted relative to the optical lines to
the southeast of the nucleus.

In addition to the primary kinematic component seen in optical emission
lines, \citet{bow95} identify a second kinematical component, most clearly
seen in [OIII].  The [OIII] emission is slightly blueshifted relative to the
other lines northwest of the nucleus and redshifted by as much as 200~\kms\
relative to the other lines southeast of the nucleus.  This structure is
qualitatively similar to the enhanced wings seen in the infrared lines
discussed in section~\ref{sec:mk1066wings}.  Specifically, the infrared
lines show an enhancement blueward of the peak on the nucleus and to the
northwest, and they show an enhancement redward of the peak to the
southeast.  It is likely that some of the wings seen on the infrared
emission lines are associated with the kinematic component \citet{bow95}
identify in [OIII].

\subsubsection{Extinction}
\label{mk1066extinction}

Since \pab\ and \brg\ are both hydrogen recombination lines (n=5--3 and
n=7--4 respectively), they should arise in the same physical locations.
Their intrinsic Case B flux ratio is well known (assuming radiative
transfer effects are negligible).  Because of reddening, in dusty
regions the observed \pab\ line will be suppressed relative to the
observed \brg\ line.  If the dust is intermixed with the line emitting
clouds, and there are strong velocity gradients, the two lines can
appear to have different velocity structures.  In Mk~1066, both the
infrared and optical data indicate that extinction effects are stronger
to the southeast than they are to the northwest.

A difference in both the spatial and velocity distributions of \pab\ and
\brg\ is visible in Figure~\ref{mk1066lv}.  The \brg\ line is peaked
close to the continuum peak, and is stronger relative to \pab\ toward
the southeast than it is toward the northwest.  The simplest explanation
of this difference is a larger extinction in the southeast.  Spatially
variable extinction may also be responsible for the slightly different
\pab\ and \brg\ velocity gradients seen in Figure~\ref{mk1066lv}.  To
the southeast, where dust extinction appears greater, the discrepancy
between the velocity gradients for the \pab\ and \brg\ lines is larger.
Extinction could also explain the difference between the velocities of
the infrared lines and the optical H$\alpha$ line southeast of the
nucleus \citep{bow95}.  The H$\alpha$-infrared slope discrepancy is in
the same sense as the \pab--\brg\ slope discrepancy, i.e., lines at
shorter wavelengths have shallower gradients; see
Figure~\ref{mk1066vvspos}.

This same extinction effect is seen in the red wings of the \pab\ and
\brg\ lines southeast of the nucleus as discussed in
section~\ref{sec:mk1066wings}; the wing is relatively more prominent in
\brg\ than in \pab\, suggesting that the extinction to the region where
the wings originate is greater than the extinction to the region where
the bulk of the line emission originates.  If the wings represent
outflowing gas, then the redshifted receding gas, which may be partially
obscured by dust in the disk of the galaxy, would show higher
extinction.  There is also evidence from optical wavelengths that the
outflowing gas in the southeast is heavily extinguished.  The optical
line emission is seen emerging from the nucleus in ionization cones at
position angles of 135$^\circ$/315$^\circ$.  The [OIII]+H$\beta$
narrowband image \citep{bow95} is much stronger to the northwest,
suggesting suppressed emission to the southeast.  The 5~GHz radio
emission \citep{ulv89} indicates that the strength of the outflow is
slightly asymmetric, but the difference in the strength of the radio
emission northwest and southeast of the nucleus is not as great as the
difference in the strength of the [OIII] emission which \citet{bow95}
associate with the outflow.

Extinction cannot explain the discrepancies between the \htwo\ and \brg\
lines, since the differential extinction between the two is negligible.  The
\htwo\ line shows a different velocity gradient than \brg\
(Figure~\ref{mk1066lv}), and the peak of the \htwo\ emission is shifted to
the northwest relative to the peak of the \brg\ emission.  The differences
in the structure of these lines may indicate that the emission seen from
each species originates from different clouds along the same line of sight.
Additionally, there may be a systematic variation in the physical conditions
along the cone axis of the galaxy.

\subsubsection{A Picture of the Circumnuclear Regions of Mk~1066}

The infrared data in combination with the data in \citet{bow95} indicate
that there are two dominant kinematic components in near the nucleus of
Mk~1066.  One component shows up in infrared \htwo, in the peaks of the
infrared lines, and in optical H$\alpha$, [\ion{N}{2}], and other
species.  This component shows a clear velocity gradient along the
135$^\circ$ slit, is blueshifted to the southeast, and is redshifted to the
northwest.  The second, higher-excitation component shows up in the
wings on the infrared lines and in optical [\ion{O}{3}], and has the
opposite velocity trend from the first component.

\citet{bow95} identify the first, low-excitation component with a
rotating disk.  The velocity gradient seen in the infrared line emission
for this component along the 135$^\circ$ slit is consistent with a
rotation.  The much smaller (though non-zero) velocity gradient along
the 45$^\circ$ slit indicates that the major axis of the rotating disk
must be closer to 135$^\circ$ than 45$^\circ$; a fit of a uniformly
rotating disk to the peaks of the infrared line emission along both
slits suggests that the major axis of the disk is at $\simeq120^\circ$.
Following \citet{bow95}, we identify the second component with
high-excitation outflowing gas.  This second component is not seen off
of the nucleus in the infrared spectra along the 45$^\circ$ slit, and
the optical [\ion{O}{3}] emission is clearly extended along an angle of
$\sim135^\circ$.  Consequently, the data indicate that the outflow is
\emph{not} perpendicular to the rotating disk; rather, the two physical
systems are at low relative inclination when projected on to the plane
the sky.

\ifthenelse{\boolean{ispreprint}}{
\begin{figure}[htb]
\begin{center}
\epsscale{1.0}
\plotone{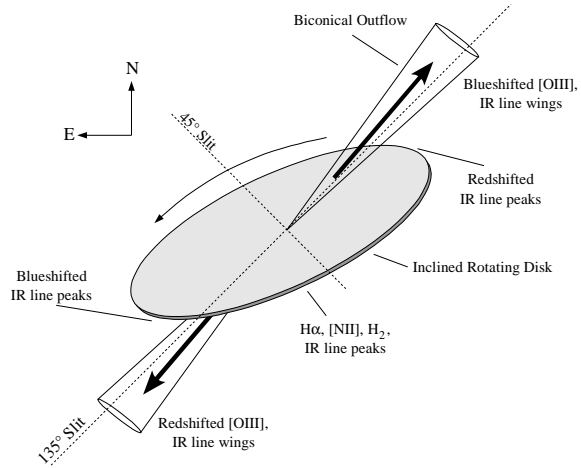}
\caption[Mk~1066 Cartoon]{Cartoon drawing of the nuclear regions of
Mk~1066, from the point of view of the observer.  The rotating disk is
the source of low excitation emission, and the source of the primary
velocity gradient seen in most optical and infrared lines.  The outflow
is higher excitation emission, and shows up as wings in the infrared
lines.  The dusty disk extinguishes features from the southeast cone of
the outflow.  Superimposed on the cartoon are the angles of the
135$^\circ$ and 45$^\circ$ infrared slits.}
\label{mk1066art}
\end{center}
\end{figure}
}{}

Figure~\ref{mk1066art} is a cartoon drawing of the system based on the
optical and infrared data.  In this picture, there is an inclined disk with
a major axis of 120$^\circ$ projected on the sky.  This orientation explains
both the greater spatial extent and the larger velocity gradient in the
infrared lines along the 135$^\circ$ slit.  The disk is dusty, explaining
the extinction effects discussed in section~\ref{mk1066extinction}.  There
is a biconical outflow associated with the higher-excitation gas, as drawn.
To the northwest, this outflow is relatively unobscured, as it is in front
of the plane of the galaxy.  The blue wings seen on the infrared lines to the
northwest are associated with this outflow.  To the southeast, the outflow
is obscured by the plane of the galaxy, suppressing the [OIII] emission
observed in that direction.  Extinction resulting from this obscuration
enhances the \brg/\pab\ ratio on the red side of the line profile southeast
of the nucleus (see Figure~\ref{mk1066closeups}).

\subsubsection{\feii\ and \htwo\ Emission Processes}

The fact that various features of the \feii\ emission line profiles
correspond to those identified with the active nucleus in other species,
such as the wings seen in [OIII] \citep{bow95}, suggests that the \feii\
emission is dominated by processes directly associated with the Seyfert
nucleus.  The radio emission of Mk~1066 \citep{bow95} appears to represent two
oppositely directed radio jets aligned with the outflow seen in [OIII].  In
the wings on the infrared line profiles we associate with this outflow,
\feii\ emission is enhanced relative to \pab\ (see
Figure~\ref{mk1066closeups}).  Shocks resulting from the interaction of
outflowing radio plasma with the ambient gas are a natural mechanism for the
production of \feii\ emission.  Fast shocks can both destroy dust grains,
increasing the gas phase abundance of iron, and create large partially
ionized regions, two conditions which enhance \feii\ emission.

The most striking feature of the \htwo\ emission is that its strength does
not drop off as quickly with distance as the other infrared lines from the
nucleus perpendicular to the outflow axis.  While \brg\ is only detected out
to 150pc from the nucleus at a position angle of 45$^\circ$, \htwo\ is
detected to $>$300pc.  The presence of \htwo\ emisison along both slits
shows that warm molecular gas is ubiquitous in the near-nuclear disk of
Mk~1066.  Since the \htwo\ is not confined to the cone axis, it is evidently
not directly dependent on ionization from the active nucleus or the radio
jets for its excitation.  There is probably little \htwo\ emission
associated with the high excitation outflow, since the wings seen in the
\feii\ and \pab\ lines associated with this outflow were not detected in the
\htwo\ line profiles.

Heating by X-rays from the Seyfert nucleus likely cannot explain the extent
of the \htwo\ emission perpendicular to the ionization cone axis.  Slow
(v$<25$~\kms) shocks originating in the nucleus may heat molecular clouds in
the disk of the galaxy immediately surounding the nucleus, but the
relatively flatter distribution of \htwo\ relative to the other species
along the 45$^\circ$ slit argues for an extended source of heating for the
molecular gas in the regions within $\sim500$~pc of the nucleus of Mk~1066.

\subsubsection{Summary}

\begin{enumerate}
\item
The infrared \pab, \brg, and \feii\ lines are extended over 5$''$ (1.2~kpc
projected on the plane of the sky) along a 135$^\circ$ position angle
coincident with the direction of the optically identified ionization cones.
Along a perpendicular angle, their intensity drops off faster with distance
from the nucleus; 1.3$''$ (0.3~kpc projected) away from the nucleus at
45$^\circ$, the intensity of the lines relative to the nucleus is 20\% of
their value along a position angle of 135$^\circ$.

\item
The infrared lines have strongly peaked narrow components.  The peaks of
the lines have a velocity gradient similar to the optical H$\alpha$,
H$\beta$, [NII], [SII], and [OI] lines along a position angle of
135$^\circ$, which \citet{bow95} identify as part of a rotating,
LINER-like disk.

\item
There are secondary components in the form of enhanced wings in the infrared
lines whose kinematics differ from the strong, narrow components in a manner
similar to the difference between the kinematics of [OIII] emission and the
other optical lines.  The [OIII] emission line and the near infrared line
wings are likely associated with a nuclear outflow which produces
blueshifted emission to the northwest of the nucleus.

\item
The relative spatial distributions of \brg\ and \pab\ indicate that there is
greater extinction to the southeast of the nucleus than there is to the
northwest.  This may explain differences seen in the velocity centers of
hydrogen recombination lines to the southeast, which is most likely due to
obscuration of the gas by an inclined disk in Mk~1066.

\item
The \htwo\ line is more extended than the other lines along an angle
perpendicular to the ionization cone.  Along the axis of the cone, it
shows a steeper velocity gradient than \pab, \brg\, or \feii.

\item
We propose that there is a measurable contribution to the \feii\ emission
from shocks associated with the nuclear outflow.  It may have a direct link
to the radio jets seen at 3.6cm.  The \htwo\ emission, on the other hand,
arises primarily from the inner (500pc) disk of Mk~1066, most of which is
presumably shielded from direct illumination of the nucleus.  This warm
molecular gas may be heated by extended local sources such as star
formation.
\end{enumerate}


\subsection{NGC~2110}
\label{sec:2110}

NGC~2110 is an S0 galaxy with a systemic velocity of 2342~\kms\
\citep{dev91}, corresponding to a calculated distance of 31~Mpc, implying
a spatial scale of 150~pc~arcsec$^{-1}$.  HST narrowband imaging in
optical H$\alpha$+[NII] and [OIII] emission shows a narrow jetlike
feature near the nucleus extending to the north at a position angle of
$\sim$340$^\circ$, and a weaker opposing feature extending to the south
at $\sim$160$^\circ$ \citep{mul94}.  The feature curves into an ``S''
shape 4$''$ away from the nucleus in each direction.  At 1.49~GHz,
NGC~2110 shows a triple radio source extended over 4$''$ with a position
angle of $\sim$0$^\circ$ \citep{ulv84a}, slightly different from the
position angle of the optical line emission.  The curvature of the ``S''
of the optical line emission is towards the direction of the radio
emission.  In X-rays, NGC~2110 has a 2-10~keV luminosity of
1.4$\times10^{43}$~erg~s$^{-1}$ with a comparatively high absorbing
column density of $7.4\times10^{22}$~cm$^{22}$ \citep{rei85}.  This X-ray
luminosity is closer to that typical for a moderate luminosity Seyfert 1
galaxy than for a Seyfert 2 galaxy ($\sim5\times10^{41}$~erg~s$^{-1}$)
\citep{kri80}.  ROSAT High Resolution Imager (HRI) data show a spatially
extended X-ray source, with secondary X-ray emission 4$''$ north of the
nucleus which may be associated with the soft X-ray excess seen by
\citet{wea95}.

Our longslit spectra of NGC~2110 along the ionization cones at a
position angle of 160$^\circ$ (Figure~\ref{ngc2110spec}) show strong
emission lines in \feii, \pab, \htwo~$\nu$=1--0 S(1), and \brg.  No
\six\ (\lsix) is detected on the nucleus (with an upper limit on the
equivalent width of $1\times10^{-4}\mu$m).
Figure~\ref{ngc2110spec} shows the summed spectra in spatially adjacent
1$''$ bins along the slit.  These lines are visible on the nucleus, and
are seen to be extended by $\gtrsim6''$ (900~pc).  In comparison, the
FWHM of the continuum is 2.4$''$ at 1.2$\mu$m and 1.5$''$ at 2.2$\mu$m,
while the seeing at the time of the observations was $\sim1.1''$.  The
most striking feature of the NGC~2110 J-band spectrum is the strength of
the \feii\ line.  On the nucleus, the \feii/\pab\ flux ratio is
8.1$\pm$0.8, which is greater than that observed in any other Seyfert
galaxy in this survey \citep{kno96supp}.

\ifthenelse{\boolean{ispreprint}}{
\begin{figure}[htbp]
\begin{center}
\epsscale{0.8}
\plotone{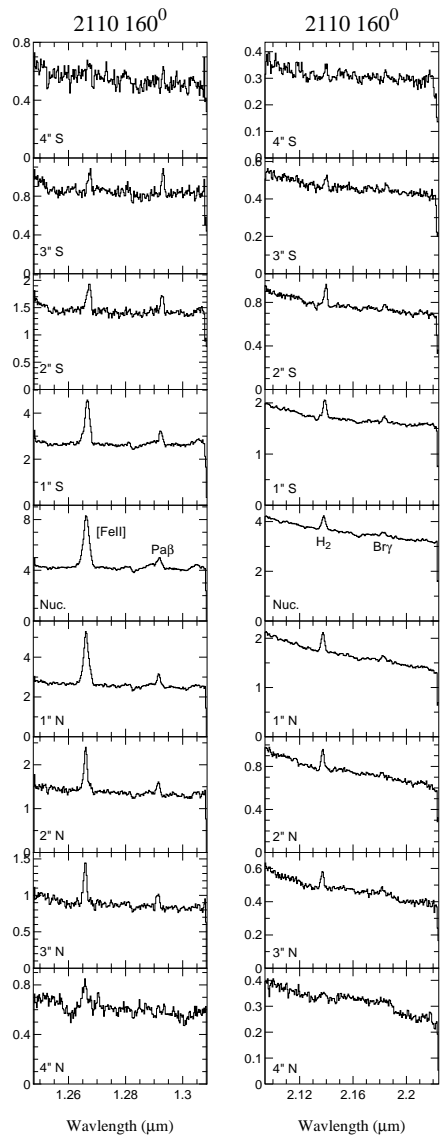}
\caption[Spectra of NGC~2110]{Spectra of NGC~2110 along the 160$^\circ$
slit.  Each spectrum is produced from a $1.0\times0.6''$ rectangular
beam.  Spatial bins are adjacent.  The distance of each spatial bin from
the nucleus in a southerly direction is indicated in each panel.  Note
the strength of \feii\ and the large \feii/\pab\ flux ratio on the
nucleus.  Flux units (F$_\lambda$) are arbitrary.}\label{ngc2110spec}
\end{center}
\end{figure}
}{}

\subsubsection{Integrated Line Flux Ratios}
\label{sec:2110fluxrat}

Table~\ref{ngc2110eqw} in Appendix~\ref{sec:tabappndx} lists the
equivalent widths and flux ratios of the lines observed in NGC~2110.
The \feii/\pab\ and \htwo/\brg\ line ratios are very large on and near
the nucleus.  These flux ratios are plotted as a function of position
along the slit in Figure~\ref{ngc2110fluxrat}.  Near the nucleus, the
\htwo/\brg\ flux ratio is close to 3, locally rising at a position $2''$
north of the nucleus, and dropping off in the furthest spatial bins.
The \feii/\pab\ flux ratio peaks on the nucleus, falling rapidly with
radius away from the nucleus.

\ifthenelse{\boolean{ispreprint}}{
\begin{figure}[htb]
\begin{center}
\epsscale{1.0}
\plotone{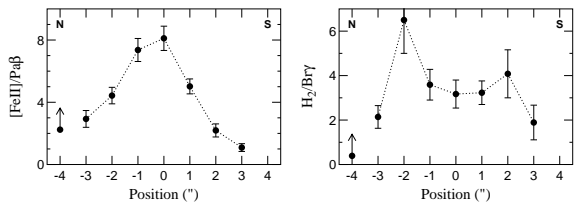}
\caption[NGC~2110 Flux Ratios]{NGC~2110 flux ratios as a function of
position along the 160$^\circ$ slit.  Positive positions are to the south.}
\label{ngc2110fluxrat}
\end{center}
\end{figure}
}{}

For most Seyfert galaxies, values of \feii/\pab\ typically range between
0.5 and 2.0 \citep[e.g.,][]{goo94,sim96b}, while the
\htwo~$\nu$=1--0~S(1)/\brg\ flux ratio typically lies between 0.5
and 1.5 \citep[e.g.,][]{moo88}.  The flux ratios observed in NGC~2110 are
significantly higher than these, with the \feii\ line in particular
being unusually strong in comparison to the \pab\ line.  Indeed, the
value of 8.1 for the \feii/\pab\ flux ratio on the nucleus is higher
than that observed in any published Seyfert galaxy spectrum
\citep[e.g.,][]{goo94,for93,mou93}.  This high \feii/\pab\ ratio is due to
both a strong \feii\ line and a weak \pab\ line in comparison to other
Seyfert galaxies. The equivalent width of the \feii\ line on the
nucleus, 25$\times10^{-3}\mu$m, is larger by a factor of 2-4 than that
typically seen in Seyfert galaxies, although it is not the largest known
equivalent width of \feii\ in a Seyfert 2 galaxy \citep[e.g.,
NGC~449;][]{kno96supp}.

\subsubsection{Velocity structure}
\label{sec:2110velocitystruc}

In Figure~\ref{ngc2110lv}, we show a set of position versus velocity plots
for each of the four near-infrared lines seen in NGC~2110.  A smoothed
continuum has been subtracted from the data in each case.  Velocity is
plotted relative to  the systemic velocity of NGC~2110.

\ifthenelse{\boolean{ispreprint}}{
\begin{figure}[htb]
\begin{center}
\epsscale{1.0}
\plotone{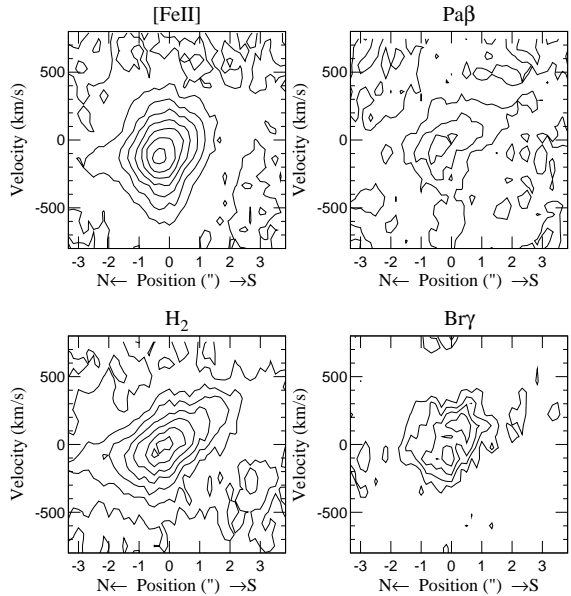}
\caption[NGC~2110 Longitude-Velocity Contour Plots]{Velocity versus position
along 160$^\circ$ slit for NGC~2110. Velocity is relative to 2340~\kms, the
systemic velocity of NGC~2110.}
\label{ngc2110lv}
\end{center}
\end{figure}
}{}

All of the lines in Figure~\ref{ngc2110lv} show a gradient with velocity
increasing from north to south.  Unlike Mk~1066, the velocity gradients
appear to be quite similar for the four infrared lines seen in NGC~2110.
Figure~\ref{ngc2110vvspos} is a plot of the velocity centroid of each
line profile as a function of position for each of the four lines.  The
centroids for \pab\ have excluded the flux due to the contamination from
\hei\ and \feii\ on the blue side of \pab\ (as discussed in
section~\ref{sec:mk1066}).  Table~\ref{ngc2110kinematics} in
Appendix~\ref{sec:tabappndx} lists these velocity centroids, as well a
the full widths at half maximum of each of the line profiles.  The
velocity structure in Figure~\ref{ngc2110vvspos} is dominated by the
gradient from north to south.  \feii\ appears different from the other
species in Figure~\ref{ngc2110lv} because it is much wider on the
nucleus than it is off of the nucleus.  The shape of the contours
convolves the spatial variation of the line width and the gradient in
the line's velocity.  The slope of the gradient in all species is
steeper to the south than to the north (Figure~\ref{ngc2110vvspos}),
reaching velocities of about $-$100 to $-$150~\kms\ at 3$''$.  South of
the nucleus and $+$250 to $+$300~\kms\ at 3$''$ South of the Nucleus.

\ifthenelse{\boolean{ispreprint}}{
\begin{figure}[htb]
\begin{center}
\epsscale{1.0}
\plotone{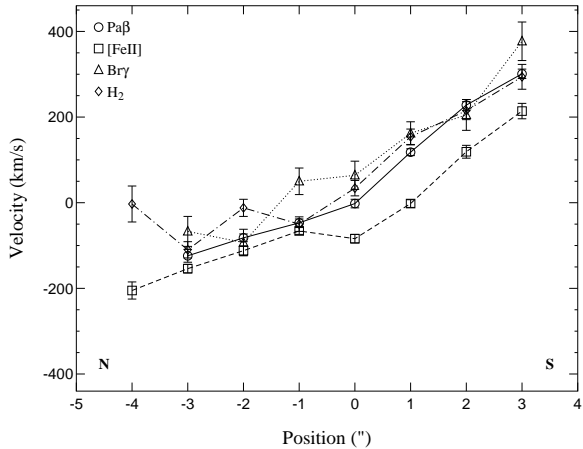}
\caption[NGC~2110 Position Versus Velocity]{Velocity centroid of the whole
line profile versus position for each line at each position along the
160$^\circ$ slit.  Velocity is relative to 2340~\kms, the systemic velocity
of NGC~2110.}
\label{ngc2110vvspos}
\end{center}
\end{figure}
}{}

A second feature apparent in Figure~\ref{ngc2110vvspos} is that the \feii\
line is systematically at a lower velocity than the other lines.  This
effect is strongest south of the nucleus, although here the strong blue
wings in \feii\ pull the centroids to lower velocities.  In each spatial bin
the velocity peak of the \feii\ line is shifted to the blue with respect to
the other lines by about 50--100~\kms, which is several times the
uncertainty in the wavelength calibration.

In many cases, the lines are asymmetric, although the asymmetries are mostly
small enough that the velocity centroids are very close to the velocities of
the peaks of the lines.  On the nucleus and 1$''$ south of the nucleus,
\pab, \feii, and \htwo\ show a blue wing in places, while north of the
nucleus, \brg, \feii, and \htwo\ show red wings.

The \feii\ line is significantly broader than the other lines, particularly
close to the nucleus (see Figure~\ref{ngc2110lv}).  On the nucleus, the FWHM
of the \feii\ line is 500~\kms, as compared to 340~\kms\ for the \pab\ line.
In addition, the \feii\ line has wings with FWHM $\sim850$~\kms\ which are
not visible in the \pab\ line.  The \feii\ line is narrower at larger radii,
having a FWHM of $<400$~\kms, 3$''$ away from the nucleus.  At 3$''$ south
of the nucleus, the \feii\ line remains $\lesssim100$~\kms\ broader than the
\pab\ line, while north of the nucleus the line widths are comparable.

\subsubsection{Comparison with Previous IR Spectroscopy}
\label{sec:ngc2110ircompare}

\citet{sb99} (hereafter cited as S99) performed spatially extended
infrared spectroscopy of NGC~2110 with resolutions of $\sim700$ and
$\sim2000$.  They have data along two slit angles: one at 170$^\circ$,
similar to our slit angle of 160$^\circ$, and one at a perpendicular
angle of 80$^\circ$.  The features of their spectra along the
170$^\circ$ slit are also seen in this work, including the notable
increase in the \feii/\pab\ ratio on the nucleus.  Any quantitative
differences between the results of the two studies are easily explained
by differences in slit widths and seeings.  One exception is the slope
of the continuum in the 2.15$\mu$m spectra, where the slope seen in this
work is opposite that seen in S99.  This is probably also a result of
differences in the sampled spatial region, but may be partially a result
of differences in data reduction, indicating the difficulty in absolute
calibration of the continuum in infrared spectra.  Since this work is
primarily concerned with the spatial and velocity distribution of
emission line radiation, it is important to consider how these
differences affect calculated line flux ratios.  For the differences
seen between the two nuclear continuum spectra, the \htwo/\brg\ flux
ratio should not be misestimated by more than $\sim$15\%, comparable to
our quoted error on the nuclear flux ratio.  S99 saw a nuclear
\htwo/\brg\ flux ratio of 3.0$\pm$2.0, consistent with our value of
3.2$\pm$0.6; this indicates that the flux ratios observed in each case
are in fact robust.

Along the perpendicular 80$^\circ$ slit, S99 have data in a wavelength
range including \brg\ and \htwo\ (\lhtwo).  \htwo\ is well detected to a
greater radius along this 80$^\circ$ slit, which is perpendicular to the
ionization cones.  Although S99 have weak or no detection of \brg\ off
of the nucleus in either direction, the S/N of the off-nuclear \htwo\
detection is better along the 80$^\circ$ slit than it is along the
170$^\circ$ slit.  This, coupled with our observations of the
\htwo/\brg\ ratio along our 160$^\circ$ slit, suggests that also as with
Mk~3, the \htwo/\brg\ ratio off of the nucleus is larger in the slit
perpendicular to the ionization cone than it is along the cone axis.

\subsubsection{Comparison With Optical Spectra and Radio Morphology}
\label{sec:ngc2110compare}

S99 report the detection of a broad \pab\ line in NGC~2110.  This
line has a width of 1200~\kms, and is highly asymmetric to the blue.
Although the same spectral feature is present in the data here (see the
nuclear spectrum in Figure~\ref{ngc2110spec}), we interpret the enhanced
blue wing as contamination from \hei\ and \feii\ (see
section~\ref{sec:mk1066}), rather than as evidence for high velocity
blueshifted \pab\ emission.  The \feii~($\lambda=1.2788\mu$m) line which
contributes to this contamination is generally only $<0.1$ the strength of
the \feii\ (\lfeii) line \citep{goo94}.  However, given that \feii\ (\lfeii)
is $\sim8$ times as strong as \pab\ on the nucleus of NGC~2110, it is likely
that the \feii\ ($\lambda=1.2788\mu$m) line contributes more significantly
to the emission blueward of \pab\ than is the case in most Seyfert~2
galaxies.

The velocity structure seen in H$\alpha$, H$\beta$, and [OIII] \citep{wil85}
is quite similar to the velocity structure seen in the infrared lines
plotted in Figures~\ref{ngc2110lv} and \ref{ngc2110vvspos}.
\citet{wil85} and \citet{wil85_2} describe this velocity field as a
rotating disk with a projected major axis of 161$^\circ$.  (This paper
only presents infrared data for NGC~2110 along one slit position angle.
Although the rotation identified by \citet{wil85_2} is evident in this
data, the lack of data along two spatial dimensions precludes any
independent conclusions about the major axis of the rotation.)

Near the nucleus, the [OIII] line has a higher root mean square (RMS)
width (equivalent to 1$\sigma$ for a Gaussian line profile) than the
H$\beta$ line by 100-200~\kms\ \citep{wil85}, although the higher [OIII]
signal-to-noise allows weak wings to be visible in the former which are
not visible in the latter.  South of the nucleus, [OIII] shows a
tendency toward enhanced blue wings, similar to what is observed in
\feii, and to a lesser extent in \htwo.  These enhanced wings show some
spatial similarities with the 2~cm emission \citep{ulv84a}, which shows
three peaks aligned along a roughly north-south axis.  The strongest
peak is on the nucleus.  The blue wings are strongest in \feii, \htwo,
and [OIII] 2$''$ south of the nucleus, which is the approximate position
of one of the secondary radio peaks.  The other off-nuclear radio peak
is 2$''$ north of the nucleus, where we see a red shoulder in \feii\ and
\htwo.  This suggests that, as with Mk~1066, there may be two
kinematical components: a rotating disk, and a second velocity field,
possibly an outflow associated with the radio jets (see below).  Note
that even though the \emph{projected} position angle of these outflowing
jets appears similar to the major axis of the rotation identified by
\citet{wil85_2}, the two physical systems need not be aligned in space.

\subsubsection{\feii\ and \htwo\ Emission Processes}

The fact that the \feii/\pab\ ratio is very large on the nucleus and rapidly
drops with radius suggests that the \feii\ emission is associated with the
Seyfert nucleus itself.  There is indirect evidence that the physical
process responsible for the \feii\ emission in NGC~2110 is X-ray heating and
ionization.  NGC~2110 has a 2--10~keV X-ray luminosity which is an order of
magnitude greater than that typical for a Seyfert~2 galaxy.  On the nucleus,
the \feii/\pab\ ratio is also nearly an order of magnitude greater than the
typical value for Seyfert~2 galaxies.  Out to radii 3--4$''$ north of the
nucleus, the \feii/\pab\ ratio remains atypically large.  About $\sim4''$
north of the nucleus, at the position of the secondary 0.1--2.4~keV X-ray
emission peak identified by \citet{wea95}, there is a lower limit on the
\feii/\pab\ flux ratio of $\geq2$ (due to an undetected \pab\ line).  South
of the nucleus, opposite the direction of the extended X-ray emission, the
\feii/\pab\ ratio drops to unity.

Although the \feii\ excitation may be dominated by X-ray ionization, there
is likely to be an additional contribution due to fast shocks resulting from
the interaction of outflowing radio plasma with surrounding gas.  This would
explain the enhanced red and blue wings respectively north and south of the
nucleus which are spatially coincident with the radio peaks seen by
\citet{ulv84a} at 1.49~GHz.

Nuclear X-ray heating of molecular clouds may also be the primary mechanism
responsible for the \htwo\ emission, although the enhanced \htwo\ emission
is not as dramatic as is the case with \feii\ in this galaxy.  The
\htwo/\brg\ ratio in NGC~2110 is large for Seyfert galaxies, though the ratio
is not strongly peaked on the nucleus as is the case for the \feii/\pab\
ratio.  The \htwo/\brg\ flux ratio is nearly constant over the 6$''$ (900pc)
where the lines are detected.  In contrast to Mk~1066, the presence of the
wings in the \htwo\ line profiles indicates that there may be warm molecular
gas in the outflowing clouds.

\subsubsection{Summary}

\begin{enumerate}
\item
In NGC~2110, the \feii/\pab\ flux ratio is unusually large, reaching a
maximum value of 8.1 on the nucleus and dropping off rapidly with radius.
The \htwo/\brg\ ratio is also large, being $\gtrsim3$ within 2$''$ of the
nucleus.

\item
In general, the infrared lines show a smooth velocity gradient similar to
the velocity gradient seen in optical emission by \citet{wil85}.

\item
The centroid of the infrared \feii\ line is systematically shifted with
respect to the other lines by about 50-100~\kms\ to the blue.  Additionally,
the \feii\ line is $\sim160$~\kms\ broader than the other infrared lines
near the nucleus.

\item
There are enhanced wings on the infrared lines.  To the south, where line
centroids are redshifted with respect to the systemic velocity, there are
enhanced blue wings.  To the north, where line centroids are blueshifted
with respect to the systemic velocity, there are enhanced red wings.  These
wings are strongest in \feii\ and \htwo\ at positions coincient with the
secondary radio peaks of \citet{ulv84a}.
\end{enumerate}

We suggest that the \feii\ excitation is dominated by nuclear X-ray
photoionization.  However, the enhanced off-nuclear wings which are
spatially coincident with the nonthermal 2cm radio peaks indicate that there
is some emission associated with outflowing gas.  Fast shocks may contribute
to the \feii\ emission at these locations.

While the \htwo\ emission is likely to be enhanced by the strong X-ray flux
in NGC~2110, the distribution of the warm molecular gas is not as centrally
peaked as is the \feii\ emission.  This may indicate a large molecular
reservoir surrounding the nucleus.  In addition, the wings seen in the
\htwo\ line profile suggest that at least some of the \htwo\ emitting gas
takes part in the outflow driven by radio jets.


\subsection{NGC 4388}
\label{sec:4388}

NGC 4388 is a highly inclined spiral galaxy at a systemic velocity of
2525~\kms\ \citep{pet93}.  Although its Hubble type is uncertain
\citep{dev91}, \citet{cor88} classify it as SB(s)b~pec.  It is generally
considered to be a member of the Virgo cluster \citep{hel79}, implying
that the distance to the galaxy is probably $\sim$14~Mpc \citep{pie88}.
This yields a spatial scale of 68~pc~arcsec$^{-1}$.  Optical narrowband
imaging \citep{cor88} shows an H$\alpha$+[NII] extent of $>40''$ east and
west of the nucleus, aligned with the major axis of the galaxy at
position angle $\sim90^\circ$.  This emission originates in the spiral
arms of the galaxy, and is probably due to HII region complexes.  Close
to the nucleus on a spatial scale of $\sim5''$, the [OIII] emission is
extended about the nucleus in a halo-like distribution that has a major
axis at a position angle of $\sim30^\circ$.  High resolution radio maps
at 4.86~GHz show emission which is elongated southwest of the nucleus at
a similar position angle of 200$^\circ$ \citep{hum91}.  At 4.86~GHz,
NGC~4388 is double peaked, with a primary peak on the nucleus and a
secondary peak 2.5$''$ southwest of the nucleus.  There is a weak
tertiary peak about 1$''$ northeast of the nucleus.  NGC~4388 has also
been detected in X-rays, with a 2-10~keV luminosity of
$\sim2\times10^{42}$~erg~s$^{-1}$ \citep{han90}.  The soft (0.1-2.4~keV)
X-ray emission of NGC 4388 is extended over $\gtrsim15''$ \citep{mat94}.

Because H$\alpha$ emission is extended over $>40''$ along a position
angle of 120$^\circ$ \citep{cor88}, we obtained a ``sky'' frame distant
(200$''$ east) from the center of the galaxy.  Subtracting this sky
frame from spectra of the galaxy verified that no infrared emission
lines were detected at a radius greater than $5''$ in this direction to
a 3$\sigma$ flux limit of $\sim2\times10^{-16}$~erg~cm$^{-2}~$s$^{-1}$
in a $1.0''\times0.6''$ spatial beam.

Figures~\ref{ngc4388spec30} and \ref{ngc4388spec120} are one-dimensional
projections in various spatial bins of the data along the slits oriented
at position angles of 30$^\circ$ and 120$^\circ$ respectively.  The size
of each spatial bin used in the extractions is 6 pixels (1$''$),
slightly larger than the psf at the time of the observations, which was
0.7--0.8$''$ for the spectra at both position angles.  The nuclear
spectra for the two slit position angles show qualitative differences,
which are likely attributed to vagaries of spatial sampling with the
rectangular synthetic aperture.  The 30$^\circ$ slit was chosen to align
with the major axis of the distribution of [OIII] emission near the
nucleus \citep{cor88}.  The 120$^\circ$ slit was chosen to be
perpendicular to the [OIII] axis.

In NGC~4388, the \feii, \pab, \brg, and \htwo\ lines are detected with a
spatial extent up to 8$''$ (540~pc) along the 30$^\circ$ slit.  Along
the perpendicular 120$^\circ$ slit, the lines diminish quickly with
distance from the nucleus, but are still visible out to a radius of
2--3$''$.  In addition to these lines, there is a narrow line seen on
the nucleus blueward of \feii, which we identify as the fine structure
line \six\ at a rest wavelength of 1.25249$\mu$m \citep{kuh96}.
\citet{oli94} identify this line as \six\ in the Circinus galaxy.
However, \citet{sim96b} and \citet{tho95} identify this line as \hei\
($\lambda=1.2528\mu$m), with possibly some contamination from \six\ in
NGC~1068 and NGC~4151 respectively.

The arguments in favor of this line being primarily \six\ are threefold.
First, the \hei\ line is not strong enough to account for the observed flux in
the objects where this line is seen.  Recombination cascade calculations
\citep{rob68} indicate that the strength of \hei\ ($\lambda=1.2528\mu$m) should
be $\lesssim20$\% of the strength of the \hei\ ($\lambda=1.2785\mu$m) line.
The latter line, blended with \hei\ ($\lambda=1.2791\mu$m) and \feii\
($\lambda=1.2788\mu$m), comprises emission which is frequently seen as a
weak blue wing on the \pab\ line profile.  However, when seen (see also
\citet{kno96supp}), the line blueward of \feii\ is of comparable strength
or stronger than the blend blueward of \pab, arguing against its
identification as \hei.

Secondly, the observed wavelength centroid is more consistent with \six\
than with \hei.  If the line is identified as \six, it is blueshifted by
$\sim100$~\kms\ relative to \pab.  If the line is \hei, its blueshift
would be 174~\kms.  Since the ionization potential of He is only
24.5~eV, we would expect He to show a similar spatial and velocity
structure to the H$^+$ lines.  S$^{7+}$, on the other hand, has a much
larger ionization potential of 328~eV~\citep{crc}, and may arise from
very different physical regions than does most of the \pab\ and \brg\
emission.

Finally, the line we identify as \six\ is only seen in a small number of
galaxies for which infrared spectra are available \citep{kno96supp}.  A \hei\
line would be expected to be seen with a similar strength relative to \pab\
in most Seyfert galaxies, as the two lines should arise from similar
physical regions. The strength of a high ionization coronal line may vary
greatly between objects, however, due to variations in the ionization
conditions, as well as differences in extinction to the regions responsible
for the H$^+$ emission and the coronal emission.  We therefore conclude that
the line observed at a rest wavelength of 1.252$\mu$m is \six.

The \six\ line is very strong on the nucleus.  Its equivalent width of
4.5$\times10^{-4}\mu$m is nearly half that of \feii.  It is not as
spatially extended as the other infrared lines along the position angle
of either slit.  It is marginally resolved spatially along the
30$^\circ$ slit, showing a stronger flux (0.39$\pm$0.08 of the nuclear
\six\ flux) in the $\pm1''$ spatial bins than would be expected for
atmospheric seeing of 0.8$''$ (0.08 of the nuclear flux).  It is not
resolved along the 120$^\circ$ slit.

\ifthenelse{\boolean{ispreprint}}{
\begin{figure}[htbp]
\begin{center}
\epsscale{0.9}
\plotone{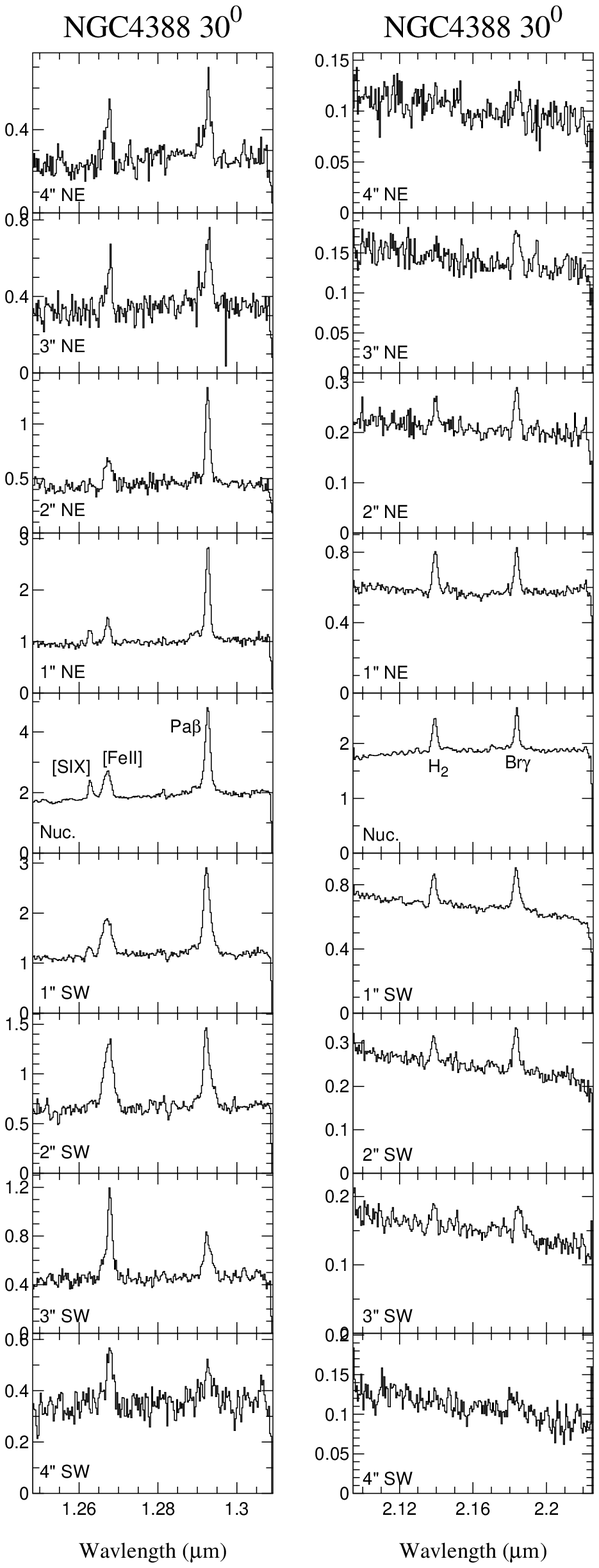}
\caption[Spectra of NGC~4388 at 30$^\circ$]{Spectra of NGC~4388 along
the 30$^\circ$ slit.  Each spectrum is within a rectangular beam which
is 1$''$ along the slit by the width of the slit ($\sim0.6''$).  Spatial
bins are adjacent.  The distance of each spatial bin from the nucleus
along a position angle of 30$^\circ$ is indicated in each spectrum.
Flux units (F$_\lambda$) are arbitrary.}\label{ngc4388spec30}
\end{center}
\end{figure}
}{}

\ifthenelse{\boolean{ispreprint}}{
\begin{figure}[htb]
\begin{center}
\epsscale{0.9}
\plotone{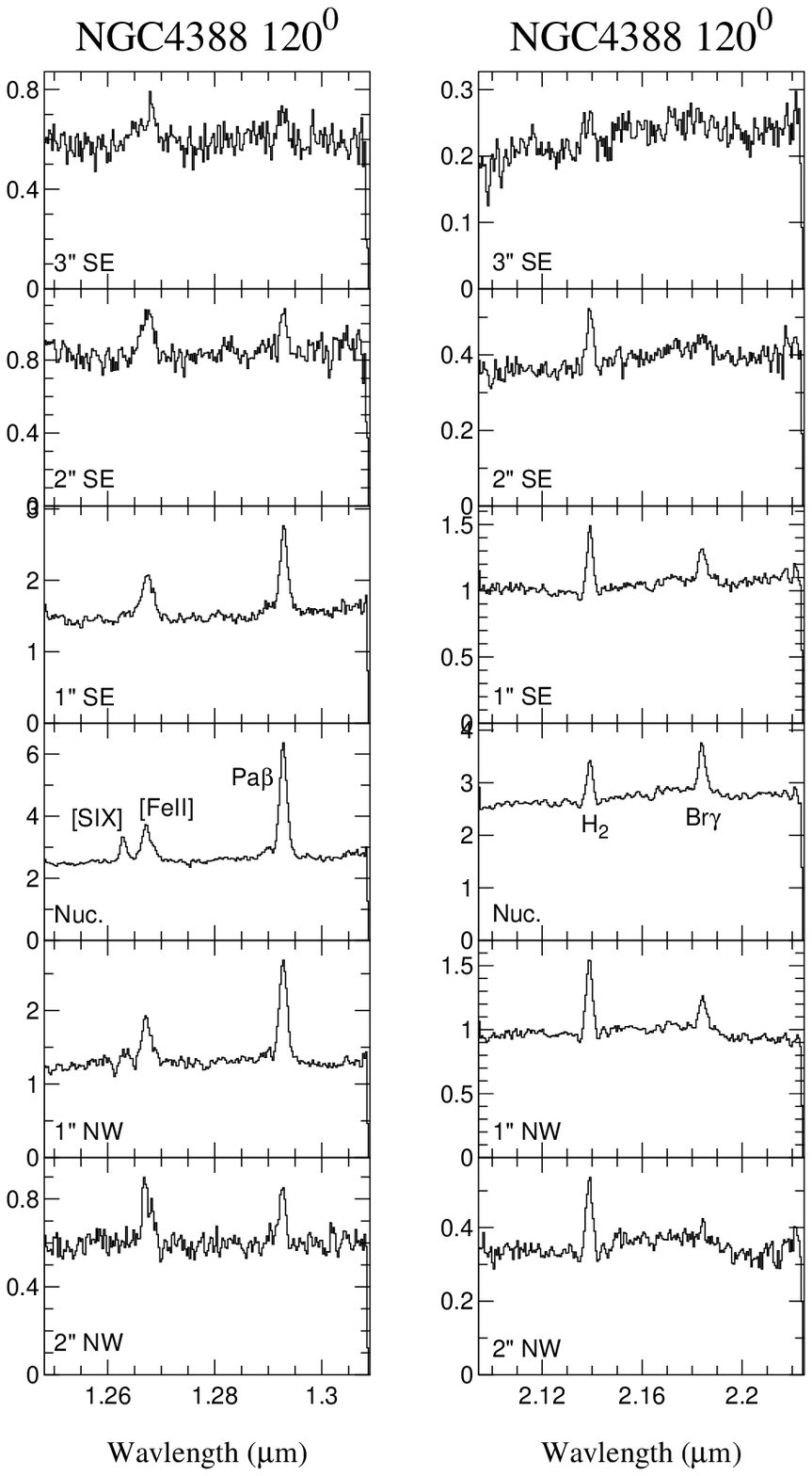}
\caption[Spectra of NGC~4388 at 120$^\circ$]{Spectra of NGC~4388 along
the 120$^\circ$ slit.  Each spectrum is within a rectangular beam which
is 1$''$ along the slit by the width of the slit ($\sim0.6''$).  Spatial
bins are adjacent.  The distance of each spatial bin from the nucleus
along a position angle of 120$^\circ$ is indicated in each spectrum.
Flux units (F$_\lambda$) are arbitrary.}\label{ngc4388spec120}
\end{center}
\end{figure}
}{}

\subsubsection{Integrated Line Flux Ratios}

Table~\ref{ngc4388eqw} in Appendix~\ref{sec:tabappndx} lists the
equivalent widths and flux ratios of the infrared lines detected in
NGC~4388 along both slits.  Figure~\ref{ngc4388fluxrats} shows the
\feii/\pab\ and \htwo/\brg\ flux ratios along both slits.  On the
nucleus, the \feii/\pab\ flux ratio is about 0.4, a relatively low value
for a Seyfert 2 galaxy; however, the \feii/\pab\ flux ratio rises off
the nucleus along both slits.  The \feii/\pab\ ratio is largest where
the \feii\ velocity centers are most discrepant from the \pab\ velocity
centers, specifically to the southwest along the 30$^\circ$ slit and to
the southeast along the 120$^\circ$ slit (see
section~\ref{sec:4388velocitystruc}).

\ifthenelse{\boolean{ispreprint}}{
\begin{figure}[htb]
\begin{center}
\epsscale{1.0}
\plotone{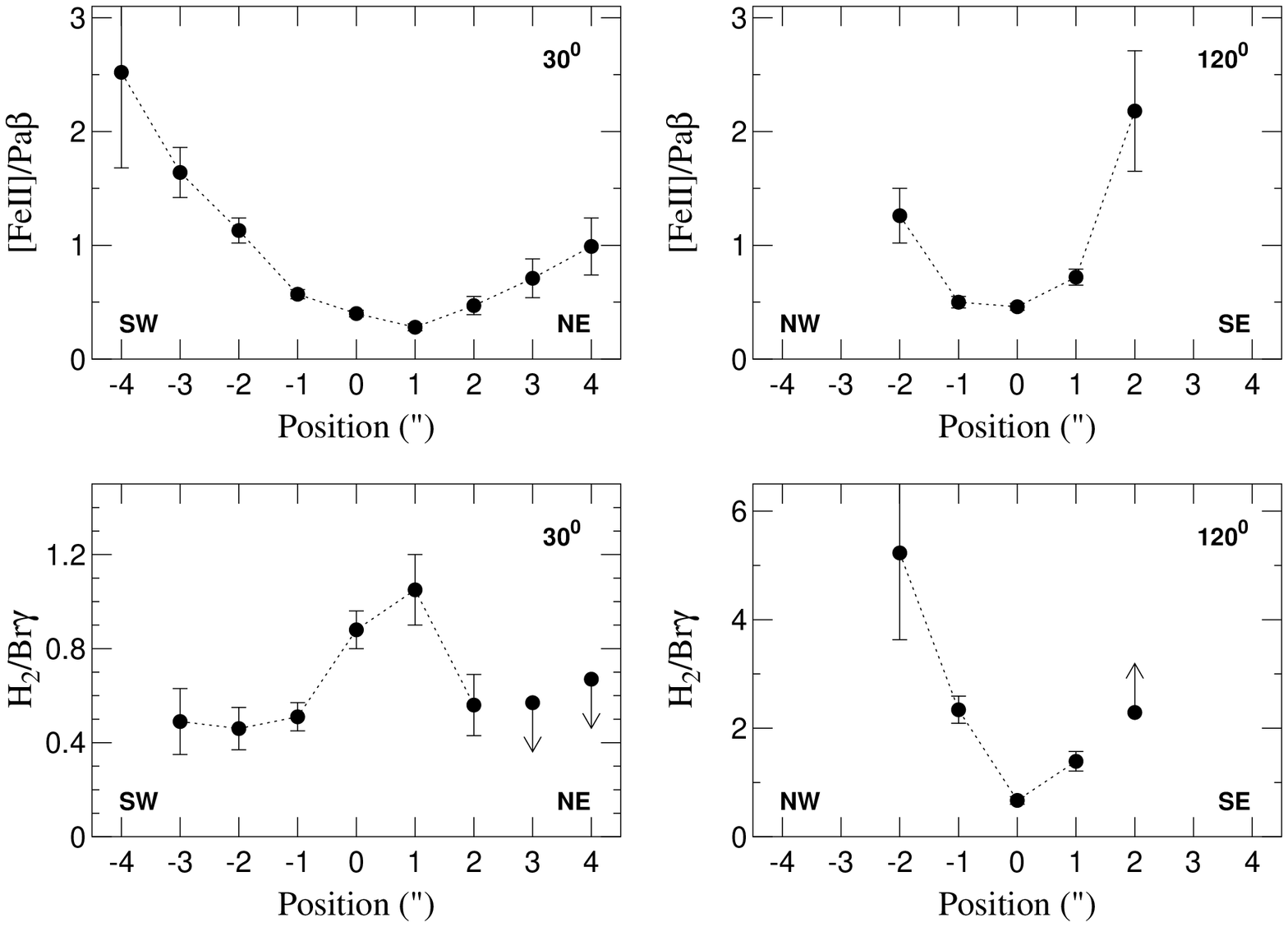}
\caption[NGC~4388 Flux Ratios]{Flux ratios for NGC 4388.  The two plots on
the left are for the 30$^\circ$ slit; the two plots on the right, for the
120$^\circ$ slit.}
\label{ngc4388fluxrats}
\end{center}
\end{figure}
}{}

Along the 30$^\circ$ slit, which is coincident with the major axis of
the optical [OIII] emission \citep{cor88}, the \htwo/\brg\ ratio is
$\lesssim$1 on the nucleus, decreasing slowly with radius.  Along the
120$^\circ$ slit, however, the flux of the \htwo\ line does not diminish
nearly as quickly as does the flux from the \brg\ line.  At a distance
2$''$ southeast and northwest of the nucleus, the \brg\ line is $<1/25$
of its nuclear strength, while the \htwo\ line is $1/8$ of its nuclear
strength (see Figure~\ref{ngc4388spec120}).  This behavior is
qualitatively similar to that seen in Mk~1066.

\subsubsection{Velocity structure}
\label{sec:4388velocitystruc}

Figure~\ref{ngc4388lv30} is set of continuum-subtracted position/velocity
contour plots for the emission lines observed along the 30$^\circ$ slit for
NGC 4388.  There is an enhanced blue wing of \pab\ on the nucleus, that is
probably dominated by a combination of \hei\ ($\lambda$=1.2786$\mu$m) and
\feii\ ($\lambda$=1.2788$\mu$m), as discussed in section~\ref{sec:mk1066}.

\ifthenelse{\boolean{ispreprint}}{
\begin{figure}[htb]
\begin{center}
\epsscale{1.0}
\plotone{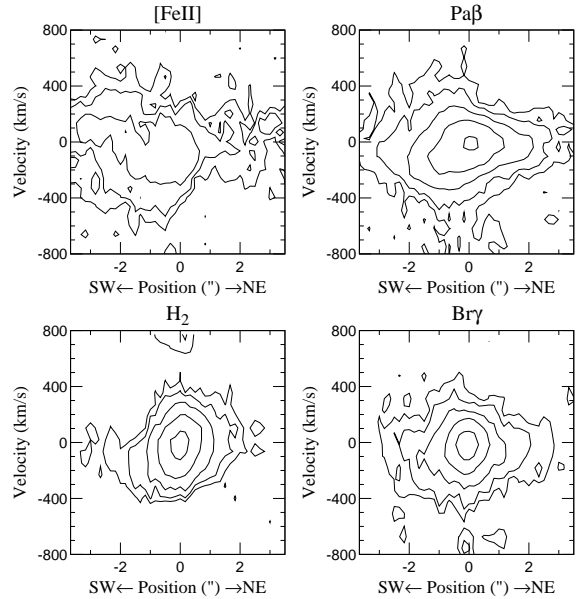}
\caption[NGC~4388 Longitude-Velocity Contour Plots]{Position versus velocity
plots for NGC 4388.  Position is along a 30$^\circ$ slit.  Velocity is
relative to 2525~\kms, the systemic velocity of NGC 4388. Contours
in this figure are logarithmic and separated by a factor of two.}
\label{ngc4388lv30}
\end{center}
\end{figure}
}{}

\ifthenelse{\boolean{ispreprint}}{
\begin{figure}[htb]
\begin{center}
\epsscale{1.0}
\plotone{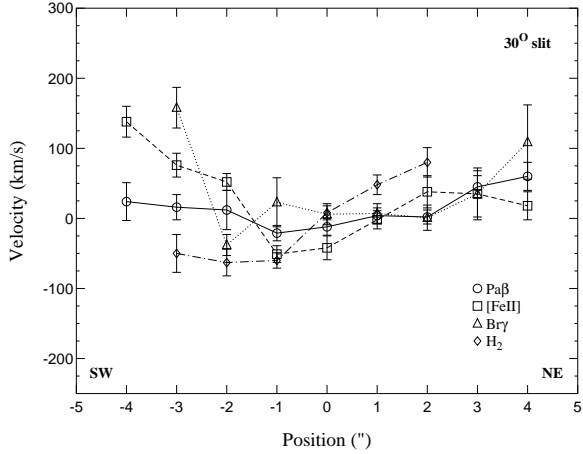}
\caption[NGC~4388 Position Versus Velocity at 30$^\circ$]{Velocity centroid
of each line versus position along the 30$^\circ$ slit for NGC~4388.
Positive distance is to the northeast.}
\label{ngc4388vvspos30}
\end{center}
\end{figure}
}{}

Table~\ref{ngc4388kinematics} in Appendix~\ref{sec:tabappndx} lists the
centroid and width of each line in each spatial bin along both slits.
Figures~\ref{ngc4388vvspos30} and \ref{ngc4388vvspos120} show the
velocity centroids of the line profiles as a function of position along
the 30$^\circ$ and 120$^\circ$ slits respectively.  Northeast of the
nucleus along the 30$^\circ$ slit, the \pab, \brg, and \feii\ lines
agree in velocity centroid, and show only a small increase in velocity
with distance from the nucleus.  The \htwo\ line is slightly redshifted
relative to the other lines.  Southwest of the nucleus, however, the
lines show very different velocity structures.  The \pab, \brg, and
\feii\ lines turn towards increasing velocity with radius to the
southwest, while the velocity gradient of \htwo\ flattens.  The \feii\
line curves back more sharply than the \pab\ line, until its peak is
about 100~\kms\ redward of \pab\ 3$''$ south of the nucleus.
Figure~\ref{ngc4388closeups} shows profiles of the \pab\ and \feii\
lines southwest of the nucleus, and indicates that the difference
between \feii\ and \pab\ is not the result of errors in velocity
measurements, but that the centroid and profile of \feii\ is genuinely
redshifted relative to \pab.  \brg\ also differs from \pab\ southwest of
the nucleus, increasing in velocity closer to the nucleus than does
\pab.

Along the 120$^\circ$ slit, the infrared line fluxes drop off faster with
radius than they do along the 30$^\circ$ slit.  Along this slit, there is no
evidence for significant velocity changes in the \pab\ line.  Where it was
detected, the velocity peak of the \brg\ line is marginally consistent with
that of the \pab\ line, as is the \feii\ line everywhere except for 1$''$
southeast of the nucleus, where \feii\ is blueshifted by $80$~\kms\ relative
to \pab.  The \htwo\ line is always blueshifted relative to the \brg\ line
by 50--100~\kms.

\ifthenelse{\boolean{ispreprint}}{
\begin{figure}[htb]
\begin{center}
\epsscale{1.0}
\plotone{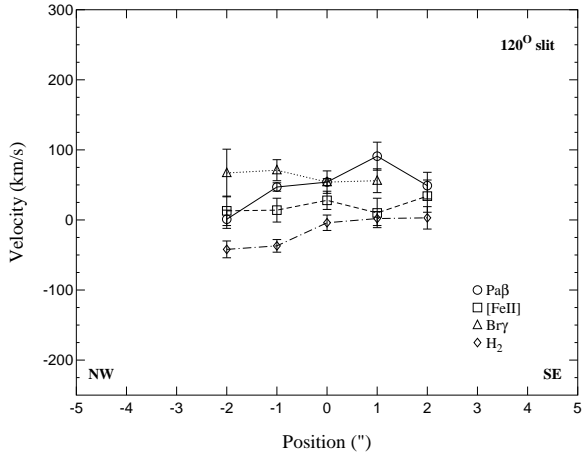}
\caption[NGC~4388 Position Versus Velocity at 120$^\circ$]{Velocity centroid
of each line versus position along the 120$^\circ$ slit for NGC~4388.
Positive distance is to the southeast.}
\label{ngc4388vvspos120}
\end{center}
\end{figure}
}{}

In addition to different variations in velocity centroid with position,
the lines have differing profiles.  Along the 30$^\circ$ slit, not only
is \feii\ generally wider than \pab, and at a different velocity
centroid southwest of the nucleus, but the asymmetries in the lines are
different.  For instance, 2--3$''$ southwest of the nucleus the
\feii/\pab\ ratio is preferentially enhanced at redshifted velocities
(Figure~\ref{ngc4388closeups}).  These asymmetries suggest that systems
of clouds with different flux ratios at different velocities are
superimposed along the line of sight.  Along the 120$^\circ$ slit, the
\feii\ line has an enhanced red shoulder to the northwest of the
nucleus.  Again, the \pab\ line is redshifted relative to the \feii\
line, so this red shoulder may correspond to a component which is
associated with the \pab\ peak.

\ifthenelse{\boolean{ispreprint}}{
\begin{figure}[htb]
\begin{center}
\epsscale{1.0}
\plotone{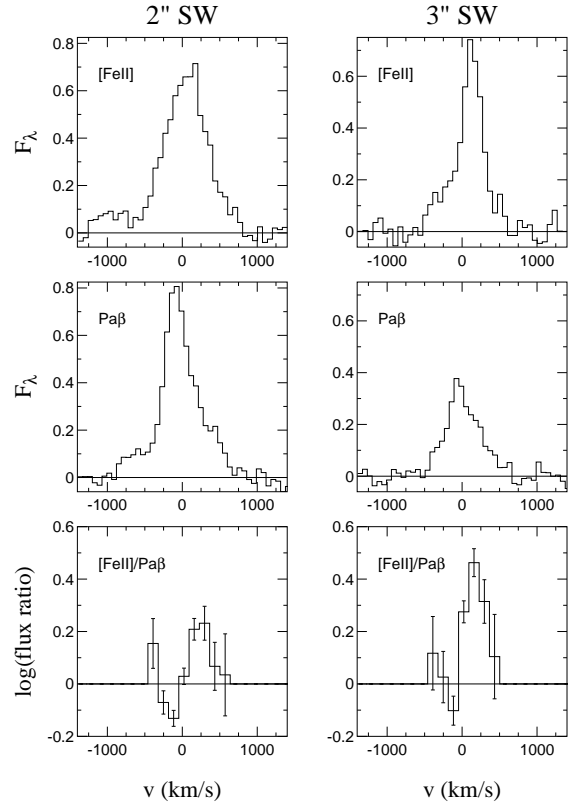}
\caption[NGC~4388 Line Profiles]{Line profiles of \feii\ and \pab\ in
NGC~4388 2-3$''$ southwest of the nucleus on the 30$^\circ$ slit, and the log
of the \feii/\pab\ ratio as a function of velocity.  Points with greater
than 50\% uncertainty in the flux ratio have been omitted from the latter
plot.}\label{ngc4388closeups}
\end{center}
\end{figure}
}{}

\subsubsection{Comparison with other observations}

In their nuclear spectrum which was marginally resolved spatially along
a 13$^\circ$ slit, \citet{win00} observed \pab\ and \feii\ with similar
profiles.  This is in contrast to our data, where \feii\ is clearly
wider than \pab\ near the nucleus along a 30$^\circ$ slit.  Narrowband
optical imaging and spectroscopy \citep[e.g.][]{cor88,pet93}, radio
imaging \citep{hum91}, as well as this study and that of
\citeauthor{win00} indicate that NGC~4388 is a complicated system with
much spatial structure at all wavelengths.  Consequently, the
differences in observed line profiles almost certainly arise from
differences in the spatial region sampled due to slit sizes and angles
as well as seeing conditions.  Similarly, \citeauthor{win00} observed a
\feii/\pab\ flux ratio of $\sim$0.7, in comparison to our nuclear ratio
of 0.40$\pm$0.03.  Since we observe a rapidly changing flux ratio with
position, the difference between the two measurements of the nuclear
ratio may also be easily ascribed to the different spatial region
sampled.

%
%
\citet{fil85} report a faint broad (Full Width at ``Zero Intensity'' of
6000~\ifthenelse{\boolean{ispreprint}}{\linebreak}{}\kms) H$\alpha$ line
on the nucleus, while \citeauthor{shi88} (1988; 1996) report the
detection of a broad line in scattered H$\alpha$ flux within 5$''$ of
the nucleus.  Figures~\ref{ngc4388spec30} and \ref{ngc4388spec120} show
no convincing evidence of a broad Pa$\beta$ or Br$\gamma$ line.  The
upper limit on the flux of a broad (FWHM$\sim2700$\kms) \pab\ line on
the nucleus is $<3\times10^{-15}$~erg~cm$^{-2}$~s$^{-1}$, corresponding
to an upper limit in its equivalent width of $1.6\times10^{-3}\mu$m.

\citet{pet93} report on optical spectroscopy in slits oriented at a
number of position angles, including two angles, 0$^\circ$ and
60$^\circ$, which bracket our 30$^\circ$ slit.  To the north and east,
within 5$''$ of the nucleus, the optical lines show a shallow gradient
to the red with all line velocities within 100~\kms\ of the systemic
velocity, consistent with the infrared emission lines.  Optically, this
gradient continues to the south and west, with lines more distant from
the nucleus at lower velocity, in contrast to the increasing velocity
seen in infrared \brg\ and \feii\ emission.

Based on the offset between the centroid of the IR continuum image of
NGC~4388 obtained with the spectrometer in imaging mode and a star from
the Hubble Guide Star Catalog 3$'$ away, the peak of the infrared
continuum emission corresponds to the brighter, northeastern radio peak
of two peaks separated by $\sim2''$ in the 4.86~GHz map of
\citet{hum91}.  As with Mk~1066 (section \ref{sec:mk1066}), there is a
spatial correspondence in NGC~4388 between enhanced \feii\ emission and
extended radio emission.  To the southwest close to the second peak of
radio emission \citep{hum91}, the \feii/\pab\ ratio increases to values
$>1.5$ (see Figure~\ref{ngc4388fluxrats}), whereas within 1$''$ of the
nucleus the \feii/\pab\ ratio is $\lesssim0.5$.  To the northeast, the
\feii/\pab\ ratio is also higher than it is on the nucleus, but is still
not as large as the values to the southwest.  Additionally, to the
southwest the \feii\ line profiles are most disparate from the \pab\
lines profiles, where the peak of the \feii\ emission is redshifted
relative to the peak of the \pab\ emission.

\subsubsection{\feii\ and \htwo\ emission mechanisms}

The velocity field of NGC~4388 is known to be complicated.  There is a
rotating disk with major axis at position angle $\sim60^\circ$ and
inclination $\sim-70^\circ$ \citep{pet93}.  However, in addition to this,
there is blueshifted gas out of the plane of the galaxy to the north and
south, as well as other localized sites within 20$''$ where the velocities
of the gas are discrepant from pure rotation \citep{cor88}.

In the infrared spectra, we are probably observing at least two
kinematic components within 4$''$ of the nucleus of NGC~4388.  One
kinematic component is the rotating disk, which is responsible for the
velocities of the lines northeast of the nucleus, and for the velocity
of the \htwo\ line southwest of the nucleus
(Figure~\ref{ngc4388vvspos30}).  These velocities match most closely the
velocities of the optical lines which \citet{pet93} identify with the
rotation curve of the galaxy.  In addition to this rotating component,
there may be a receding outflow to the southwest, associated with the
radio emission \citep{hum91}.  The shoulders associated with this outflow
pull the \pab, \brg, and \feii\ line centroids to higher velocities than
would be the case from pure rotation.  The velocity difference is
greater for \feii\ than it is for \pab\ because of the higher
\feii/\pab\ ratio in the outflow than in the disk.  This receding
outflow has a higher \feii/\pab\ ratio than the gas in the disk of the
galaxy. \citet{col87} observed red components in the [OIII] line
profiles along a 23$^\circ$ slit; to the southwest, the mean [OIII]
velocities are redshifted from the peak [OIII] velocities by 50--80~\kms.
The \brg\ line southwest of the nucleus probably includes contributions
from both the disk and outflow components.  The velocities of \brg\ here
are redshifted relative to \pab, and relative to optical emission lines
that correspond to the rotation of the disk.  Additionally, the
\brg/\pab\ ratio is higher to the southwest than it is to the northeast
(although it is highest on the nucleus).  This suggests that the
emission associated with the outflow is viewed through substantial
extinction in the disk of the galaxy.

The \feii\ emission also increases off the nucleus along the 120$^\circ$
slit which is perpendicular to the putative outflow axis.  This increased
emission does not correspond to notable differences in emission
at other wavelengths.  It is likely that this strong \feii\ emission is
associated with the disk of the galaxy, as it shares a similar (although not
identical) velocity gradient with the other lines
(Figure~\ref{ngc4388vvspos120}).

The \htwo\ emission appears to be originating primarily in the disk of
the galaxy, as its velocity as plotted in Figures~\ref{ngc4388vvspos30}
and \ref{ngc4388vvspos120} follows closely the \pab\ and optical line
velocities identified with the rotating disk \citep{pet93}.  The
\htwo/\brg\ ratio diminishes away from the nucleus along the
direction of the radio emission, suggesting that the \htwo\ emission is
not associated with the radio jets or the outflow.  Similar to Mk~1066,
the \htwo/\brg\ ratio shows a striking increase along the slit
perpendicular to the outflow axis.  Along this direction, the \htwo\
emission appears to share a similar spatial structure to the \feii\
emission.  The ratios of both lines to hydrogen recombination lines
(Figure~\ref{ngc4388fluxrats}) and the velocity curve of both lines
(Figure~\ref{ngc4388vvspos120}) along a position angle of 120$^\circ$
share the same shape (although the \htwo\ emission increases faster
relative to \brg\ than the \feii\ emission does relative to \pab).  The
mechanism of excitation of the \feii\ and \htwo\ lines in the disk of
NGC~4388 at radii of $>$150pc are probably related.

\subsubsection{Summary}

\begin{enumerate}
\item
Strong \pab, \brg, \feii, and \htwo\ emission is seen from NGC~4388.  These
lines are observed to be extended by $\sim8''$ along the direction of
greatest extent of optical [OIII] line emission.  Infrared line emission is
seen extended by $\sim5''$ perpendicular to this direction.

\item
We identify a line at the rest wavelength of 1.252$\mu$m with the infrared
coronal line \six.  This line is only marginally spatially resolved at a
position angle of 30$^\circ$.  On the nucleus, its flux is 1/6 the flux of
\pab.

\item
The \feii/\pab\ flux ratio is lowest on the nucleus with a value of 0.4.  It
rises to values $\gtrsim2$ southwest of the nucleus where the velocity
centers of the \pab\ and \feii\ lines are most different, spatially
coincident with a secondary peak of radio emission.  The \htwo/\brg\ flux
ratio is highest ($>$2-3) away from the nucleus along the 120$^\circ$ slit,
perpendicular to the axis of the extended [OIII] emission.  Along this
direction, the \htwo/\brg\ and \feii/\pab\ ratios both increase away from
the nucleus.  Along the 30$^\circ$ slit, the two line ratios show opposing
trends, with the \htwo/\brg\ ratio increasing and the \feii/\pab\ ratio
decreasing with distance from the nucleus.

\item
Northeast of the nucleus, the infrared lines in NGC~4388 show structure
which is similar to the optical line measured by \citet{pet93}.  Southwest
of the nucleus, the \brg\ and \feii\ lines are redshifted relative to \pab\
by 50--150~\kms.

\item
In some cases where different lines have different velocity centroids,
there is substructure suggestive that the same velocity components may be
present in every line with different strengths.  Specifically, southwest of
the nucleus where the \feii\ emission is redshifted relative to the \pab\
emission, there are enhanced blue wings in \feii\ and enhanced red wings in
\pab.

\item
A possible explanation for the velocity structures seen in the infrared
lines is that there are at least two kinematical components.  Most
velocities are governed primarily by a rotating disk.  There may
additionally be an obscured redshifted outflow southwest of the nucleus.
The coincidence of strong, redshifted \feii\ emission to the southwest may
be evidence of shock-excited \feii\ emitting gas associated with this
outflow.
\end{enumerate}

\subsection{Mk~3}
\label{sec:mk3}

Mk~3 is an S0 galaxy at a systemic velocity of 4046~\kms\ \citep{whi88}.
This gives a calculated distance of 54~Mpc, which in turn yields a
spatial scale of 262~pc~arcsec$^{-1}$.  Narrowband optical imaging with
the HST in several lines include [OIII] and H$\alpha$ shows that the NLR
of the galaxy is S-shaped \citep{cap95}.  On small spatial scales of
$\sim1''$, the NLR is at a position angle of 86$^\circ$.  On larger
spatial scales of a few arcseconds, the apparent position angle of the
NLR is 113$^\circ$ \citep{pog93}.  Radio observations at 14.9~GHz show a
linear structure oriented at a position angle of 86$^\circ$
\citep{kuk93}.  There are two radio peaks, the stronger one 1.1$''$ west
of the nucleus, the weaker one 0.4$''$ east of the nucleus.  Higher
resolution maps of the source at 5~GHz resolve the linear structure into
a number of components.  In particular, the eastern lobe is resolved
into three components aligned along the axis of the source, together
with a third component coincident with the core of the galaxy.  Mk~3 is
also quite bright in X-rays, with an X-ray luminosity of
$3\times10^{42}$~erg~s$^{-1}$ (with H$_0=75$~\kms~Mpc$^{-1}$), and a
very high absorbing column of $6\times10^{23}$~cm$^{-2}$ \citep{awa90}.
The 0.1--2~keV X-ray source is unresolved with the ROSAT HRI on a scale
of $\sim4''$ \citep{mor95}.

The reduced 1.29$\mu$m and 2.21$\mu$m spectra of Mk~3 along a 113$^\circ$
slit are plotted in Figure~\ref{mk3spec}, in spatial bins along the slit
separated by 1$''$.  The position angle of the 113$^\circ$ slit was chosen
to align with the major axis of the [OIII] distribution on scales of
$\sim1''$ \citep{mul96}.  The \feii\ and \pab\ emission lines are visible in
the 1.2$\mu$m spectra, and the \htwo\ and \brg\ emission lines are visible
in the 2.2$\mu$m spectra.  Note that for these observations, the \brg\ line
was centered in the K-band spectrum so that the \htwo\ line is near the blue
edge of the spectrum.  This may lead to inaccuracies in placing the
continuum on the blue side of the \htwo, resulting in underestimates of the
flux of the \htwo\ line and any blue wings in the profile of the \htwo\
line.

\ifthenelse{\boolean{ispreprint}}{
\begin{figure}[htbp]
\begin{center}
\epsscale{1.0}
\plotone{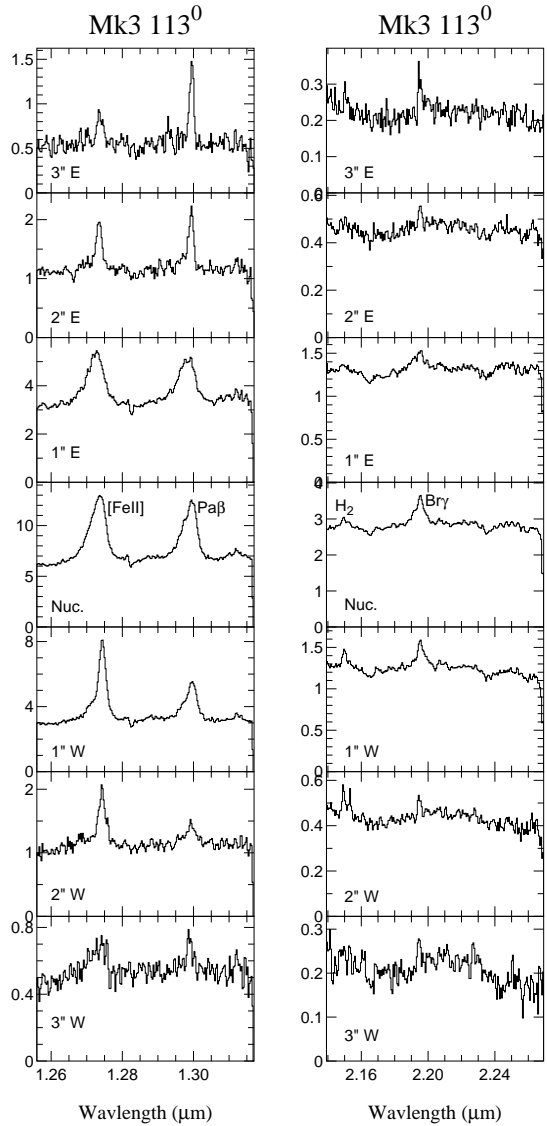}
\caption[Spectra of Mk~3]{Spectra of Mk~3 along the 113$^\circ$ slit.
Each spectrum is within a rectangular beam which is 1.0$''$ along the
slit by the width of the slit ($\sim0.6''$).  Spatial bins are adjacent.
The distance of each spatial bin from the nucleus in an easterly
direction is indicated in the spectrum.  Flux units (F$_\lambda$) are
arbitrary.}\label{mk3spec}
\end{center}
\end{figure}
}{}

The \feii, \pab, \htwo, and \brg\ lines are extended along the slit by
$\lesssim7''$ (1.8~kpc).  The equivalent widths of the lines on the
nucleus, given in Table~\ref{mk3eqw} in Appendix~\ref{sec:tabappndx},
are larger by a factor of $\gtrsim2$ than those seen in most Seyfert 2
nuclei \citep{kno96supp}.

Most of the extended line emission is characteristically narrow (FWHM$<500$
\kms), yet on the nucleus the infrared lines are quite broad (see
Figure~\ref{mk3nucprof}).  The FWHM of the nuclear lines are \brg\
$\sim$700~\kms, \pab\ $\sim$1000~\kms, and \feii\ $\sim$1150~\kms.  These
lines are also markedly asymmetric, with strong blue wings.  The peak of the
\pab\ line on the nucleus is at a velocity $\sim200$~\kms\ greater than the
centroid of the full line provile.  Although \six\ is not obvious, it could
easily be hidden within the broad blue wings of the \feii\ line profile

\ifthenelse{\boolean{ispreprint}}{
\begin{figure}[htb]
\begin{center}
\epsscale{1.0}
\plotone{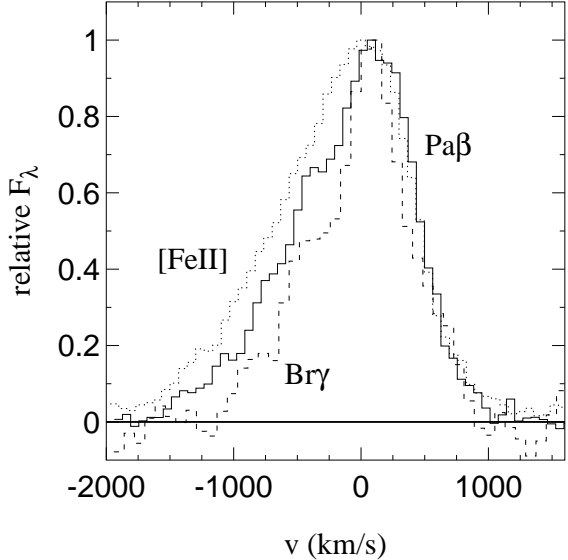}
\caption[Mk~3 Line Profiles]{Profiles of nuclear lines in Mk~3.  The lines
are normalized to the same peak flux.  The solid line is \pab, the dotted
line \feii, and the dashed line \brg.}\label{mk3nucprof}
\end{center}
\end{figure}
}{}

While the observed width of the \pab\ line on the nucleus is comparable
to the FWHM of broad \pab\ lines seen in some Seyfert 1 galaxies
\citep[e.g.,][]{kno96supp}, the observed line profiles in Mk~3 probably
do not indicate the detection of a canonical broad line region, since
\feii\ and \pab\ lines are both broad and have similar profiles.
Because the critical density of \feii\ is only
n$_\mathrm{e}\sim10^5\mathrm{cm}^{-3}$, the broad \feii\ emission cannot
be produced in a high density ($\mathrm{n_e\sim10^8-10^{10}cm^{-3}}$)
broad line region.  Some or all of the high velocity \pab\ line flux is
therefore probably coming from the NLR.

\subsubsection{Flux Ratios}
\label{sec:mk3fluxrat}

Table~\ref{mk3eqw} in Appendix~\ref{sec:tabappndx} lists the equivalent
widths and flux ratios of the lines observed in Mk~3.
Figure~\ref{mk3fluxrat} shows the flux ratios of \feii/\pab\ and
\htwo/\brg\ as a function of position along the slit.  The \feii/\pab\
ratio changes by a factor of nearly 4 along the slit.  This change is a
monotonic increase from east to west, although it is poorly determined
3$''$ west of the nucleus due to the low signal to noise ratio in the
\pab\ line.  The \htwo/\brg\ flux ratio, close to $0.3-0.4$ in all but
one spatial bin, is relatively low for Seyfert galaxies.  The
\htwo/\brg\ flux ratio is anomalously large 2$''$ west of the nucleus.
The strong, double-peaked \htwo\ emission responsible for this enhanced
flux ratio is visible in Figure~\ref{mk3spec}.  The higher velocity peak
is unresolved, and is at a velocity of 450~\kms\ relative to the
systemic velocity of Mk~3.  This peak is not well detected in every
individual image that goes into the summed spectrum of Mk~3, and so we
suspect that this peak is an artifact associated with poor background
subtraction.  Because \htwo\ at the redshift of Mk~3 falls right at the
wavelength of an atmospheric OH line, the spatial bins in which the
detection of \htwo\ is of low significance may be containmated by poor
subtraction of this line.  If this is the case, then the \htwo/\brg\
ratio would be $\lesssim1.5$, which is still higher than in the other
spatial bins, although the caveat about background subtraction applies
to the full \htwo\ profile in this spatial bin.

\ifthenelse{\boolean{ispreprint}}{
\begin{figure}[htb]
\begin{center}
\epsscale{1.0}
\plotone{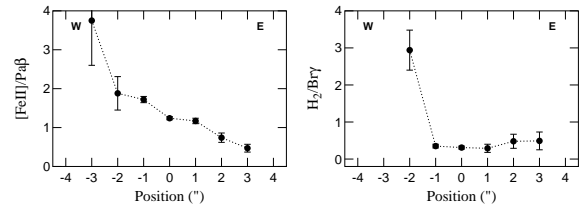}
\caption[Mk~3 Flux Ratios]{Mk~3 flux ratios as a function of position along
the 113$^\circ$ slit.  Positive positions are to the east.  Fluxes are
integrated over the entire line profile in each spatial bin.}
\label{mk3fluxrat}
\end{center}
\end{figure}
}{}

\subsubsection{Velocity Structure}
\label{sec:mk3velocitystruc}
\label{sec:mk3velsubstruc}

Figure~\ref{mk3lv} shows position versus velocity plots for each of the
four emission lines seen in Mk~3.  A smoothed continuum has been
subtracted from the data in each case.  Table~\ref{mk3kinematics} in
Appendix~\ref{sec:tabappndx} lists the velocity centroids and total full
widths at half maximum of the lines in each of the 1$''$ spatial bins
for which spectra are plotted in Figure~\ref{mk3spec}.

\ifthenelse{\boolean{ispreprint}}{
\begin{figure}[htb]
\begin{center}
\epsscale{1.0}
\plotone{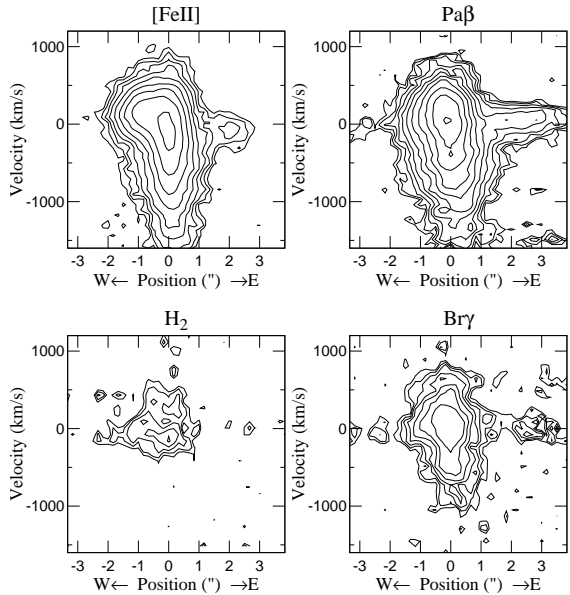}
\caption[Mk~3 Longitude-Velocity Contour Plots]{Position versus velocity
plots for Mk~3.  Position is along a 113$^\circ$ slit.  Velocity is relative
to 4046~\kms, the systemic velocity of Mk~3 \protect\citep{whi88}.  Contours
are logarithmic, separated by a factor of $\protect\sqrt{2}$.}
\label{mk3lv}
\end{center}
\end{figure}
}{}

To the east, the \pab\ line is narrow and stronger than any narrow \pab\
emission to the west.  A very shallow velocity gradient at the limit of
our resolution is marginally visible in this narrow component.  The
\pab\ line is redshifted by $\sim100$~\kms\ $\gtrsim3''$ east of the
nucleus.  The \brg\ line is not as well detected as the \pab\ line, but
appears to share a similar velocity structure, although the shallow
velocity gradient is not apparent in \brg.  The \brg\ line is slightly
blueshifted relative to the \pab\ line east of the nucleus.  The \pab\
and \brg\ line peaks appear to share a shallow velocity decrease from
east to west.  The total blueshift of the lines is $\lesssim100$~\kms\
west of the nucleus.

The velocity structure of the \feii\ line to the east is different from that
of the hydrogen recombination lines.  To the east of the nucleus and on the
nucleus, the peak of the \feii\ line is blueshifted by $\sim$100--200~\kms\
relative to the \pab\ and \brg\ lines.  West of the nucleus, the velocity of
the \feii\ line is closer to that of the hydrogen recombination lines.  The
\htwo\ line is not confidently detected east of the nucleus.  West of
nucleus by 2$''$, the \htwo\ line shows the aforementioned peculiar
double-peaked shape (see Figure~\ref{mk3spec}).

On the nucleus, the \feii, \pab, and \brg\ lines all have qualitatively
similar profiles.  All three lines show strong enhanced emission to the blue
of the peak.  However, as is visible in Figure~\ref{mk3nucprof}, the
profiles differ in detail.  In \brg, the blue wing is not as strong
relative to the peak as it is for \pab\ and \feii. The peak of the \feii\
line profile is shifted to the blue of the \pab\ line profile by 80~\kms,
and the FWHM of \feii\ is about 100~\kms\ larger than the FWHM of \pab,
apperently due to enhanced
emission on the blue wing of \feii\ (see Figure~\ref{mk3nucprof}).

While the profiles to the east of the nucleus are narrow, Figure~\ref{mk3lv}
clearly shows that 1--2$''$ west of the nucleus the lines are broader, with
the FWHM of \pab\ line equal to 500--600~\kms.  Figure~\ref{mk3closeups}
shows the profiles of the \brg, \pab, and \feii\ lines 1$''$ west of the
nucleus.  Two features are apparent from this plot.  First, the red wing of
the \brg\ line is stronger relative to the line peak than the red wing of
\pab.  This may indicate that there is receeding gas obscured behind dust,
as the longer wavelength \brg\ is less affected by extinction than \pab.
Second, the \feii/\pab\ ratio is largest with a value of $\sim2.2$ at a
redshifted velocity near $\sim150$~\kms.  The blue hump on the \pab\ and
\feii\ line profiles has an \feii/\pab\ ratio of about 1, close to the
nuclear flux ratio (see section~\ref{sec:mk3fluxrat}).

\ifthenelse{\boolean{ispreprint}}{
\begin{figure}[htbp]
\begin{center}
\epsscale{0.7}
\plotone{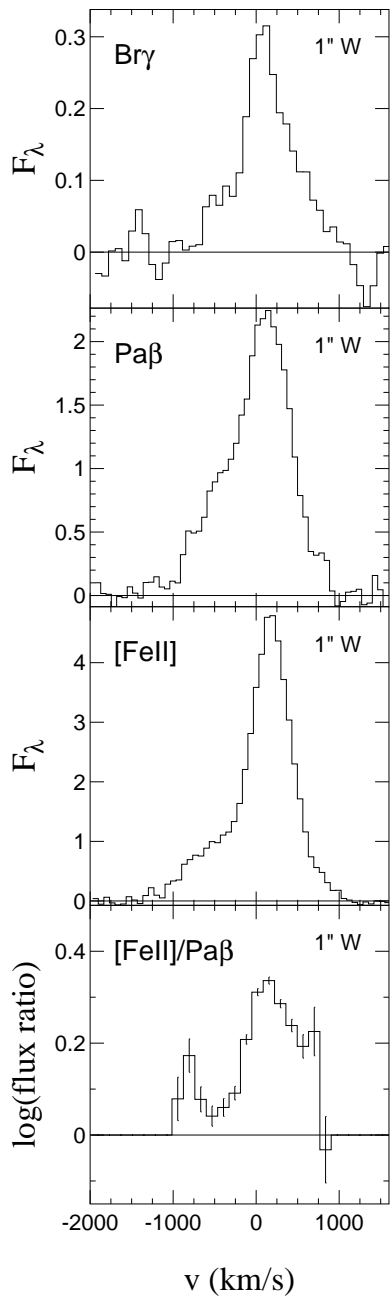}
\caption[Mk~3 Line Profiles]{Line profiles of \brg, \feii, and \pab, and the
log of the \feii/\pab\ flux ratio as a function of velocity, 1$''$ west of
the nucleus of Mk~3 at a position angle of 293$^\circ$.  Points with
uncertainty greater than 20\% have been omitted from the latter
plot.}\label{mk3closeups}
\end{center}
\end{figure}
}{}

\subsubsection{Comparison with Other Observations}

The velocity gradient seen in the off-nuclear narrow \pab\ line is generally
consistent with the rotation curve seen in optical H$\beta$ and [OIII]
emission by \citet{whi88}.  However, the optical slit position angle is
86$^\circ$, while the infrared data presented here was obtained through a
slit at a position angle of 113$^\circ$.  East of the nucleus, the [OIII]
line peaks at 120~\kms, while west of the nucleus it peaks at -120~\kms,
similar to the limiting off-nuclear velocities of $\pm100$~\kms\ in the
infrared lines.  \citet{whi88} also saw an increase in the FWHM of the
optical H$\beta$ and [OIII] lines near the nucleus, similar to that seen in
the \pab, \brg, and \feii\ lines.  This broader emission is blueshifted
relative to the peak of the lines, again consistent with that seen in the
infrared lines.  In addition, a red wing in the [OIII] and H$\beta$ profiles
is similar to the enhanced red emission seen in the \brg\ and \feii\ lines
in Figure~\ref{mk3closeups}, which was also strongest 1$''$ west of the
nucleus.  A Gaussian decomposition of the optical lines placed the relative
velocity of the redshifted component at $\sim300$~\kms; our Gaussian
decomposition of the infrared lines places the velocity of the reddened
component of the infrared lines at somewhat higher velocity, closer to
$\sim500$~\kms, although this is uncertain given the complex blended line
profiles (see Figure~\ref{mk3closeups}).

The spatial location of this red wing is coincident with the stronger of two
radio peaks seen at 2cm \citep{kuk93}.  The relatively high extinction seen
for this reddened component in the infrared \brg\ and \pab\ lines is
consistent with the picture of this radio peak as a lobe coincidental with a
receding outflow of ionized gas which is obscured behind the plane of the
galaxy.  The weaker radio lobe is 0.4$''$ east of the nucleus along the same
position angle.  \citet{whi88} do not identify any component in the optical
spectra with the eastern radio lobe.  However, the enhanced blue wings seen
on the nucleus in both the infrared and [OIII] emission lines may be part of
an approaching outflow associated with the eastern radio lobe.

\subsubsection{\feii\ and \htwo\ Emission Processes}

As with Mk~1066 and NGC~4388, and to a lesser extent NGC~2110, the \feii\
emission in Mk~3 shows spatial correlations with the radio emission which
suggests that processes involving the radio jet's interaction with
circumnuclear gas are significantly enhancing the \feii\ emission.  West of
the nucleus at the position of the western radio peak, the \feii\ line shows
enhanced redshifted emission, which is also seen in the \pab\ and \brg\
lines as well as in the optical lines \citep{whi88}.  At the position of this
radio peak west of the nucleus, the \feii/\pab\ ratio is higher than it is
to the east of the nucleus where the radio emission is much weaker.  This
enhanced redshifted emission 1$''$ west of the nucleus also has a larger
\brg/\pab\ flux ratio, suggesting higher extinction to the region which is
the source of the enhanced \feii\ emission.  This is consistent with the
picture of the region as a receding outflow which is partially obscured by
the galactic disk.

The eastern radio peak is too close (0.4$''$) to the nucleus to distinguish
infrared line emission spatially coincident with it from emission directly
from the nucleus.  However, if the two knots of radio emission represent the
radio lobes associated with a biconical outflow, the enhanced blue wings in
the infrared and optical lines on the nucleus may in fact be correlated with
the second radio lobe.  Higher resolution 5~GHz maps of Mk~3 \citep{kuk93}
resolve the eastern lobe into a number of individual linearly spaced
components.  The infrared lines near the nucleus show a complicated,
evidently multi-component structure in velocity (see Figure~\ref{mk3spec}).
The multi-component spatial nature of the radio emission and the multiple
velocity components in the infrared line emission suggest the identification
of the broad blue wings on the infrared lines near the nucleus with the
eastern radio lobe.  The \feii\ emission on the nucleus is blueshifted
relative to the \pab\ emission, suggesting that the \feii/\pab\ flux ratio
may be enhanced at the position of the eastern radio peaks.  Given this
spatial correspondence, it is likely that the \feii\ emission is enhanced by
fast shocks resulting from the interaction of the radio jet and the
circumnuclear gas.

Except for the anomalously high value 2$''$ west of the nucleus, the
\htwo/\brg\ ratio is lower in Mk~3 than it is in most Seyfert galaxies.  West
of the nucleus, the narrow \pab\ and perhaps \brg\ emission may be
associated with the rotation of a galactic disk, based on similarities with
optical lines \citep{whi88}.  At these positions, the \htwo\ emission is
suppressed relative to the \brg\ emission, indicating that Mk~3 continues
relatively little warm molecular gas.

\subsubsection{Summary}

\begin{enumerate}
\item
At a position angle of 113$^\circ$, the infrared line emission is observed
to be extended by $\lesssim7''$.

\item
There is a FWHM$\sim$1000~\kms\ blue wing visible on the nucleus of Mk~3,
most most prominent in \feii\ and \pab\, but also seen in \brg.
There is a red wing on the line profiles, arising from a region of
relatively high extinction, $\sim1''$ west of the nucleus.  This red wing
corresponds to a redshifted component seen in optical emission
\citep{whi88}.  Both wings are spatially consistent with the position of two
radio peaks \citep{kuk93}.

\item
The \pab\ and \brg\ narrow line peaks may follow a barely resolved smooth
velocity gradient which lies at $+100$~\kms\ east of the nucleus and
$-100$~\kms\ west of the nucleus.  East of the nucleus, the \brg\ line is
marginally blueshifted relative to \pab.  The \feii\ emission is blueshifted
relative to the \pab\ emission both east of the nucleus, and significantly
so on the nucleus.

\item
The \feii/\pab\ flux ratio monotonically increases from $\sim0.5$ to $\sim2$
over 5$''$ from east to west across the nucleus.  The \htwo/\brg\ flux ratio
is close to $0.3-0.4$ everywhere except where it is anamalously high,
possibly as a result of an artifact, 2$''$ west of the nucleus.
\end{enumerate}

We suggest that emission associated with an outflow of ionized gas
contributes significantly to the infrared line emission in Mk~3.  The
\feii/\pab\ ratio is enhanced in components identified with this outflow and
coincident with the 2cm radio lobes, suggesting a possible shock
contribution to the \feii\ emission.  The low \htwo/\brg\ ratio indicates
that in contrast to other Sefyert galaxies, the circumnuclear disk is
relatively depleted in warm molecular gas.

\section{General Discussion}

\subsection{\feii\ Emission Processes}
\label{sec:sey2feiisumm}

One obvious conclusion from the data presented here is that the \feii/\pab\
ratio in Seyfert galaxies varies significantly with position on scales of
$\leq1''$.  Of the four galaxies observed, each showed a different
dependence of \feii/\pab\ on position, whether it is monotonically
increasing across the nucleus (Mk~3), highest at the nucleus (NGC~2110),
lowest at the nucleus (NGC~4388), or close to constant with point-to-point
variations (Mk~1066).  These data show that flux ratios integrated over large
nuclear beams may be of limited value for drawing conclusions about the
excitation processes of the infrared lines.  The flux ratios are observed to
vary significantly on scales of $\lesssim10^2$~pc, and to show significant
velocity structure.

Circumnuclear starbursts and processes associated with the central AGN
directly have both been implicated in the generation of \feii\ emission
lines seen in Seyfert galaxies \citep[e.g.,][]{moo88,mou93}.  Our data in
general support the active nucleus as the power source of \feii\ in the
Seyfert galaxies.  Despite the differences in the details of the velocity
structure of the infrared lines, the velocity gradients of \feii\ tend to
match those of \pab\ and \brg.  The \feii\ line can generally be
kinematically associated with other spectral features linked with the
nuclear activity (e.g., [OIII]), or sites of enhanced radio emission.  The
one exception is the extended \feii\ emission along position angle
120$^\circ$ in NGC~4388.

There are two primary proposed mechanisms for \feii\ emission associated
with the AGN itself.  Because the ionization potential of Fe$^+$ is only
16.2~eV, the \feii\ line will be strong in partially ionized regions.
If clouds illuminated by the Seyfert nucleus have a high enough column
density, only the side of the cloud facing the nuclear ionizing source
will be fully ionized.  Because the ionization cross section for
high-energy X-rays is relatively small, X-rays can penetrate deeply into
these dense clouds, creating large warm partially ionized regions
suitable for \feii\ emission \citep{gra90}.  Fast shocks can also
produce partially ionized regions heated to 6000~K \citep{shu87}.
Models indicate that these shocks may also destroy dust grains
\citep{sea83}.  Because much of the iron in the local interstellar
medium is locked up in dust grains, the destruction of those grains
would enhance the gas phase Fe abundance, thereby increasing the
infrared \feii\ emission.

\citet{sim96b} report a slightly stronger correlation between infrared
\feii\ and 6cm radio emission than between \pab\ and 6cm emission.
Based on the difference in the correlations, \citeauthor{sim96b}
conclude that $\sim20$\% of the \feii\ emission in Seyfert galaxies is
due to the action of fast shocks, and that the bulk of it is due to
X-ray photoionization.

%
%
The data presented here for Mk~1066,
\ifthenelse{\boolean{ispreprint}}{\linebreak}{} NGC~2110, NGC~4388, and
Mk~3, combined with the calculations of \citeauthor{sim96b}, provide
spatial and kinematic evidence that indicate that X-ray photoionization
is responsible for most of the \feii\ emission, but that the \feii\
emission is probably enhanced in some places by a contribution from
shocks.  In most cases the \feii\ emission tends to be spatially
associated with the \pab\ emission, and the primary component of \feii\
often shares the same velocity structure as the \pab\ line.  Thus, most
of the \feii\ emission probably comes from gas physically and
kinematically associated with the NLR, as traced by the \pab\ emission.
Except for the nucleus of NGC~2110, the flux ratios observed at various
positions in the galaxies are within the range of those seen in the
spectra of \citeauthor{sim96b}.

While the bulk of the \feii\ appears to be photoionized, there are clear
signatures of local enhancements of the \feii\ emission from shocks
associated with the expanding radio plasma.  In all four of the
galaxies, spectral features were identified in the \feii\ emission line
profile which occured at or near the position of peaks in the
non-thermal radio emission.  In many cases, these features showed
relatively high \feii/\pab\ ratios.  In Mk~3 and NGC~4388, the \feii\
emission was enhanced relative to the \pab\ emission, or showed the
largest kinematical differences from the \pab\ emission, at positions
spatially coincident with radio peaks.  This suggests that processes
associated with radio jets are contributing measurably to the \feii\
emission.  The most probable explanation of this is fast shocks
resulting from the interaction of outflowing radio plasma with the
circumnuclear gas.

\subsection{\htwo\ Emission Processes}

Molecular hydrogen emission can result from ultraviolet fluorescence of
hydrogen molecules in photodissociation regions illuminated by photons from
OB stars or by a nuclear UV continuum source.  Alternatively, the \htwo\
emission may be excited thermally, through either slow (v$<25$~\kms) shocks
\citep{shu87}, or through X-ray heating of dense molecular clouds
\citep{dra90}.  Observations of the flux ratio between \htwo\ ($\nu$=1--0 S(1))
(\lhtwo) and \htwo\ ($\nu$=2--1 S(1)) (2.2471$\mu$m) rule out ultraviolet
fluorescence as a significant mechanism in most Seyfert galaxies
\citep{moo90}.

In all four of the galaxies discussed here, we have identified features
in \feii\ and hydrogen recombination line profiles which we associate
with outflows or other interactions with radio lobes and jets.  Except
for NGC~2110, these features are weaker or invisible in the \htwo\ line
profile; the kinematics of the \htwo\ are usually better explained by
galactic rotation.  In addition, in those two galaxies where we have
data perpendicular to the outflow axis (Mk~1066 and NGC~4388), and
possibly as well in NGC~2110 (S99), the \htwo\ is significantly more
extended than the atomic hydrogen recombination lines, with the
\htwo/\brg\ flux ratio increasing by a factor of $\sim$4 or more at
200--300~pc away from the nucleus.

Although X-ray heating of molecular clouds that cross into the ionization
cones can give rise to the extended \htwo\ emission along the cone axes, it
is unlikely that the nucleus is responsible for the enhanced \htwo\ emission
along the axis perpendicular to the cones.  According to the standard model
of Seyfert galaxies, the radiation from the central source is obscured
except along the axis of the ionization cones; thus, nuclear X-rays cannot
penetrate significantly into the gas perpendicular to the cone axis to
produce the extended \htwo\ emission.  While these X-rays would heat the
inner face of the obscuring molecular torus, the size of that torus is
believed to be on the order of a few parsecs, so this cannot explain the
extended \htwo\ emission that we observe. Also, if molecular hydrogen
emission does emerge from photodissociation regions heated by X-rays from a
central source, we would expect strong \brg\ emission illuminated by
ultraviolet radiation from the same central source.

Shocks driven directly or indirectly by the nucleus are another possible
explanation for the excitation of the \htwo\ emission.  Slow shocks moving
through high density molecular clouds in the disk of the galaxy could excite
molecular hydrogen emission.  Where the shocks encounter lower density
clouds, they should be faster, and produce substantial \feii\ emission.  The
\feii/\pab\ ratio does rise with distance from the nucleus, at least
initially, perpendicular to the primary direction of extended
high-excitation gas in both Mk~1066 and NGC~4388 .  Alternatively, secondary
shocks in molecular clouds circumscribing radio outflows might give rise to
enhanced \htwo\ emission.  This process might produce the differing \feii\
and \htwo\ spatial morphologies observed in NGC~1068 by \citet{bli94}.  One
problem with this scenario is that, in contrast to the axis of ioniziation
cones where outflowing radio plasma may drive shocks into the surrounding
gas, there is no obvious mechansim to drive shocks to distances of
$\sim$1~kpc perpendicular to the cone axis.

The relatively large extent of the \htwo\ emission perpendicular to the
ionization cones in Mk~1066 and NGC~4388 suggest that there are local source
of heating which is giving rise to much of the \htwo\ emission.  In both
galaxies, there may be contribution to the \htwo\ emission from young stars
heating molecular clouds.  While this may contribute to the \htwo\ emission,
it is unlikely to be the only source, as the \htwo/\brg\ ratios observed in
the extended \htwo\ gas tend to be higher than the values $\lesssim0.5$
typically observed in starburst galaxies \citep{moo88,moo90,kaw88}.  NGC~4388
has a bar oriented at position angle $\sim90^\circ$ \citep{cor88}.  Dynamical
processes associated with this bar may be creating shocks which give rise to
the observed \htwo\ emission.  In Mk~1066, it is also possible that the bulk
motions of clouds which do not move completely with the rotation of the disk
are colliding with other clouds, creating slow shocks which excite the cold
molecular gas.  In Mk~3, the lower \htwo/\brg\ ratio combined with the
narrower \htwo\ line profile suggest that local star formation activity may
be able to explain most of the \htwo\ emission, with no need for a
contribution from outflowing gas associated with the nucleus.  By contrast,
in NGC~2110, where wings associated with the outflow are seen in \htwo,
nuclear X-ray heating may be a more significant contributor.

\section{Summary and Conclusions}

\begin{enumerate}
\item
We have observed strong infrared \pab, \brg, \feii, and \htwo\ line emission
which is spatially extended on scales of a few hundred parsecs in Mk~1066,
NGC~2110, NGC~4388, and Mk~3.

\item
We have detected the near infrared coronal line \six\ (\lsix) in the
nucleus of NGC~4388.  In contrast to the other infrared lines, this line
is only marginally spatially extended.  There is no apparent \six\
emission from Mk~1066, NGC~2110, or Mk~3.  This is similar to what is
seen optically, that high ionization species are visible in some but not
all Seyfert galaxies.

\item
While often showing similar velocity trends over the central few arcseconds,
the \feii, \htwo, and H lines have profiles which differ in detail from each
other over the full spatial range probed by these observations.  The peaks
of lines are shifted relative to each other at some spatial positions in
some galaxies, and some species show enhanced blue and red wings relative to
other species.  These differences are usually due to the effects of
extinction, an enhancement to the \feii\ emission associated with other
indicators of the action of an outflow, or a possible local source of
heating of \htwo.

\item
The \feii/\pab\ and \htwo/\brg\ flux ratios can vary dramatically on scales
of 10$^2$~pc near the nuclei of Seyfert 2 galaxies.  The details of the
spatial structure of the \feii/\pab\ ratio frequently support the notion
that partially ionized regions created by X-ray photoionization from the
nucleus, or shocks from the interaction of outflowing gas with ambient gas,
or both, are enhancing the circumnuclear \feii\ emission.  The spatial
variation of the \htwo/\brg\ ratio is again frequently ascribable to local
heating of \htwo.

\item
The \feii\ emission appears to be associated with the nuclear activity, not
requiring the invocation of a circumnuclear starburst in any of these four
galaxies.  The \feii\ can be kinematically and spatially linked to processes
dominated by the active nucleus.  In all of these galaxies, there are
correlations between features in the \feii\ emission and both radio jets and
optically identified outflows, suggesting that some \feii\ emission is
produced by fast shocks resulting from interactions between radio jets and
circumnuclear gas.

\item
The \htwo\ emission, although peaked on the nucleus, is relatively strong
compared to hydrogen recombination and \feii\ emission perpendicular to the
axis of ionization cones or the major axis of high-excitation gas in Mk~1066
and NGC~4388.  The spatial distribution and line profile of \htwo\ generally
differs from that of \feii.  A local source such as bar-driven shocks or
circumnuclear star formation may be partially responsible for the \htwo\
emission in these galaxies.

\end{enumerate}

\acknowledgements

The authors would like to thank the staff and night assistants at
Palomar Observatory for assistance in completing these observations.
Additionaly, they would like to thank James Larkin for discussion and
assistance in the earlier phases of the project.

\appendix

\section{Measurements of Equivalent Widths, Fluxes, and Line Ratios}
\label{sec:tabappndx}

In this appendix are tables with data from the infrared spectra of
Mk~1066, NGC~2110, NGC~4388, and Mk~3.  Tables \ref{mk1066eqw} through
\ref{mk3eqw} list the equivalent widths for the integrated line profile
for each species observed in each spatial bin.  They additionally list
relevant flux ratios.  Tables \ref{mk1066kinematics} through
\ref{mk3kinematics} give the centroids and line widths (FWHM) for each
species in each spatial bin.

\ifthenelse{\boolean{ispreprint}}{
\begin{table}[htb]
\renewcommand{\baselinestretch}{1.0}
\scriptsize
\settowidth{\digsp}{0}
\newcommand{\dig}{\hspace*{\digsp}}
\settowidth{\dotsp}{.}
\newcommand{\nodot}{\hspace*{\dotsp}}
\settowidth{\pmsp}{$\pm$0.00}
\newcommand{\nopm}{\hspace*{\pmsp}}
\begin{lrbox}{\thisbox}
\begin{tabular}{r r@{\hspace{0.08in}}r@{\hspace{0.08in}}r %
                   @{\hspace{0.08in}}r %
                  r@{\hspace{0.08in}}r}
\hline
\hline
        & \multicolumn{4}{c}{Equivalent Width (10$^{-4}\mu$m)}
        & \multicolumn{2}{c}{Flux Ratios} \\
Pos$^a$ & \multicolumn{4}{c}{\hrulefill}
        & \multicolumn{2}{c}{\hrulefill} \\
($''$)  & \cnc{[Fe II]} & \cnc{Pa$\beta$}
        & \cnc{H$_2$} & \cnc{Br$\gamma$} 
        & \cnc{\feii/Pa$\beta$} & \cnc{H$_2$/Br$\gamma$} \\
\hline \\[5pt]
\multicolumn{7}{c}{\small Mk~1066, 135$^\circ$ slit} \\[4pt]
2.67
        & 6.7$\pm$4.5\dig\dig
        & 17.1$\pm$12.4\dig
        & \ldots\dig\dig\dig\dig
        & \ldots\dig\dig\dig\dig
        & 0.43$\pm$0.14 
        & \ldots\dig\dig\dig\dig \\
2\nodot\dig\dig
        & 4.7$\pm$1.7\dig\dig
        & 5.8$\pm$1.9\dig\dig
        & 4.0$\pm$1.9\dig\dig
        & $<$4.3\nopm
        & 0.79$\pm$0.21 
        & $>$0.87\nopm \\
1.33
        & 18.4$\pm$3.2\dig\dig
        & 15.3$\pm$2.6\dig\dig
        & 9.4$\pm$2.0\dig\dig
        & 10.3$\pm$2.7\dig\dig
        & 1.17$\pm$0.08 
        & 0.86$\pm$0.09 \\
0.67
        & 23.0$\pm$2.6\dig\dig
        & 29.0$\pm$2.6\dig\dig
        & 12.6$\pm$1.7\dig\dig
        & 18.7$\pm$2.8\dig\dig
        & 0.75$\pm$0.03 
        & 0.63$\pm$0.05 \\
0\nodot\dig\dig
        & 20.8$\pm$1.9\dig\dig
        & 27.4$\pm$2.1\dig\dig
        & 13.0$\pm$1.2\dig\dig
        & 14.0$\pm$1.5\dig\dig
        & 0.75$\pm$0.04 
        & 0.89$\pm$0.06 \\
-0.67
        & 39.0$\pm$3.8\dig\dig
        & 33.0$\pm$4.6\dig\dig
        & 17.8$\pm$2.6\dig\dig
        & 20.6$\pm$3.8\dig\dig
        & 1.20$\pm$0.04 
        & 0.85$\pm$0.06 \\
-1.33
        & 21.6$\pm$3.4\dig\dig
        & 23.2$\pm$4.6\dig\dig
        & 9.0$\pm$2.7\dig\dig
        & 16.7$\pm$6.3\dig\dig
        & 0.99$\pm$0.05 
        & 0.56$\pm$0.09 \\
-2\nodot\dig\dig
        & 5.3$\pm$2.2\dig\dig
        & 6.7$\pm$2.5\dig\dig
        & $<$4.4\nopm
        & 11.5$\pm$6.1\dig\dig
        & 0.85$\pm$0.17 
        & $<$0.40\nopm \\[12pt]
\multicolumn{7}{c}{\small Mk~1066, 45$^\circ$ slit} \\[4pt]
1.33
        & 5.0$\pm$1.3\dig\dig
        & 4.2$\pm$1.1\dig\dig
        & 13.0$\pm$5.3\dig\dig
        & $<$2.7\nopm
        & 1.30$\pm$0.32 
        & $>$4.41\nopm \\
0.67
        & 11.2$\pm$1.6\dig\dig
        & 10.3$\pm$1.3\dig\dig
        & 12.6$\pm$2.7\dig\dig
        & 6.7$\pm$1.1\dig\dig
        & 1.06$\pm$0.09 
        & 1.74$\pm$0.21 \\
0\nodot\dig\dig
        & 17.2$\pm$1.4\dig\dig
        & 29.1$\pm$2.7\dig\dig
        & 11.5$\pm$1.5\dig\dig
        & 13.7$\pm$1.4\dig\dig
        & 0.57$\pm$0.03 
        & 0.76$\pm$0.06 \\
-0.67
        & 15.9$\pm$2.6\dig\dig
        & 17.9$\pm$2.7\dig\dig
        & 12.6$\pm$2.3\dig\dig
        & 10.0$\pm$2.0\dig\dig
        & 0.87$\pm$0.07 
        & 1.21$\pm$0.10 \\
-1.33
        & 5.2$\pm$1.4\dig\dig
        & 8.9$\pm$2.6\dig\dig
        & 9.0$\pm$4.0\dig\dig
        & $<$4.7\nopm
        & 0.64$\pm$0.16 
        & $>$1.83\nopm \\
-2\nodot\dig\dig
        & \ldots\dig\dig\dig\dig
        & \ldots\dig\dig\dig\dig
        & 5.6$\pm$3.7\dig\dig
        & $<$5.3\nopm
        & \ldots\dig\dig\dig\dig 
        & $>$0.91\nopm \\[5pt]
\hline \\
\end{tabular}
\end{lrbox}
\settowidth{\thiswid}{\usebox{\thisbox}}
\begin{center}
\caption{Mk~1066 Equivalent Widths and Flux Ratios}\label{mk1066eqw}
\begin{minipage}{\thiswid}
\usebox{\thisbox}
\tiny
$a$ Positive positions are towards the position angle of the slit as quoted.
\end{minipage}
\end{center}
\end{table}
}{
\placetable{mk1066eqw}
}

\ifthenelse{\boolean{ispreprint}}{
\begin{table}[htb]
\renewcommand{\baselinestretch}{1.0}
\scriptsize
\settowidth{\digsp}{0}
\newcommand{\dig}{\hspace*{\digsp}}
\settowidth{\dotsp}{.}
\newcommand{\nodot}{\hspace*{\dotsp}}
\settowidth{\pmsp}{$\pm$0.00}
\newcommand{\nopm}{\hspace*{\pmsp}}
\begin{lrbox}{\thisbox}
\begin{tabular}{r r@{\hspace{0.08in}}r@{\hspace{0.08in}}r %
                   @{\hspace{0.08in}}r %
                  r@{\hspace{0.08in}}r}
\hline
\hline
        & \multicolumn{4}{c}{Equivalent Width (10$^{-4}\mu$m)}
        & \multicolumn{2}{c}{Flux Ratios} \\
Pos$^a$ & \multicolumn{4}{c}{\hrulefill}
        & \multicolumn{2}{c}{\hrulefill} \\
($''$)  & \cnc{[Fe II]} & \cnc{Pa$\beta$}
        & \cnc{H$_2$} & \cnc{Br$\gamma$} 
        & \cnc{\feii/Pa$\beta$} & \cnc{H$_2$/Br$\gamma$} \\
\hline \\[5pt]
3\nodot\dig\dig
        & 3.5$\pm$1.0\dig\dig
        & 3.3$\pm$0.8\dig\dig
        & 3.8$\pm$1.3\dig\dig
        & 2.1$\pm$0.9\dig\dig
        & 1.09$\pm$0.25 
        & 1.89$\pm$0.78 \\
2\nodot\dig\dig
        & 6.2$\pm$1.3\dig\dig
        & 2.9$\pm$0.7\dig\dig
        & 7.2$\pm$1.4\dig\dig
        & 1.9$\pm$0.6\dig\dig
        & 2.19$\pm$0.42 
        & 4.08$\pm$1.08 \\
1\nodot\dig\dig
        & 15.4$\pm$1.4\dig\dig
        & 3.1$\pm$0.5\dig\dig
        & 6.2$\pm$0.8\dig\dig
        & 2.0$\pm$0.4\dig\dig
        & 5.02$\pm$0.48 
        & 3.23$\pm$0.53 \\
0\nodot\dig\dig
        & 24.7$\pm$1.6\dig\dig
        & 3.1$\pm$0.4\dig\dig
        & 4.8$\pm$0.4\dig\dig
        & 1.6$\pm$0.3\dig\dig
        & 8.11$\pm$0.78 
        & 3.17$\pm$0.63 \\
-1\nodot\dig\dig
        & 22.5$\pm$2.3\dig\dig
        & 3.2$\pm$0.5\dig\dig
        & 6.6$\pm$0.7\dig\dig
        & 2.0$\pm$0.4\dig\dig
        & 7.36$\pm$0.74 
        & 3.59$\pm$0.69 \\
-2\nodot\dig\dig
        & 11.7$\pm$2.3\dig\dig
        & 2.8$\pm$0.6\dig\dig
        & 8.4$\pm$1.7\dig\dig
        & 1.4$\pm$0.4\dig\dig
        & 4.43$\pm$0.53 
        & 6.50$\pm$1.50 \\
-3\nodot\dig\dig
        & 8.7$\pm$2.0\dig\dig
        & 3.1$\pm$0.9\dig\dig
        & 5.5$\pm$1.2\dig\dig
        & 2.8$\pm$1.0\dig\dig
        & 2.93$\pm$0.54 
        & 2.14$\pm$0.51 \\
-4\nodot\dig\dig
        & 5.8$\pm$2.8\dig\dig
        & $<$2.7\nopm
        & 3.6$\pm$1.5\dig\dig
        & $<$9.7\nopm
        & $>$2.24\nopm
        & $>$0.39\nopm\\[5pt]
\hline \\
\end{tabular}
\end{lrbox}
\settowidth{\thiswid}{\usebox{\thisbox}}
\begin{center}
\caption{NGC~2110 Equivalent Widths and Flux Ratios}\label{ngc2110eqw}
\begin{minipage}{\thiswid}
\usebox{\thisbox}
\tiny
$a$ Positive positions are toward the southeast at a position angle of
160$^\circ$.
\end{minipage}
\end{center}
\end{table}
}{
\placetable{ngc2110eqw}
}

\ifthenelse{\boolean{ispreprint}}{
\begin{table}[htb]
\renewcommand{\baselinestretch}{1.0}
\scriptsize
\settowidth{\digsp}{0}
\newcommand{\dig}{\hspace*{\digsp}}
\settowidth{\dotsp}{.}
\newcommand{\nodot}{\hspace*{\dotsp}}
\settowidth{\pmsp}{$\pm$0.00}
\newcommand{\nopm}{\hspace*{\pmsp}}
\begin{lrbox}{\thisbox}
\begin{tabular}{r r@{\hspace{0.08in}}r@{\hspace{0.08in}}r %
                   @{\hspace{0.08in}}r@{\hspace{0.08in}}r %
                  r@{\hspace{0.08in}}r@{\hspace{0.08in}}r}
\hline
\hline
        & \multicolumn{5}{c}{Equivalent Width (10$^{-4}\mu$m)}
        & \multicolumn{3}{c}{Flux Ratios} \\
Pos$^a$ & \multicolumn{5}{c}{\hrulefill}
        & \multicolumn{3}{c}{\hrulefill} \\
($''$)  & \cnc{[Fe II]} & \cnc{Pa$\beta$}
        & \cnc{H$_2$} & \cnc{Br$\gamma$} & \cnc{[S IX]}
        & \cnc{\feii/Pa$\beta$} & \cnc{H$_2$/Br$\gamma$}
        & \cnc{\six/Pa$\beta$} \\
\hline \\[5pt]
\multicolumn{9}{c}{\small NGC~4388, 30$^\circ$ slit} \\[4pt]
4\nodot\dig\dig
        & 22.8$\pm$19.1\dig
        & 19.0$\pm$14.9\dig
        & $<$7.4\nopm
        & 13.0$\pm$12.8\dig
        & \ldots\dig\dig\dig\dig
        & 0.99$\pm$0.25 
        & $<$0.67\nopm
        & \ldots\dig\dig\dig\dig \\
3\nodot\dig\dig
        & 15.5$\pm$9.5\dig\dig
        & 18.2$\pm$9.1\dig\dig
        & $<$7.0\nopm
        & 13.3$\pm$9.9\dig\dig
        & \ldots\dig\dig\dig\dig
        & 0.71$\pm$0.17 
        & $<$0.57\nopm
        & \ldots\dig\dig\dig\dig \\
2\nodot\dig\dig
        & 13.4$\pm$6.0\dig\dig
        & 26.1$\pm$7.0\dig\dig
        & 6.5$\pm$3.1\dig\dig
        & 12.7$\pm$6.7\dig\dig
        & \ldots\dig\dig\dig\dig
        & 0.47$\pm$0.08 
        & 0.56$\pm$0.13 
        & \ldots\dig\dig\dig\dig \\
1\nodot\dig\dig
        & 7.7$\pm$1.9\dig\dig
        & 25.7$\pm$5.5\dig\dig
        & 13.6$\pm$4.1\dig\dig
        & 12.7$\pm$3.2\dig\dig
        & 3.6$\pm$1.0\dig\dig
        & 0.28$\pm$0.03 
        & 1.05$\pm$0.15 
        & 0.13$\pm$0.02 \\
0\nodot\dig\dig
        & 11.2$\pm$1.7\dig\dig
        & 26.3$\pm$4.2\dig\dig
        & 9.5$\pm$1.2\dig\dig
        & 10.6$\pm$1.5\dig\dig
        & 4.5$\pm$0.8\dig\dig
        & 0.40$\pm$0.03 
        & 0.88$\pm$0.08 
        & 0.16$\pm$0.02 \\
-1\nodot\dig\dig
        & 18.8$\pm$4.6\dig\dig
        & 32.2$\pm$6.8\dig\dig
        & 8.4$\pm$1.6\dig\dig
        & 17.1$\pm$3.9\dig\dig
        & 2.4$\pm$0.8\dig\dig
        & 0.57$\pm$0.04 
        & 0.51$\pm$0.06 
        & 0.07$\pm$0.01 \\
-2\nodot\dig\dig
        & 31.0$\pm$9.4\dig\dig
        & 26.8$\pm$7.4\dig\dig
        & 6.3$\pm$2.7\dig\dig
        & 14.7$\pm$6.5\dig\dig
        & \ldots\dig\dig\dig\dig
        & 1.13$\pm$0.11 
        & 0.46$\pm$0.09 
        & \ldots\dig\dig\dig\dig \\
-3\nodot\dig\dig
        & 29.7$\pm$11.8\dig
        & 17.8$\pm$7.1\dig\dig
        & 6.1$\pm$3.6\dig\dig
        & 13.6$\pm$10.5\dig
        & \ldots\dig\dig\dig\dig
        & 1.64$\pm$0.22 
        & 0.49$\pm$0.14 
        & \ldots\dig\dig\dig\dig \\
-4\nodot\dig\dig
        & 12.5$\pm$6.6\dig\dig
        & 4.8$\pm$2.7\dig\dig
        & $<$6.9\nopm
        & $<$11.7\nopm
        & \ldots\dig\dig\dig\dig
        & 2.52$\pm$0.84 
        & \ldots\dig\dig\dig\dig 
        & \ldots\dig\dig\dig\dig \\[12pt]
\multicolumn{9}{c}{\small NGC~4388, 120$^\circ$ slit} \\[4pt]
2\nodot\dig\dig
        & 9.0$\pm$2.5\dig\dig
        & 4.0$\pm$1.3\dig\dig
        & 12.4$\pm$5.3\dig\dig
        & $<$4.9\nopm
        & \ldots\dig\dig\dig\dig
        & 2.18$\pm$0.53 
        & $>$2.29\nopm
        & \ldots\dig\dig\dig\dig \\
1\nodot\dig\dig
        & 12.7$\pm$2.1\dig\dig
        & 16.8$\pm$3.1\dig\dig
        & 14.1$\pm$2.9\dig\dig
        & 9.4$\pm$2.0\dig\dig
        & \ldots\dig\dig\dig\dig
        & 0.72$\pm$0.07 
        & 1.39$\pm$0.18 
        & \ldots\dig\dig\dig\dig \\
0\nodot\dig\dig
        & 11.5$\pm$1.4\dig\dig
        & 24.1$\pm$2.7\dig\dig
        & 8.4$\pm$1.1\dig\dig
        & 11.6$\pm$1.8\dig\dig
        & 4.4$\pm$0.6\dig\dig
        & 0.46$\pm$0.03 
        & 0.67$\pm$0.07 
        & 0.17$\pm$0.01 \\
-1\nodot\dig\dig
        & 10.1$\pm$2.0\dig\dig
        & 19.9$\pm$3.5\dig\dig
        & 19.6$\pm$4.6\dig\dig
        & 8.0$\pm$1.7\dig\dig
        & \ldots\dig\dig\dig\dig
        & 0.50$\pm$0.05 
        & 2.34$\pm$0.25 
        & \ldots\dig\dig\dig\dig \\
-2\nodot\dig\dig
        & 8.9$\pm$3.0\dig\dig
        & 7.2$\pm$2.2\dig\dig
        & 18.0$\pm$5.9\dig\dig
        & 3.2$\pm$1.5\dig\dig
        & \ldots\dig\dig\dig\dig
        & 1.26$\pm$0.24 
        & 5.23$\pm$1.60 
        & \ldots\dig\dig\dig\dig \\[12pt]
\hline \\
\end{tabular}
\end{lrbox}
\settowidth{\thiswid}{\usebox{\thisbox}}
\begin{center}
\caption{NGC~4388 Equivalent Widths and Flux Ratios}\label{ngc4388eqw}
\begin{minipage}{\thiswid}
\usebox{\thisbox}
\tiny
$a$ Positive positions are towards the position angle of the slit as quoted.
\end{minipage}
\end{center}
\end{table}
}{
\placetable{ngc4388eqw}
}

\ifthenelse{\boolean{ispreprint}}{
\begin{table}[htb]
\renewcommand{\baselinestretch}{1.0} \scriptsize \settowidth{\digsp}{0}
\newcommand{\dig}{\hspace*{\digsp}} \settowidth{\dotsp}{.}
\newcommand{\nodot}{\hspace*{\dotsp}} \settowidth{\pmsp}{$\pm$0.00}
\newcommand{\nopm}{\hspace*{\pmsp}}
\begin{lrbox}{\thisbox}
\begin{tabular}{r r@{\hspace{0.08in}}r@{\hspace{0.08in}}r %
                   @{\hspace{0.08in}}r %
                  r@{\hspace{0.08in}}r}
\hline
\hline
        & \multicolumn{4}{c}{Equivalent Width (10$^{-4}\mu$m)}
        & \multicolumn{2}{c}{Flux Ratios} \\
Pos$^a$ & \multicolumn{4}{c}{\hrulefill}
        & \multicolumn{2}{c}{\hrulefill} \\
($''$)  & \cnc{[Fe II]} & \cnc{Pa$\beta$}
        & \cnc{H$_2$} & \cnc{Br$\gamma$} 
        & \cnc{\feii/Pa$\beta$} & \cnc{H$_2$/Br$\gamma$} \\
\hline \\[5pt]
3\nodot\dig\dig
        & 11.6$\pm$4.3\dig\dig
        & 24.5$\pm$8.6\dig\dig
        & 5.8$\pm$4.2\dig\dig
        & 12.0$\pm$8.4\dig\dig
        & 0.47$\pm$0.10 
        & 0.49$\pm$0.24  \\
2\nodot\dig\dig
        & 13.0$\pm$3.5\dig\dig
        & 17.7$\pm$4.7\dig\dig
        & \ldots\dig\dig\dig\dig
        & 4.0$\pm$1.4\dig\dig
        & 0.74$\pm$0.12 
        & \ldots\dig\dig\dig\dig  \\
1\nodot\dig\dig
        & 37.6$\pm$4.2\dig\dig
        & 31.3$\pm$3.5\dig\dig
        & 3.4$\pm$1.2\dig\dig
        & 11.5$\pm$2.6\dig\dig
        & 1.17$\pm$0.07 
        & 0.29$\pm$0.11  \\
0\nodot\dig\dig
        & 52.2$\pm$5.1\dig\dig
        & 39.3$\pm$3.2\dig\dig
        & 5.3$\pm$0.7\dig\dig
        & 16.7$\pm$2.1\dig\dig
        & 1.24$\pm$0.04 
        & 0.31$\pm$0.04  \\
-1\nodot\dig\dig
        & 46.3$\pm$6.2\dig\dig
        & 26.4$\pm$3.0\dig\dig
        & 4.6$\pm$0.8\dig\dig
        & 12.8$\pm$2.2\dig\dig
        & 1.72$\pm$0.08 
        & 0.35$\pm$0.05  \\
-2\nodot\dig\dig
        & 17.9$\pm$5.7\dig\dig
        & 9.5$\pm$2.9\dig\dig
        & 13.1$\pm$4.5\dig\dig
        & 4.4$\pm$1.6\dig\dig
        & 1.88$\pm$0.43 
        & 2.94$\pm$0.54  \\
-3\nodot\dig\dig
        & 17.6$\pm$7.7\dig\dig
        & 4.3$\pm$1.7\dig\dig
        & \ldots\dig\dig\dig\dig
        & \ldots\dig\dig\dig\dig
        & 3.75$\pm$1.15 
        & \ldots\dig\dig\dig\dig \\[5pt]
\hline
\end{tabular}
\end{lrbox}
\settowidth{\thiswid}{\usebox{\thisbox}}
\begin{center}
\caption{Mk~3 Equivalent Widths and Flux Ratios}\label{mk3eqw}
\begin{minipage}{\thiswid}
\usebox{\thisbox}
\tiny
$a$ Positive positions are east towards the 113$^\circ$ position angle of
the slit.
\end{minipage}
\end{center}
\end{table}
}{
\placetable{mk3eqw}
}

\ifthenelse{\boolean{ispreprint}}{
\begin{table}[htb]
\renewcommand{\baselinestretch}{1.0}
\tiny
\settowidth{\digsp}{0}
\newcommand{\dig}{\hspace*{\digsp}}
\settowidth{\dotsp}{.}
\newcommand{\nodot}{\hspace*{\dotsp}}
\settowidth{\pmsp}{$\pm$00}
\newcommand{\nopm}{\hspace*{\pmsp}}
\begin{lrbox}{\thisbox}
\begin{tabular}{r  r@{\hspace*{0.05in}}r  r@{\hspace*{0.05in}}r  r@{\hspace*{0.05in}}r  r@{\hspace*{0.05in}}r  r@{\hspace*{0.05in}}r}
\hline
\hline
        & \multicolumn{2}{c}{[Fe II]}
        & \multicolumn{2}{c}{Pa$\beta$}
        & \multicolumn{2}{c}{H$_2$}
        & \multicolumn{2}{c}{Br$\gamma$} \\
        & \multicolumn{2}{c}{\hrulefill}
        & \multicolumn{2}{c}{\hrulefill}
        & \multicolumn{2}{c}{\hrulefill}
        & \multicolumn{2}{c}{\hrulefill}  \\
\cnc{\tiny Pos$^a$}
        & \cnc{\tiny Centroid$^b$}    & \cnc{\tiny FWHM$^c$}
        & \cnc{\tiny Centroid}        & \cnc{\tiny FWHM}
        & \cnc{\tiny Centroid}        & \cnc{\tiny FWHM}
        & \cnc{\tiny Centroid}        & \cnc{\tiny FWHM} \\
\cnc{$''$}
        & \cnc{\tiny\kms}            & \cnc{\tiny\kms}
        & \cnc{\tiny\kms}            & \cnc{\tiny\kms}
        & \cnc{\tiny\kms}            & \cnc{\tiny\kms}
        & \cnc{\tiny\kms}            & \cnc{\tiny\kms} \\
\hline \\
\multicolumn{9}{c}{\small Mk~1066, 135$^\circ$ slit} \\[4pt]
2.67
        & -118$\pm$24\dig                 & $<154$\nopm
        & -233$\pm$41\dig                 &  677$\pm$194
        & \ldots\dig\dig                  & \ldots\dig\dig 
        & \ldots\dig\dig                  & \ldots\dig\dig  \\
2\nodot\dig\dig
        &  -68$\pm$25\dig                 &  267$\pm$109
        &  -58$\pm$20\dig                 &  221$\pm$83\dig
        &  -72$\pm$35\dig                 & $<190$\nopm
        & \ldots\dig\dig                  & \ldots\dig\dig  \\
1.33
        &  -59$\pm$14\dig                 &  294$\pm$53\dig
        &  -77$\pm$19\dig                 & $<154$\nopm
        &  -45$\pm$15\dig                 &  328$\pm$46\dig
        &  -90$\pm$19\dig                 &  383$\pm$34\dig \\
0.67
        &  -38$\pm$14\dig                 &  240$\pm$32\dig
        &  -67$\pm$14\dig                 & $<154$\nopm
        &  -45$\pm$15\dig                 &  321$\pm$50\dig
        &  -73$\pm$18\dig                 &  237$\pm$46\dig \\
0\nodot\dig\dig
        & -125$\pm$16\dig                 &  296$\pm$41\dig
        &  -20$\pm$15\dig                 &  205$\pm$50\dig
        &    3$\pm$13\dig                 &  353$\pm$29\dig
        &  -19$\pm$18\dig                 &  329$\pm$27\dig \\
-0.67
        &  -54$\pm$14\dig                 &  289$\pm$54\dig
        &   51$\pm$14\dig                 & $<154$\nopm
        &   89$\pm$13\dig                 &  223$\pm$37\dig
        &   51$\pm$19\dig                 &  269$\pm$32\dig \\
-1.33
        &   -8$\pm$15\dig                 &  341$\pm$62\dig
        &   32$\pm$16\dig                 &  167$\pm$74\dig
        &  143$\pm$20\dig                 &  360$\pm$78\dig 
        &   70$\pm$20\dig                 &  308$\pm$44\dig \\
-2\nodot\dig\dig
        &  109$\pm$18\dig                 &  218$\pm$75\dig
        &   70$\pm$18\dig                 &  202$\pm$65\dig
        & \ldots\dig\dig                  & \ldots\dig\dig 
        &  133$\pm$46\dig                 &  715$\pm$196    \\[12pt]
\multicolumn{9}{c}{\small Mk~1066, 45$^\circ$ slit} \\[4pt]
1.33
        &  -34$\pm$17\dig                 &  222$\pm$73\dig
        &  -50$\pm$19\dig                 &  179$\pm$93\dig
        &   -6$\pm$15\dig                 &  491$\pm$59\dig
        & \ldots\dig\dig                  & \ldots\dig\dig  \\
0.67
        &  -52$\pm$14\dig                 &  179$\pm$52\dig
        &  -20$\pm$10\dig                 & $<150$\nopm
        &  -12$\pm$11\dig                 &  392$\pm$35\dig
        &  -43$\pm$17\dig                 &  429$\pm$62\dig \\
0\nodot\dig\dig
        &  -74$\pm$12\dig                 &  179$\pm$38\dig
        &  -41$\pm$13\dig                 & $<150$\nopm
        &  -14$\pm$12\dig                 &  192$\pm$61\dig
        &  -83$\pm$17\dig                 &  250$\pm$44\dig \\
-0.67
        &  -54$\pm$15\dig                 &  219$\pm$46\dig
        &  -24$\pm$11\dig                 & $<150$\nopm
        &  -21$\pm$10\dig                 &  277$\pm$33\dig
        &  -49$\pm$14\dig                 &  313$\pm$51\dig \\
-1.33
        &  -10$\pm$19\dig                 &  247$\pm$83\dig
        &   64$\pm$49\dig                 & $<150$\nopm
        &   39$\pm$18\dig                 &  329$\pm$82\dig
        & \ldots\dig\dig                  & \ldots\dig\dig  \\
-2\nodot\dig\dig
        & \ldots\dig\dig                  & \ldots\dig\dig 
        & \ldots\dig\dig                  & \ldots\dig\dig 
        &  107$\pm$23\dig                 & $<190$\nopm
        & \ldots\dig\dig                  & \ldots\dig\dig  \\[5pt]
\hline \\
\end{tabular}
\end{lrbox}
\settowidth{\thiswid}{\usebox{\thisbox}}
\begin{center}
\begin{minipage}{\thiswid}
\caption{Mk~1066 Line Centroids and Widths}\label{mk1066kinematics}
\usebox{\thisbox}
\tiny\begin{tabular}{l l}
$a$ & Positive position is towards the position angle of the slit as
quoted. \\
$b$ & Centroids are relative to 3625~\kms, the systemic velocity of Mk~1066. \\
$c$ & The resolution has been subtracted in quadrature. \\
\end{tabular}
\end{minipage}
\end{center}
\end{table}
}{
\placetable{mk1066kinematics}
}

\ifthenelse{\boolean{ispreprint}}{
\begin{table}[htb]
\renewcommand{\baselinestretch}{1.0}
\tiny
\settowidth{\digsp}{0}
\newcommand{\dig}{\hspace*{\digsp}}
\settowidth{\dotsp}{.}
\newcommand{\nodot}{\hspace*{\dotsp}}
\settowidth{\pmsp}{$\pm$00}
\newcommand{\nopm}{\hspace*{\pmsp}}
\begin{lrbox}{\thisbox}
\begin{tabular}{r  r@{\hspace*{0.05in}}r  r@{\hspace*{0.05in}}r  r@{\hspace*{0.05in}}r  r@{\hspace*{0.05in}}r  r@{\hspace*{0.05in}}r}
\hline
\hline
        & \multicolumn{2}{c}{[Fe II]}
        & \multicolumn{2}{c}{Pa$\beta$}
        & \multicolumn{2}{c}{H$_2$}
        & \multicolumn{2}{c}{Br$\gamma$} \\
        & \multicolumn{2}{c}{\hrulefill}
        & \multicolumn{2}{c}{\hrulefill}
        & \multicolumn{2}{c}{\hrulefill}
        & \multicolumn{2}{c}{\hrulefill}  \\
\cnc{\tiny Pos$^a$}
        & \cnc{\tiny Centroid$^b$}    & \cnc{\tiny FWHM$^c$}
        & \cnc{\tiny Centroid}        & \cnc{\tiny FWHM}
        & \cnc{\tiny Centroid}        & \cnc{\tiny FWHM}
        & \cnc{\tiny Centroid}        & \cnc{\tiny FWHM} \\
\cnc{$''$}
        & \cnc{\tiny\kms}            & \cnc{\tiny\kms}
        & \cnc{\tiny\kms}            & \cnc{\tiny\kms}
        & \cnc{\tiny\kms}            & \cnc{\tiny\kms}
        & \cnc{\tiny\kms}            & \cnc{\tiny\kms} \\
\hline \\
3\nodot\dig\dig
        &  212$\pm$18\dig                 &  219$\pm$88\dig
        &  299$\pm$11\dig                 & $<150$\nopm
        &  292$\pm$29\dig                 &  223$\pm$132
        &  375$\pm$45\dig                 & $<189$\nopm     \\
2\nodot\dig\dig
        &  117$\pm$15\dig                 &  334$\pm$44\dig
        &  226$\pm$13\dig                 & $<150$\nopm
        &  213$\pm$21\dig                 & $<189$\nopm
        &  203$\pm$36\dig                 &  213$\pm$169    \\
1\nodot\dig\dig
        &   -4$\pm$9\dig\dig              &  373$\pm$36\dig
        &  116$\pm$8\dig\dig              &  213$\pm$41\dig
        &  152$\pm$18\dig                 &  275$\pm$35\dig
        &  160$\pm$27\dig                 &  259$\pm$92\dig \\
0\nodot\dig\dig
        &  -86$\pm$10\dig                 &  496$\pm$34\dig
        &   -4$\pm$10\dig                 &  338$\pm$43\dig
        &   32$\pm$18\dig                 &  359$\pm$36\dig
        &   62$\pm$33\dig                 &  391$\pm$116    \\
-1\nodot\dig\dig
        &  -68$\pm$8\dig\dig              &  383$\pm$25\dig
        &  -49$\pm$9\dig\dig              &  186$\pm$51\dig
        &  -52$\pm$17\dig                 &  273$\pm$34\dig
        &   48$\pm$31\dig                 &  304$\pm$111    \\
-2\nodot\dig\dig
        & -114$\pm$11\dig                 &  171$\pm$54\dig
        &  -84$\pm$9\dig\dig              &  176$\pm$53\dig
        &  -14$\pm$20\dig                 &  238$\pm$52\dig
        &  -95$\pm$31\dig                 & $<189$\nopm     \\
-3\nodot\dig\dig
        & -156$\pm$10\dig                 &  197$\pm$31\dig
        & -126$\pm$15\dig                 &  203$\pm$82\dig
        & -112$\pm$19\dig                 &  205$\pm$61\dig
        &  -69$\pm$35\dig                 &  313$\pm$136    \\
-4\nodot\dig\dig
        & -207$\pm$20\dig                 &  347$\pm$87\dig
        & \ldots\dig\dig                  & \ldots\dig\dig 
        &   -5$\pm$42\dig                 &  546$\pm$226
        & \ldots\dig\dig                  & \ldots\dig\dig  \\[5pt]
\hline \\
\end{tabular}
\end{lrbox}
\settowidth{\thiswid}{\usebox{\thisbox}}
\begin{center}
\begin{minipage}{\thiswid}
\caption{NGC~2110 Line Centroids and Widths}\label{ngc2110kinematics}
\usebox{\thisbox}
\tiny\begin{tabular}{l l}
$a$ & Positive position is southeast along the 160$^\circ$ slit. \\
$b$ & Centroids are relative to 2342~\kms, the systemic velocity of
NGC~2110. \\
$c$ & The resolution has been subtracted in quadrature.\\
\end{tabular}
\end{minipage}
\end{center}
\end{table}
}{
\placetable{ngc2110kinematics}
}

\ifthenelse{\boolean{ispreprint}}{
\begin{table}[htb]
\renewcommand{\baselinestretch}{1.0}
\tiny
\settowidth{\digsp}{0}
\newcommand{\dig}{\hspace*{\digsp}}
\settowidth{\dotsp}{.}
\newcommand{\nodot}{\hspace*{\dotsp}}
\settowidth{\pmsp}{$\pm$00}
\newcommand{\nopm}{\hspace*{\pmsp}}
\begin{lrbox}{\thisbox}
\begin{tabular}{r  r@{\hspace*{0.05in}}r  r@{\hspace*{0.05in}}r  r@{\hspace*{0.05in}}r  r@{\hspace*{0.05in}}r  r@{\hspace*{0.05in}}r}
\hline
\hline
        & \multicolumn{2}{c}{[Fe II]}
        & \multicolumn{2}{c}{Pa$\beta$}
        & \multicolumn{2}{c}{H$_2$}
        & \multicolumn{2}{c}{Br$\gamma$}
        & \multicolumn{2}{c}{[S IX]}     \\
        & \multicolumn{2}{c}{\hrulefill}
        & \multicolumn{2}{c}{\hrulefill}
        & \multicolumn{2}{c}{\hrulefill}
        & \multicolumn{2}{c}{\hrulefill}
        & \multicolumn{2}{c}{\hrulefill}  \\
\cnc{\tiny Pos$^a$}
        & \cnc{\tiny Centroid$^b$}    & \cnc{\tiny FWHM$^c$}
        & \cnc{\tiny Centroid}        & \cnc{\tiny FWHM}
        & \cnc{\tiny Centroid}        & \cnc{\tiny FWHM}
        & \cnc{\tiny Centroid}        & \cnc{\tiny FWHM}
        & \cnc{\tiny Centroid}        & \cnc{\tiny FWHM} \\
\cnc{$''$}
        & \cnc{\tiny\kms}            & \cnc{\tiny\kms}
        & \cnc{\tiny\kms}            & \cnc{\tiny\kms}
        & \cnc{\tiny\kms}            & \cnc{\tiny\kms}
        & \cnc{\tiny\kms}            & \cnc{\tiny\kms}
        & \cnc{\tiny\kms}            & \cnc{\tiny\kms} \\
\hline \\
\multicolumn{11}{c}{\small NGC~4388, 30$^\circ$ slit} \\[4pt]
4\nodot\dig\dig
        &   18$\pm$20\dig                 &  353$\pm$60\dig
        &   60$\pm$20\dig                 &  192$\pm$75\dig
        & \ldots\dig\dig                  & \ldots\dig\dig 
        &  109$\pm$53\dig                 &  454$\pm$237
        & \ldots\dig\dig                  & \ldots\dig\dig  \\
3\nodot\dig\dig
        &   35$\pm$37\dig                 & $<150$\nopm
        &   45$\pm$16\dig                 &  334$\pm$64\dig
        & \ldots\dig\dig                  & \ldots\dig\dig 
        &   35$\pm$33\dig                 &  424$\pm$139
        & \ldots\dig\dig                  & \ldots\dig\dig  \\
2\nodot\dig\dig
        &   38$\pm$23\dig                 &  443$\pm$94\dig
        &    2$\pm$10\dig                 &  206$\pm$30\dig
        &   80$\pm$21\dig                 &  197$\pm$114
        &    1$\pm$18\dig                 &  284$\pm$71\dig
        & \ldots\dig\dig                  & \ldots\dig\dig  \\
1\nodot\dig\dig
        &   -2$\pm$13\dig                 &  249$\pm$42\dig
        &    4$\pm$11\dig                 &  168$\pm$40\dig
        &   48$\pm$14\dig                 &  338$\pm$53\dig
        &    7$\pm$14\dig                 &  286$\pm$52\dig
        &  -70$\pm$17\dig                 & $<150$\nopm \\
0\nodot\dig\dig
        &  -42$\pm$17\dig                 &  479$\pm$67\dig
        &  -12$\pm$12\dig                 &  249$\pm$43\dig
        &    8$\pm$10\dig                 &  306$\pm$34\dig
        &    6$\pm$15\dig                 & $<182$\nopm
        &   -5$\pm$28\dig                 & $<150$\nopm \\
-1\nodot\dig\dig
        &  -51$\pm$12\dig                 &  606$\pm$34\dig
        &  -21$\pm$11\dig                 &  366$\pm$48\dig
        &  -60$\pm$11\dig                 &  316$\pm$40\dig
        &   23$\pm$35\dig                 &  403$\pm$84\dig
        & -110$\pm$22\dig                 &  204$\pm$95\dig \\
-2\nodot\dig\dig
        &   52$\pm$12\dig                 &  624$\pm$31\dig
        &   12$\pm$28\dig                 &  321$\pm$62\dig
        &  -63$\pm$19\dig                 &  272$\pm$86\dig
        &  -38$\pm$15\dig                 &  392$\pm$53\dig
        & \ldots\dig\dig                  & \ldots\dig\dig  \\
-3\nodot\dig\dig
        &   76$\pm$17\dig                 &  195$\pm$64\dig
        &   16$\pm$18\dig                 &  383$\pm$70\dig
        &  -50$\pm$27\dig                 &  278$\pm$173
        &  158$\pm$29\dig                 &  524$\pm$98\dig
        & \ldots\dig\dig                  & \ldots\dig\dig  \\
-4\nodot\dig\dig
        &  138$\pm$22\dig                 &  357$\pm$91\dig
        &   24$\pm$27\dig                 & $<150$\nopm
        & \ldots\dig\dig                  & \ldots\dig\dig 
        & \ldots\dig\dig                  & \ldots\dig\dig 
        & \ldots\dig\dig                  & \ldots\dig\dig  \\[12pt]
\multicolumn{11}{c}{\small NGC~4388, 120$^\circ$ slit} \\[4pt]
2\nodot\dig\dig
        &   34$\pm$23\dig                 &  601$\pm$103
        &   49$\pm$19\dig                 &  218$\pm$81\dig
        &    3$\pm$16\dig                 &  253$\pm$80\dig
        & \ldots\dig\dig                  & \ldots\dig\dig 
        & \ldots\dig\dig                  & \ldots\dig\dig  \\
1\nodot\dig\dig
        &   10$\pm$21\dig                 &  624$\pm$75\dig
        &   91$\pm$20\dig                 &  291$\pm$54\dig
        &    2$\pm$10\dig                 &  221$\pm$34\dig
        &   56$\pm$17\dig                 &  404$\pm$41\dig
        & \ldots\dig\dig                  & \ldots\dig\dig  \\
0\nodot\dig\dig
        &   28$\pm$13\dig                 &  380$\pm$60\dig
        &   54$\pm$5\dig\dig              &  267$\pm$56\dig
        &   -4$\pm$11\dig                 &  211$\pm$40\dig
        &   54$\pm$16\dig                 &  286$\pm$62\dig
        &  -29$\pm$11\dig                 &  201$\pm$36\dig \\
-1\nodot\dig\dig
        &   14$\pm$17\dig                 &  332$\pm$100
        &   47$\pm$6\dig\dig              &  333$\pm$20\dig
        &  -37$\pm$9\dig\dig              &  284$\pm$30\dig
        &   71$\pm$15\dig                 &  331$\pm$58\dig
        & \ldots\dig\dig                  & \ldots\dig\dig  \\
-2\nodot\dig\dig
        &   13$\pm$21\dig                 &  252$\pm$140
        &    1$\pm$13\dig                 &  240$\pm$67\dig
        &  -42$\pm$12\dig                 &  249$\pm$442
        &   67$\pm$34\dig                 & $<194$\nopm
        & \ldots\dig\dig                  & \ldots\dig\dig  \\[12pt]
\hline \\
\end{tabular}
\end{lrbox}
\settowidth{\thiswid}{\usebox{\thisbox}}
\begin{center}
\begin{minipage}{\thiswid}
\caption{NGC~4388 Line Centroids and Widths}\label{ngc4388kinematics}
\usebox{\thisbox}
\tiny\begin{tabular}{l l}
$a$ & Positive position is towards the position angle of the slit as
quoted. \\
$b$ & Centroids are relative to 2525~\kms, the systemic velocity of NGC~4388. \\
$c$ & The resolution has been subtracted in quadrature. \\
\end{tabular}
\end{minipage}
\end{center}
\end{table}
}{
\placetable{ngc4388kinematics}
}

\ifthenelse{\boolean{ispreprint}}{
\begin{table}[htb]
\renewcommand{\baselinestretch}{1.0}
\tiny
\settowidth{\digsp}{0}
\newcommand{\dig}{\hspace*{\digsp}}
\settowidth{\dotsp}{.}
\newcommand{\nodot}{\hspace*{\dotsp}}
\settowidth{\pmsp}{$\pm$00}
\newcommand{\nopm}{\hspace*{\pmsp}}
\begin{lrbox}{\thisbox}
\begin{tabular}{r  r@{\hspace*{0.05in}}r  r@{\hspace*{0.05in}}r  r@{\hspace*{0.05in}}r  r@{\hspace*{0.05in}}r  r@{\hspace*{0.05in}}r}
\hline
\hline
        & \multicolumn{2}{c}{[Fe II]}
        & \multicolumn{2}{c}{Pa$\beta$}
        & \multicolumn{2}{c}{H$_2$}
        & \multicolumn{2}{c}{Br$\gamma$} \\
        & \multicolumn{2}{c}{\hrulefill}
        & \multicolumn{2}{c}{\hrulefill}
        & \multicolumn{2}{c}{\hrulefill}
        & \multicolumn{2}{c}{\hrulefill}  \\
\cnc{\tiny Pos$^a$}
        & \cnc{\tiny Centroid$^b$}    & \cnc{\tiny FWHM$^c$}
        & \cnc{\tiny Centroid}        & \cnc{\tiny FWHM}
        & \cnc{\tiny Centroid}        & \cnc{\tiny FWHM}
        & \cnc{\tiny Centroid}        & \cnc{\tiny FWHM} \\
\cnc{$''$}
        & \cnc{\tiny\kms}            & \cnc{\tiny\kms}
        & \cnc{\tiny\kms}            & \cnc{\tiny\kms}
        & \cnc{\tiny\kms}            & \cnc{\tiny\kms}
        & \cnc{\tiny\kms}            & \cnc{\tiny\kms} \\
\hline \\
3\nodot\dig\dig
        &   19$\pm$26\dig                 &  382$\pm$105
        &   73$\pm$12\dig                 &  200$\pm$65\dig
        &   72$\pm$60\dig                 & $<191$\nopm
        &   24$\pm$28\dig                 &  265$\pm$145 \\
2\nodot\dig\dig
        &  -44$\pm$16\dig                 &  367$\pm$51\dig
        &  -58$\pm$36\dig                 &  255$\pm$125
        & \ldots\dig\dig                  & \ldots\dig\dig 
        &   74$\pm$23\dig                 & $<191$\nopm  \\
1\nodot\dig\dig
        & -363$\pm$19\dig                 & 1165$\pm$102
        & -306$\pm$14\dig                 & 1201$\pm$122
        &  -22$\pm$79\dig                 &  710$\pm$335
        & -153$\pm$46\dig                 &  899$\pm$210 \\
0\nodot\dig\dig
        & -190$\pm$13\dig                 & 1137$\pm$41\dig
        & -109$\pm$11\dig                 &  991$\pm$46\dig
        &   -2$\pm$49\dig                 &  639$\pm$99\dig
        &  -15$\pm$21\dig                 &  564$\pm$169 \\
-1\nodot\dig\dig
        &   54$\pm$13\dig                 &  501$\pm$56\dig
        &   -4$\pm$13\dig                 &  695$\pm$53\dig
        &   81$\pm$34\dig                 & $<191$\nopm
        &  127$\pm$19\dig                 &  446$\pm$85\dig \\
-2\nodot\dig\dig
        &  162$\pm$16\dig                 &  394$\pm$83\dig
        &  -84$\pm$61\dig                 &  532$\pm$220
        &  165$\pm$32\dig                 &  745$\pm$430
        &    3$\pm$18\dig                 & $<191$\nopm \\
-3\nodot\dig\dig
        &  -77$\pm$45\dig                 & 1117$\pm$248
        &  -75$\pm$21\dig                 & $<155$\nopm
        & \ldots\dig\dig                  & \ldots\dig\dig 
        & \ldots\dig\dig                  & \ldots\dig\dig  \\[5pt]
\hline \\
\end{tabular}
\end{lrbox}
\settowidth{\thiswid}{\usebox{\thisbox}}
\begin{center}
\begin{minipage}{\thiswid}
\caption{Mk~3 Line Centroids and Widths}\label{mk3kinematics}
\usebox{\thisbox}
\tiny\begin{tabular}{l l}
$a$ & Positive position is east towards the 113$^\circ$ position angle of the
slit. \\
$b$ & Centroids are relative to 4046~\kms, the systemic velocity of Mk~3. \\
$c$ & The resolution has been subtracted in quadrature. \\
\end{tabular}
\end{minipage}
\end{center}
\end{table}
}{
\placetable{mk3kinematics}
}


\ifthenelse{\boolean{ispreprint}}{}{
\clearpage

\figcaption[Knop.fig1.ps]{Spectra of Mk~1066 in the $1.25-1.31\mu$m
range.  Position is distance from the nucleus in arcseconds along the
indicated position angle of the slit.\label{mk1066grey}}

\figcaption[Knop.fig2.ps]{Spectra of Mk~1066 along the 135$^\circ$
slit.  Each spectrum represents the spectrum in a rectangular beam which
is $0.67''$ (4 pixels) along the slit by the width of the slit
($\sim0.6''$).  Spatial bins are adjacent.  The position of the center
of each spatial bin, relative to the nucleus, is indicated in the lower
left of each panel.  Flux units (F$_\lambda$) are
arbitrary.\label{mk1066spec135}}

\figcaption[Knop.fig3.ps]{Spectra of Mk~1066 along the 45$^\circ$
slit.  Each spectrum represents the spectrum in a rectangular beam which
is $0.67''$ along the slit by the width of the slit $\sim0.6''$.
Spatial bins are adjacent.  Flux units (F$_\lambda$) are
arbitrary.\label{mk1066spec45}}

\figcaption[Knop.fig4.ps]{Flux ratios for Mk~1066.  The two plots
on the left are for the 135$^\circ$ slit, the two plots on the right for
the 45$^\circ$ slit.\label{mk1066fluxrat}}

\figcaption[Knop.fig5.ps]{Position versus velocity plots for
Mk~1066.  Position along a 135$^\circ$ slit.  Velocity is relative to
3625~\kms, the systemic velocity of Mk~1066
\protect\citep{bow95}.\label{mk1066lv}}

\figcaption[Knop.fig6.ps]{Velocity of the primary narrow
component of each line versus position along the 135$^\circ$ slit for
Mk~1066.  The data for \brg\ are shown as a dotted line in each plot.  The
dashed line on the \pab\ plot is the H$\alpha$ data of
\protect\citet{bow95}.\label{mk1066vvspos}}

\figcaption[Knop.fig7.ps]{Selected line profiles of Mk~1066 along
the 135$^\circ$ slit, chosen to illustrate shoulders seen in
asymmetrical line profiles.  One resolution element is approximately 3
bins on the plot.\label{mk1066closeups}}

\clearpage

\figcaption[Knop.fig8.ps]{Cartoon drawing of the nuclear regions of
Mk~1066, from the point of view of the observer.  The rotating disk is
the source of low excitation emission, and the source of the primary
velocity gradient seen in most optical and infrared lines.  The outflow
is higher excitation emission, and shows up as wings in the infrared
lines.  The dusty disk extinguishes features from the southeast cone of
the outflow.  Superimposed on the cartoon are the angles of the
135$^\circ$ and 45$^\circ$ infrared slits.\label{mk1066art}}

\figcaption[Knop.fig9.ps]{Spectra of NGC~2110 along the 160$^\circ$
slit.  Each spectrum is produced from a $1.0\times0.6''$ rectangular
beam.  Spatial bins are adjacent.  The distance of each spatial bin from
the nucleus in a southerly direction is indicated in each panel.  Note
the strength of \feii\ and the large \feii/\pab\ flux ratio on the
nucleus.  Flux units (F$_\lambda$) are arbitrary.\label{ngc2110spec}}

\figcaption[Knop.fig10.ps]{NGC~2110 flux ratios as a function of
position along the 160$^\circ$ slit.  Positive positions are to the
south.\label{ngc2110fluxrat}}

\figcaption[Knop.fig11.ps]{Velocity versus position
along 160$^\circ$ slit for NGC~2110. Velocity is relative to 2340~\kms, the
systemic velocity of NGC~2110.\label{ngc2110lv}}

\figcaption[Knop.fig12.ps]{Velocity centroid of the whole line
profile versus position for each line at each position along the
160$^\circ$ slit.  Velocity is relative to 2340~\kms, the systemic
velocity of NGC~2110.\label{ngc2110vvspos}}

\figcaption[Knop.fig13.ps]{Spectra of NGC~4388 along the 30$^\circ$
slit.  Each spectrum is within a rectangular beam which is 1$''$ along
the slit by the width of the slit ($\sim0.6''$).  Spatial bins are
adjacent.  The distance of each spatial bin from the nucleus along a
position angle of 30$^\circ$ is indicated in each spectrum.  Flux units
(F$_\lambda$) are arbitrary.\label{ngc4388spec30}}

\clearpage

\figcaption[Knop.fig14.ps]{Spectra of NGC~4388 along the
120$^\circ$ slit.  Each spectrum is within a rectangular beam which is
1$''$ along the slit by the width of the slit ($\sim0.6''$).  Spatial
bins are adjacent.  The distance of each spatial bin from the nucleus
along a position angle of 120$^\circ$ is indicated in each spectrum.
Flux units (F$_\lambda$) are arbitrary.\label{ngc4388spec120}}

\figcaption[Knop.fig15.ps]{Flux ratios for NGC 4388.  The two plots on
the left are for the 30$^\circ$ slit; the two plots on the right, for the
120$^\circ$ slit.\label{ngc4388fluxrats}}

\figcaption[Knop.fig16.ps]{Position versus velocity
plots for NGC 4388.  Position is along a 30$^\circ$ slit.  Velocity is
relative to 2525~\kms, the systemic velocity of NGC 4388. Contours
in this figure are logarithmic and separated by a factor of
two.\label{ngc4388lv30}}

\figcaption[Knop.fig17.ps]{Velocity centroid
of each line versus position along the 30$^\circ$ slit for NGC~4388.
Positive distance is to the northeast.\label{ngc4388vvspos30}}

\figcaption[Knop.fig18.ps]{Velocity centroid of each line versus
position along the 120$^\circ$ slit for NGC~4388.  Positive distance is
to the southeast.\label{ngc4388vvspos120}}

\figcaption[Knop.fig19.ps]{Line profiles of \feii\ and \pab\ in
NGC~4388 2-3$''$ southwest of the nucleus on the 30$^\circ$ slit, and
the log of the \feii/\pab\ ratio as a function of velocity.  Points with
greater than 50\% uncertainty in the flux ratio have been omitted from
the latter plot.\label{ngc4388closeups}}

\figcaption[Knop.fig20.ps]{Spectra of Mk~3 along the 113$^\circ$ slit.
Each spectrum is within a rectangular beam which is 1.0$''$ along the
slit by the width of the slit ($\sim0.6''$).  Spatial bins are adjacent.
The distance of each spatial bin from the nucleus in an easterly
direction is indicated in the spectrum.  Flux units (F$_\lambda$) are
arbitrary.\label{mk3spec}}

\clearpage

\figcaption[Knop.fig21.ps]{Profiles of nuclear lines in Mk~3.  The lines
are normalized to the same peak flux.  The solid line is \pab, the dotted
line \feii, and the dashed line \brg.\label{mk3nucprof}}

\figcaption[Knop.fig22.ps]{Mk~3 flux ratios as a function of position along
the 113$^\circ$ slit.  Positive positions are to the east.  Fluxes are
integrated over the entire line profile in each spatial
bin.\label{mk3fluxrat}}

\figcaption[Knop.fig23.ps]{Position versus velocity
plots for Mk~3.  Position is along a 113$^\circ$ slit.  Velocity is relative
to 4046~\kms, the systemic velocity of Mk~3 \protect\citep{whi88}.  Contours
are logarithmic, separated by a factor of
$\protect\sqrt{2}$.\label{mk3lv}}

\figcaption[Knop.fig24.ps]{Line profiles of \brg, \feii, and \pab, and the
log of the \feii/\pab\ flux ratio as a function of velocity, 1$''$ west of
the nucleus of Mk~3 at a position angle of 293$^\circ$.  Points with
uncertainty greater than 20\% have been omitted from the latter
plot.\label{mk3closeups}}
}


\ifthenelse{\boolean{ispreprint}}{}{
\clearpage

\begin{table}
\renewcommand{\baselinestretch}{1}\small
\begin{center}
\begin{tabular}{l l r r r r l}
\hline
\hline
Galaxy & Wavelength & \multicolumn{1}{c}{On-Source}   & 
        \multicolumn{1}{c}{Slit} & \multicolumn{1}{c}{Res.}  & 
        \multicolumn{1}{c}{psf}  & Date of    \\
       &            & \multicolumn{1}{c}{Integration} & 
        \multicolumn{1}{c}{PA}   & \multicolumn{1}{c}{\kms}  & 
        \multicolumn{1}{c}{FWHM} & Observation \\
\hline
Mk~1066  & 1.25$\mu$m-1.31$\mu$m & 2400s & 135$^\circ$
         & 242$\pm$15       & $0.7''$     & 1995/09/09 \\
Mk~1066  &                       & 2400s &  45$^\circ$
         & 228$\pm$15       & $0.6''$     & 1996/09/10 \\
Mk~1066  & 2.10$\mu$m-2.23$\mu$m & 2400s & 135$^\circ$
         & 289$\pm$19       & $0.6''$     & 1995/09/08 \\
Mk~1066  &                       & 2400s &  45$^\circ$
         & 289$\pm$19       & $0.5''$     & 1995/09/10 \\[12pt]
NGC~2110 & 1.25$\mu$m-1.31$\mu$m & 2400s & 160$^\circ$
         & 229$\pm$16       & $\sim1.1''$ & 1996/01/05 \\
NGC~2110 & 2.09$\mu$m-2.22$\mu$m & 1800s & 160$^\circ$
         & 285$\pm$19       & $\sim1.1''$ & 1996/01/05 \\[12pt]
NGC~4388 & 1.25$\mu$m-1.31$\mu$m & 2400s & 30$^\circ$
         & 229$\pm$15       & $\sim0.8''$ & 1996/01/03 \\
NGC~4388 &                       & 2400s & 120$^\circ$
         & 244$\pm$15       & $0.8''$ & 1996/04/08 \\
NGC~4388 & 2.10$\mu$m-2.23$\mu$m & 2400s & 30$^\circ$
         & 261$\pm$19       & $\sim0.7''$ & 1996/01/03 \\
NGC~4388 &                       & 1800s & 120$^\circ$
         & 303$\pm$19       & $0.8''$ & 1996/04/08 \\[12pt]
Mk~3     & 1.26$\mu$m-1.32$\mu$m & 3000s & 113$^\circ$
         & 243$\pm$15       & $\sim1.0''$ & 1995/11/06 \\
Mk~3     & 2.14$\mu$m-2.27$\mu$m & 1800s & 113$^\circ$
         & 292$\pm$19       & $\sim1.0''$ & 1995/11/07 \\[12pt]
\hline
\hline
\end{tabular}
\end{center}
\caption[Seyfert 2 Observation Log.]{Observation log.  References for
position angles are: Mk~1066: \protect\citet{bow95}; NGC~2110:
\protect\citet{mul94}; NGC~4388: \protect\citet{cor88}; Mk~3:
\protect\citet{mul96}.  The resolution column indicates the FWHM of an
unresolved line measured from OH sky lines in one of the galaxy frames.  The
point spread function (psf) of the atmospheric seeing was determined for
Mk~1066 and for the April 1996 observations of NGC~4388 by observing a star,
guided in the same manner as the spectra, in between spectral observations.
For Mk~3, the seeing was estimated based on 2.2$\mu$m imaging of a quasar
somewhat later than the observations of Mk~3.  For NGC~2110 and the January
1996 observations of NGC~4388, unguided G-star and photometric standard star
images were used to estimate the seeing.}
\label{sey2obslog}
\end{table}

\begin{table}
\renewcommand{\baselinestretch}{1.0}
\scriptsize
\settowidth{\digsp}{0}
\newcommand{\dig}{\hspace*{\digsp}}
\settowidth{\dotsp}{.}
\newcommand{\nodot}{\hspace*{\dotsp}}
\settowidth{\pmsp}{$\pm$0.00}
\newcommand{\nopm}{\hspace*{\pmsp}}
\begin{lrbox}{\thisbox}
\begin{tabular}{r r@{\hspace{0.08in}}r@{\hspace{0.08in}}r %
                   @{\hspace{0.08in}}r %
                  r@{\hspace{0.08in}}r}
\hline
\hline
        & \multicolumn{4}{c}{Equivalent Width (10$^{-4}\mu$m)}
        & \multicolumn{2}{c}{Flux Ratios} \\
Pos$^a$ & \multicolumn{4}{c}{\hrulefill}
        & \multicolumn{2}{c}{\hrulefill} \\
($''$)  & \cnc{[Fe II]} & \cnc{Pa$\beta$}
        & \cnc{H$_2$} & \cnc{Br$\gamma$} 
        & \cnc{\feii/Pa$\beta$} & \cnc{H$_2$/Br$\gamma$} \\
\hline \\[5pt]
\multicolumn{7}{c}{\small Mk~1066, 135$^\circ$ slit} \\[4pt]
2.67
        & 6.7$\pm$4.5\dig\dig
        & 17.1$\pm$12.4\dig
        & \ldots\dig\dig\dig\dig
        & \ldots\dig\dig\dig\dig
        & 0.43$\pm$0.14 
        & \ldots\dig\dig\dig\dig \\
2\nodot\dig\dig
        & 4.7$\pm$1.7\dig\dig
        & 5.8$\pm$1.9\dig\dig
        & 4.0$\pm$1.9\dig\dig
        & $<$4.3\nopm
        & 0.79$\pm$0.21 
        & $>$0.87\nopm \\
1.33
        & 18.4$\pm$3.2\dig\dig
        & 15.3$\pm$2.6\dig\dig
        & 9.4$\pm$2.0\dig\dig
        & 10.3$\pm$2.7\dig\dig
        & 1.17$\pm$0.08 
        & 0.86$\pm$0.09 \\
0.67
        & 23.0$\pm$2.6\dig\dig
        & 29.0$\pm$2.6\dig\dig
        & 12.6$\pm$1.7\dig\dig
        & 18.7$\pm$2.8\dig\dig
        & 0.75$\pm$0.03 
        & 0.63$\pm$0.05 \\
0\nodot\dig\dig
        & 20.8$\pm$1.9\dig\dig
        & 27.4$\pm$2.1\dig\dig
        & 13.0$\pm$1.2\dig\dig
        & 14.0$\pm$1.5\dig\dig
        & 0.75$\pm$0.04 
        & 0.89$\pm$0.06 \\
-0.67
        & 39.0$\pm$3.8\dig\dig
        & 33.0$\pm$4.6\dig\dig
        & 17.8$\pm$2.6\dig\dig
        & 20.6$\pm$3.8\dig\dig
        & 1.20$\pm$0.04 
        & 0.85$\pm$0.06 \\
-1.33
        & 21.6$\pm$3.4\dig\dig
        & 23.2$\pm$4.6\dig\dig
        & 9.0$\pm$2.7\dig\dig
        & 16.7$\pm$6.3\dig\dig
        & 0.99$\pm$0.05 
        & 0.56$\pm$0.09 \\
-2\nodot\dig\dig
        & 5.3$\pm$2.2\dig\dig
        & 6.7$\pm$2.5\dig\dig
        & $<$4.4\nopm
        & 11.5$\pm$6.1\dig\dig
        & 0.85$\pm$0.17 
        & $<$0.40\nopm \\[12pt]
\multicolumn{7}{c}{\small Mk~1066, 45$^\circ$ slit} \\[4pt]
1.33
        & 5.0$\pm$1.3\dig\dig
        & 4.2$\pm$1.1\dig\dig
        & 13.0$\pm$5.3\dig\dig
        & $<$2.7\nopm
        & 1.30$\pm$0.32 
        & $>$4.41\nopm \\
0.67
        & 11.2$\pm$1.6\dig\dig
        & 10.3$\pm$1.3\dig\dig
        & 12.6$\pm$2.7\dig\dig
        & 6.7$\pm$1.1\dig\dig
        & 1.06$\pm$0.09 
        & 1.74$\pm$0.21 \\
0\nodot\dig\dig
        & 17.2$\pm$1.4\dig\dig
        & 29.1$\pm$2.7\dig\dig
        & 11.5$\pm$1.5\dig\dig
        & 13.7$\pm$1.4\dig\dig
        & 0.57$\pm$0.03 
        & 0.76$\pm$0.06 \\
-0.67
        & 15.9$\pm$2.6\dig\dig
        & 17.9$\pm$2.7\dig\dig
        & 12.6$\pm$2.3\dig\dig
        & 10.0$\pm$2.0\dig\dig
        & 0.87$\pm$0.07 
        & 1.21$\pm$0.10 \\
-1.33
        & 5.2$\pm$1.4\dig\dig
        & 8.9$\pm$2.6\dig\dig
        & 9.0$\pm$4.0\dig\dig
        & $<$4.7\nopm
        & 0.64$\pm$0.16 
        & $>$1.83\nopm \\
-2\nodot\dig\dig
        & \ldots\dig\dig\dig\dig
        & \ldots\dig\dig\dig\dig
        & 5.6$\pm$3.7\dig\dig
        & $<$5.3\nopm
        & \ldots\dig\dig\dig\dig 
        & $>$0.91\nopm \\[5pt]
\hline \\
\end{tabular}
\end{lrbox}
\settowidth{\thiswid}{\usebox{\thisbox}}
\begin{center}
\caption{Mk~1066 Equivalent Widths and Flux Ratios}\label{mk1066eqw}
\begin{minipage}{\thiswid}
\usebox{\thisbox}
\tiny
$a$ Positive positions are towards the position angle of the slit as quoted.
\end{minipage}
\end{center}
\end{table}

\begin{table}
\renewcommand{\baselinestretch}{1.0}
\scriptsize
\settowidth{\digsp}{0}
\newcommand{\dig}{\hspace*{\digsp}}
\settowidth{\dotsp}{.}
\newcommand{\nodot}{\hspace*{\dotsp}}
\settowidth{\pmsp}{$\pm$0.00}
\newcommand{\nopm}{\hspace*{\pmsp}}
\begin{lrbox}{\thisbox}
\begin{tabular}{r r@{\hspace{0.08in}}r@{\hspace{0.08in}}r %
                   @{\hspace{0.08in}}r %
                  r@{\hspace{0.08in}}r}
\hline
\hline
        & \multicolumn{4}{c}{Equivalent Width (10$^{-4}\mu$m)}
        & \multicolumn{2}{c}{Flux Ratios} \\
Pos$^a$ & \multicolumn{4}{c}{\hrulefill}
        & \multicolumn{2}{c}{\hrulefill} \\
($''$)  & \cnc{[Fe II]} & \cnc{Pa$\beta$}
        & \cnc{H$_2$} & \cnc{Br$\gamma$} 
        & \cnc{\feii/Pa$\beta$} & \cnc{H$_2$/Br$\gamma$} \\
\hline \\[5pt]
3\nodot\dig\dig
        & 3.5$\pm$1.0\dig\dig
        & 3.3$\pm$0.8\dig\dig
        & 3.8$\pm$1.3\dig\dig
        & 2.1$\pm$0.9\dig\dig
        & 1.09$\pm$0.25 
        & 1.89$\pm$0.78 \\
2\nodot\dig\dig
        & 6.2$\pm$1.3\dig\dig
        & 2.9$\pm$0.7\dig\dig
        & 7.2$\pm$1.4\dig\dig
        & 1.9$\pm$0.6\dig\dig
        & 2.19$\pm$0.42 
        & 4.08$\pm$1.08 \\
1\nodot\dig\dig
        & 15.4$\pm$1.4\dig\dig
        & 3.1$\pm$0.5\dig\dig
        & 6.2$\pm$0.8\dig\dig
        & 2.0$\pm$0.4\dig\dig
        & 5.02$\pm$0.48 
        & 3.23$\pm$0.53 \\
0\nodot\dig\dig
        & 24.7$\pm$1.6\dig\dig
        & 3.1$\pm$0.4\dig\dig
        & 4.8$\pm$0.4\dig\dig
        & 1.6$\pm$0.3\dig\dig
        & 8.11$\pm$0.78 
        & 3.17$\pm$0.63 \\
-1\nodot\dig\dig
        & 22.5$\pm$2.3\dig\dig
        & 3.2$\pm$0.5\dig\dig
        & 6.6$\pm$0.7\dig\dig
        & 2.0$\pm$0.4\dig\dig
        & 7.36$\pm$0.74 
        & 3.59$\pm$0.69 \\
-2\nodot\dig\dig
        & 11.7$\pm$2.3\dig\dig
        & 2.8$\pm$0.6\dig\dig
        & 8.4$\pm$1.7\dig\dig
        & 1.4$\pm$0.4\dig\dig
        & 4.43$\pm$0.53 
        & 6.50$\pm$1.50 \\
-3\nodot\dig\dig
        & 8.7$\pm$2.0\dig\dig
        & 3.1$\pm$0.9\dig\dig
        & 5.5$\pm$1.2\dig\dig
        & 2.8$\pm$1.0\dig\dig
        & 2.93$\pm$0.54 
        & 2.14$\pm$0.51 \\
-4\nodot\dig\dig
        & 5.8$\pm$2.8\dig\dig
        & $<$2.7\nopm
        & 3.6$\pm$1.5\dig\dig
        & $<$9.7\nopm
        & $>$2.24\nopm
        & $>$0.39\nopm\\[5pt]
\hline \\
\end{tabular}
\end{lrbox}
\settowidth{\thiswid}{\usebox{\thisbox}}
\begin{center}
\caption{NGC~2110 Equivalent Widths and Flux Ratios}\label{ngc2110eqw}
\begin{minipage}{\thiswid}
\usebox{\thisbox}
\tiny
$a$ Positive positions are toward the southeast at a position angle of
160$^\circ$.
\end{minipage}
\end{center}
\end{table}

\begin{table}
\renewcommand{\baselinestretch}{1.0}
\scriptsize
\settowidth{\digsp}{0}
\newcommand{\dig}{\hspace*{\digsp}}
\settowidth{\dotsp}{.}
\newcommand{\nodot}{\hspace*{\dotsp}}
\settowidth{\pmsp}{$\pm$0.00}
\newcommand{\nopm}{\hspace*{\pmsp}}
\begin{lrbox}{\thisbox}
\begin{tabular}{r r@{\hspace{0.08in}}r@{\hspace{0.08in}}r %
                   @{\hspace{0.08in}}r@{\hspace{0.08in}}r %
                  r@{\hspace{0.08in}}r@{\hspace{0.08in}}r}
\hline
\hline
        & \multicolumn{5}{c}{Equivalent Width (10$^{-4}\mu$m)}
        & \multicolumn{3}{c}{Flux Ratios} \\
Pos$^a$ & \multicolumn{5}{c}{\hrulefill}
        & \multicolumn{3}{c}{\hrulefill} \\
($''$)  & \cnc{[Fe II]} & \cnc{Pa$\beta$}
        & \cnc{H$_2$} & \cnc{Br$\gamma$} & \cnc{[S IX]}
        & \cnc{\feii/Pa$\beta$} & \cnc{H$_2$/Br$\gamma$}
        & \cnc{\six/Pa$\beta$} \\
\hline \\[5pt]
\multicolumn{9}{c}{\small NGC~4388, 30$^\circ$ slit} \\[4pt]
4\nodot\dig\dig
        & 22.8$\pm$19.1\dig
        & 19.0$\pm$14.9\dig
        & $<$7.4\nopm
        & 13.0$\pm$12.8\dig
        & \ldots\dig\dig\dig\dig
        & 0.99$\pm$0.25 
        & $<$0.67\nopm
        & \ldots\dig\dig\dig\dig \\
3\nodot\dig\dig
        & 15.5$\pm$9.5\dig\dig
        & 18.2$\pm$9.1\dig\dig
        & $<$7.0\nopm
        & 13.3$\pm$9.9\dig\dig
        & \ldots\dig\dig\dig\dig
        & 0.71$\pm$0.17 
        & $<$0.57\nopm
        & \ldots\dig\dig\dig\dig \\
2\nodot\dig\dig
        & 13.4$\pm$6.0\dig\dig
        & 26.1$\pm$7.0\dig\dig
        & 6.5$\pm$3.1\dig\dig
        & 12.7$\pm$6.7\dig\dig
        & \ldots\dig\dig\dig\dig
        & 0.47$\pm$0.08 
        & 0.56$\pm$0.13 
        & \ldots\dig\dig\dig\dig \\
1\nodot\dig\dig
        & 7.7$\pm$1.9\dig\dig
        & 25.7$\pm$5.5\dig\dig
        & 13.6$\pm$4.1\dig\dig
        & 12.7$\pm$3.2\dig\dig
        & 3.6$\pm$1.0\dig\dig
        & 0.28$\pm$0.03 
        & 1.05$\pm$0.15 
        & 0.13$\pm$0.02 \\
0\nodot\dig\dig
        & 11.2$\pm$1.7\dig\dig
        & 26.3$\pm$4.2\dig\dig
        & 9.5$\pm$1.2\dig\dig
        & 10.6$\pm$1.5\dig\dig
        & 4.5$\pm$0.8\dig\dig
        & 0.40$\pm$0.03 
        & 0.88$\pm$0.08 
        & 0.16$\pm$0.02 \\
-1\nodot\dig\dig
        & 18.8$\pm$4.6\dig\dig
        & 32.2$\pm$6.8\dig\dig
        & 8.4$\pm$1.6\dig\dig
        & 17.1$\pm$3.9\dig\dig
        & 2.4$\pm$0.8\dig\dig
        & 0.57$\pm$0.04 
        & 0.51$\pm$0.06 
        & 0.07$\pm$0.01 \\
-2\nodot\dig\dig
        & 31.0$\pm$9.4\dig\dig
        & 26.8$\pm$7.4\dig\dig
        & 6.3$\pm$2.7\dig\dig
        & 14.7$\pm$6.5\dig\dig
        & \ldots\dig\dig\dig\dig
        & 1.13$\pm$0.11 
        & 0.46$\pm$0.09 
        & \ldots\dig\dig\dig\dig \\
-3\nodot\dig\dig
        & 29.7$\pm$11.8\dig
        & 17.8$\pm$7.1\dig\dig
        & 6.1$\pm$3.6\dig\dig
        & 13.6$\pm$10.5\dig
        & \ldots\dig\dig\dig\dig
        & 1.64$\pm$0.22 
        & 0.49$\pm$0.14 
        & \ldots\dig\dig\dig\dig \\
-4\nodot\dig\dig
        & 12.5$\pm$6.6\dig\dig
        & 4.8$\pm$2.7\dig\dig
        & $<$6.9\nopm
        & $<$11.7\nopm
        & \ldots\dig\dig\dig\dig
        & 2.52$\pm$0.84 
        & \ldots\dig\dig\dig\dig 
        & \ldots\dig\dig\dig\dig \\[12pt]
\multicolumn{9}{c}{\small NGC~4388, 120$^\circ$ slit} \\[4pt]
2\nodot\dig\dig
        & 9.0$\pm$2.5\dig\dig
        & 4.0$\pm$1.3\dig\dig
        & 12.4$\pm$5.3\dig\dig
        & $<$4.9\nopm
        & \ldots\dig\dig\dig\dig
        & 2.18$\pm$0.53 
        & $>$2.29\nopm
        & \ldots\dig\dig\dig\dig \\
1\nodot\dig\dig
        & 12.7$\pm$2.1\dig\dig
        & 16.8$\pm$3.1\dig\dig
        & 14.1$\pm$2.9\dig\dig
        & 9.4$\pm$2.0\dig\dig
        & \ldots\dig\dig\dig\dig
        & 0.72$\pm$0.07 
        & 1.39$\pm$0.18 
        & \ldots\dig\dig\dig\dig \\
0\nodot\dig\dig
        & 11.5$\pm$1.4\dig\dig
        & 24.1$\pm$2.7\dig\dig
        & 8.4$\pm$1.1\dig\dig
        & 11.6$\pm$1.8\dig\dig
        & 4.4$\pm$0.6\dig\dig
        & 0.46$\pm$0.03 
        & 0.67$\pm$0.07 
        & 0.17$\pm$0.01 \\
-1\nodot\dig\dig
        & 10.1$\pm$2.0\dig\dig
        & 19.9$\pm$3.5\dig\dig
        & 19.6$\pm$4.6\dig\dig
        & 8.0$\pm$1.7\dig\dig
        & \ldots\dig\dig\dig\dig
        & 0.50$\pm$0.05 
        & 2.34$\pm$0.25 
        & \ldots\dig\dig\dig\dig \\
-2\nodot\dig\dig
        & 8.9$\pm$3.0\dig\dig
        & 7.2$\pm$2.2\dig\dig
        & 18.0$\pm$5.9\dig\dig
        & 3.2$\pm$1.5\dig\dig
        & \ldots\dig\dig\dig\dig
        & 1.26$\pm$0.24 
        & 5.23$\pm$1.60 
        & \ldots\dig\dig\dig\dig \\[12pt]
\hline \\
\end{tabular}
\end{lrbox}
\settowidth{\thiswid}{\usebox{\thisbox}}
\begin{center}
\caption{NGC~4388 Equivalent Widths and Flux Ratios}\label{ngc4388eqw}
\begin{minipage}{\thiswid}
\usebox{\thisbox}
\tiny
$a$ Positive positions are towards the position angle of the slit as quoted.
\end{minipage}
\end{center}
\end{table}

\begin{table}
\renewcommand{\baselinestretch}{1.0} \scriptsize \settowidth{\digsp}{0}
\newcommand{\dig}{\hspace*{\digsp}} \settowidth{\dotsp}{.}
\newcommand{\nodot}{\hspace*{\dotsp}} \settowidth{\pmsp}{$\pm$0.00}
\newcommand{\nopm}{\hspace*{\pmsp}}
\begin{lrbox}{\thisbox}
\begin{tabular}{r r@{\hspace{0.08in}}r@{\hspace{0.08in}}r %
                   @{\hspace{0.08in}}r %
                  r@{\hspace{0.08in}}r}
\hline
\hline
        & \multicolumn{4}{c}{Equivalent Width (10$^{-4}\mu$m)}
        & \multicolumn{2}{c}{Flux Ratios} \\
Pos$^a$ & \multicolumn{4}{c}{\hrulefill}
        & \multicolumn{2}{c}{\hrulefill} \\
($''$)  & \cnc{[Fe II]} & \cnc{Pa$\beta$}
        & \cnc{H$_2$} & \cnc{Br$\gamma$} 
        & \cnc{\feii/Pa$\beta$} & \cnc{H$_2$/Br$\gamma$} \\
\hline \\[5pt]
3\nodot\dig\dig
        & 11.6$\pm$4.3\dig\dig
        & 24.5$\pm$8.6\dig\dig
        & 5.8$\pm$4.2\dig\dig
        & 12.0$\pm$8.4\dig\dig
        & 0.47$\pm$0.10 
        & 0.49$\pm$0.24  \\
2\nodot\dig\dig
        & 13.0$\pm$3.5\dig\dig
        & 17.7$\pm$4.7\dig\dig
        & \ldots\dig\dig\dig\dig
        & 4.0$\pm$1.4\dig\dig
        & 0.74$\pm$0.12 
        & \ldots\dig\dig\dig\dig  \\
1\nodot\dig\dig
        & 37.6$\pm$4.2\dig\dig
        & 31.3$\pm$3.5\dig\dig
        & 3.4$\pm$1.2\dig\dig
        & 11.5$\pm$2.6\dig\dig
        & 1.17$\pm$0.07 
        & 0.29$\pm$0.11  \\
0\nodot\dig\dig
        & 52.2$\pm$5.1\dig\dig
        & 39.3$\pm$3.2\dig\dig
        & 5.3$\pm$0.7\dig\dig
        & 16.7$\pm$2.1\dig\dig
        & 1.24$\pm$0.04 
        & 0.31$\pm$0.04  \\
-1\nodot\dig\dig
        & 46.3$\pm$6.2\dig\dig
        & 26.4$\pm$3.0\dig\dig
        & 4.6$\pm$0.8\dig\dig
        & 12.8$\pm$2.2\dig\dig
        & 1.72$\pm$0.08 
        & 0.35$\pm$0.05  \\
-2\nodot\dig\dig
        & 17.9$\pm$5.7\dig\dig
        & 9.5$\pm$2.9\dig\dig
        & 13.1$\pm$4.5\dig\dig
        & 4.4$\pm$1.6\dig\dig
        & 1.88$\pm$0.43 
        & 2.94$\pm$0.54  \\
-3\nodot\dig\dig
        & 17.6$\pm$7.7\dig\dig
        & 4.3$\pm$1.7\dig\dig
        & \ldots\dig\dig\dig\dig
        & \ldots\dig\dig\dig\dig
        & 3.75$\pm$1.15 
        & \ldots\dig\dig\dig\dig \\[5pt]
\hline
\end{tabular}
\end{lrbox}
\settowidth{\thiswid}{\usebox{\thisbox}}
\begin{center}
\caption{Mk~3 Equivalent Widths and Flux Ratios}\label{mk3eqw}
\begin{minipage}{\thiswid}
\usebox{\thisbox}
\tiny
$a$ Positive positions are east towards the 113$^\circ$ position angle of
the slit.
\end{minipage}
\end{center}
\end{table}

\begin{table}
\renewcommand{\baselinestretch}{1.0}
\tiny
\settowidth{\digsp}{0}
\newcommand{\dig}{\hspace*{\digsp}}
\settowidth{\dotsp}{.}
\newcommand{\nodot}{\hspace*{\dotsp}}
\settowidth{\pmsp}{$\pm$00}
\newcommand{\nopm}{\hspace*{\pmsp}}
\begin{lrbox}{\thisbox}
\begin{tabular}{r  r@{\hspace*{0.05in}}r  r@{\hspace*{0.05in}}r  r@{\hspace*{0.05in}}r  r@{\hspace*{0.05in}}r  r@{\hspace*{0.05in}}r}
\hline
\hline
        & \multicolumn{2}{c}{[Fe II]}
        & \multicolumn{2}{c}{Pa$\beta$}
        & \multicolumn{2}{c}{H$_2$}
        & \multicolumn{2}{c}{Br$\gamma$} \\
        & \multicolumn{2}{c}{\hrulefill}
        & \multicolumn{2}{c}{\hrulefill}
        & \multicolumn{2}{c}{\hrulefill}
        & \multicolumn{2}{c}{\hrulefill}  \\
\cnc{\tiny Pos$^a$}
        & \cnc{\tiny Centroid$^b$}    & \cnc{\tiny FWHM$^c$}
        & \cnc{\tiny Centroid}        & \cnc{\tiny FWHM}
        & \cnc{\tiny Centroid}        & \cnc{\tiny FWHM}
        & \cnc{\tiny Centroid}        & \cnc{\tiny FWHM} \\
\cnc{$''$}
        & \cnc{\tiny\kms}            & \cnc{\tiny\kms}
        & \cnc{\tiny\kms}            & \cnc{\tiny\kms}
        & \cnc{\tiny\kms}            & \cnc{\tiny\kms}
        & \cnc{\tiny\kms}            & \cnc{\tiny\kms} \\
\hline \\
\multicolumn{9}{c}{\small Mk~1066, 135$^\circ$ slit} \\[4pt]
2.67
        & -118$\pm$24\dig                 & $<154$\nopm
        & -233$\pm$41\dig                 &  677$\pm$194
        & \ldots\dig\dig                  & \ldots\dig\dig 
        & \ldots\dig\dig                  & \ldots\dig\dig  \\
2\nodot\dig\dig
        &  -68$\pm$25\dig                 &  267$\pm$109
        &  -58$\pm$20\dig                 &  221$\pm$83\dig
        &  -72$\pm$35\dig                 & $<190$\nopm
        & \ldots\dig\dig                  & \ldots\dig\dig  \\
1.33
        &  -59$\pm$14\dig                 &  294$\pm$53\dig
        &  -77$\pm$19\dig                 & $<154$\nopm
        &  -45$\pm$15\dig                 &  328$\pm$46\dig
        &  -90$\pm$19\dig                 &  383$\pm$34\dig \\
0.67
        &  -38$\pm$14\dig                 &  240$\pm$32\dig
        &  -67$\pm$14\dig                 & $<154$\nopm
        &  -45$\pm$15\dig                 &  321$\pm$50\dig
        &  -73$\pm$18\dig                 &  237$\pm$46\dig \\
0\nodot\dig\dig
        & -125$\pm$16\dig                 &  296$\pm$41\dig
        &  -20$\pm$15\dig                 &  205$\pm$50\dig
        &    3$\pm$13\dig                 &  353$\pm$29\dig
        &  -19$\pm$18\dig                 &  329$\pm$27\dig \\
-0.67
        &  -54$\pm$14\dig                 &  289$\pm$54\dig
        &   51$\pm$14\dig                 & $<154$\nopm
        &   89$\pm$13\dig                 &  223$\pm$37\dig
        &   51$\pm$19\dig                 &  269$\pm$32\dig \\
-1.33
        &   -8$\pm$15\dig                 &  341$\pm$62\dig
        &   32$\pm$16\dig                 &  167$\pm$74\dig
        &  143$\pm$20\dig                 &  360$\pm$78\dig 
        &   70$\pm$20\dig                 &  308$\pm$44\dig \\
-2\nodot\dig\dig
        &  109$\pm$18\dig                 &  218$\pm$75\dig
        &   70$\pm$18\dig                 &  202$\pm$65\dig
        & \ldots\dig\dig                  & \ldots\dig\dig 
        &  133$\pm$46\dig                 &  715$\pm$196    \\[12pt]
\multicolumn{9}{c}{\small Mk~1066, 45$^\circ$ slit} \\[4pt]
1.33
        &  -34$\pm$17\dig                 &  222$\pm$73\dig
        &  -50$\pm$19\dig                 &  179$\pm$93\dig
        &   -6$\pm$15\dig                 &  491$\pm$59\dig
        & \ldots\dig\dig                  & \ldots\dig\dig  \\
0.67
        &  -52$\pm$14\dig                 &  179$\pm$52\dig
        &  -20$\pm$10\dig                 & $<150$\nopm
        &  -12$\pm$11\dig                 &  392$\pm$35\dig
        &  -43$\pm$17\dig                 &  429$\pm$62\dig \\
0\nodot\dig\dig
        &  -74$\pm$12\dig                 &  179$\pm$38\dig
        &  -41$\pm$13\dig                 & $<150$\nopm
        &  -14$\pm$12\dig                 &  192$\pm$61\dig
        &  -83$\pm$17\dig                 &  250$\pm$44\dig \\
-0.67
        &  -54$\pm$15\dig                 &  219$\pm$46\dig
        &  -24$\pm$11\dig                 & $<150$\nopm
        &  -21$\pm$10\dig                 &  277$\pm$33\dig
        &  -49$\pm$14\dig                 &  313$\pm$51\dig \\
-1.33
        &  -10$\pm$19\dig                 &  247$\pm$83\dig
        &   64$\pm$49\dig                 & $<150$\nopm
        &   39$\pm$18\dig                 &  329$\pm$82\dig
        & \ldots\dig\dig                  & \ldots\dig\dig  \\
-2\nodot\dig\dig
        & \ldots\dig\dig                  & \ldots\dig\dig 
        & \ldots\dig\dig                  & \ldots\dig\dig 
        &  107$\pm$23\dig                 & $<190$\nopm
        & \ldots\dig\dig                  & \ldots\dig\dig  \\[5pt]
\hline \\
\end{tabular}
\end{lrbox}
\settowidth{\thiswid}{\usebox{\thisbox}}
\begin{center}
\begin{minipage}{\thiswid}
\caption{Mk~1066 Line Centroids and Widths}\label{mk1066kinematics}
\usebox{\thisbox}
\tiny\begin{tabular}{l l}
$a$ & Positive position is towards the position angle of the slit as
quoted. \\
$b$ & Centroids are relative to 3625~\kms, the systemic velocity of Mk~1066. \\
$c$ & The resolution has been subtracted in quadrature. \\
\end{tabular}
\end{minipage}
\end{center}
\end{table}

\begin{table}
\renewcommand{\baselinestretch}{1.0}
\tiny
\settowidth{\digsp}{0}
\newcommand{\dig}{\hspace*{\digsp}}
\settowidth{\dotsp}{.}
\newcommand{\nodot}{\hspace*{\dotsp}}
\settowidth{\pmsp}{$\pm$00}
\newcommand{\nopm}{\hspace*{\pmsp}}
\begin{lrbox}{\thisbox}
\begin{tabular}{r  r@{\hspace*{0.05in}}r  r@{\hspace*{0.05in}}r  r@{\hspace*{0.05in}}r  r@{\hspace*{0.05in}}r  r@{\hspace*{0.05in}}r}
\hline
\hline
        & \multicolumn{2}{c}{[Fe II]}
        & \multicolumn{2}{c}{Pa$\beta$}
        & \multicolumn{2}{c}{H$_2$}
        & \multicolumn{2}{c}{Br$\gamma$} \\
        & \multicolumn{2}{c}{\hrulefill}
        & \multicolumn{2}{c}{\hrulefill}
        & \multicolumn{2}{c}{\hrulefill}
        & \multicolumn{2}{c}{\hrulefill}  \\
\cnc{\tiny Pos$^a$}
        & \cnc{\tiny Centroid$^b$}    & \cnc{\tiny FWHM$^c$}
        & \cnc{\tiny Centroid}        & \cnc{\tiny FWHM}
        & \cnc{\tiny Centroid}        & \cnc{\tiny FWHM}
        & \cnc{\tiny Centroid}        & \cnc{\tiny FWHM} \\
\cnc{$''$}
        & \cnc{\tiny\kms}            & \cnc{\tiny\kms}
        & \cnc{\tiny\kms}            & \cnc{\tiny\kms}
        & \cnc{\tiny\kms}            & \cnc{\tiny\kms}
        & \cnc{\tiny\kms}            & \cnc{\tiny\kms} \\
\hline \\
3\nodot\dig\dig
        &  212$\pm$18\dig                 &  219$\pm$88\dig
        &  299$\pm$11\dig                 & $<150$\nopm
        &  292$\pm$29\dig                 &  223$\pm$132
        &  375$\pm$45\dig                 & $<189$\nopm     \\
2\nodot\dig\dig
        &  117$\pm$15\dig                 &  334$\pm$44\dig
        &  226$\pm$13\dig                 & $<150$\nopm
        &  213$\pm$21\dig                 & $<189$\nopm
        &  203$\pm$36\dig                 &  213$\pm$169    \\
1\nodot\dig\dig
        &   -4$\pm$9\dig\dig              &  373$\pm$36\dig
        &  116$\pm$8\dig\dig              &  213$\pm$41\dig
        &  152$\pm$18\dig                 &  275$\pm$35\dig
        &  160$\pm$27\dig                 &  259$\pm$92\dig \\
0\nodot\dig\dig
        &  -86$\pm$10\dig                 &  496$\pm$34\dig
        &   -4$\pm$10\dig                 &  338$\pm$43\dig
        &   32$\pm$18\dig                 &  359$\pm$36\dig
        &   62$\pm$33\dig                 &  391$\pm$116    \\
-1\nodot\dig\dig
        &  -68$\pm$8\dig\dig              &  383$\pm$25\dig
        &  -49$\pm$9\dig\dig              &  186$\pm$51\dig
        &  -52$\pm$17\dig                 &  273$\pm$34\dig
        &   48$\pm$31\dig                 &  304$\pm$111    \\
-2\nodot\dig\dig
        & -114$\pm$11\dig                 &  171$\pm$54\dig
        &  -84$\pm$9\dig\dig              &  176$\pm$53\dig
        &  -14$\pm$20\dig                 &  238$\pm$52\dig
        &  -95$\pm$31\dig                 & $<189$\nopm     \\
-3\nodot\dig\dig
        & -156$\pm$10\dig                 &  197$\pm$31\dig
        & -126$\pm$15\dig                 &  203$\pm$82\dig
        & -112$\pm$19\dig                 &  205$\pm$61\dig
        &  -69$\pm$35\dig                 &  313$\pm$136    \\
-4\nodot\dig\dig
        & -207$\pm$20\dig                 &  347$\pm$87\dig
        & \ldots\dig\dig                  & \ldots\dig\dig 
        &   -5$\pm$42\dig                 &  546$\pm$226
        & \ldots\dig\dig                  & \ldots\dig\dig  \\[5pt]
\hline \\
\end{tabular}
\end{lrbox}
\settowidth{\thiswid}{\usebox{\thisbox}}
\begin{center}
\begin{minipage}{\thiswid}
\caption{NGC~2110 Line Centroids and Widths}\label{ngc2110kinematics}
\usebox{\thisbox}
\tiny\begin{tabular}{l l}
$a$ & Positive position is southeast along the 160$^\circ$ slit. \\
$b$ & Centroids are relative to 2342~\kms, the systemic velocity of
NGC~2110. \\
$c$ & The resolution has been subtracted in quadrature.\\
\end{tabular}
\end{minipage}
\end{center}
\end{table}

\begin{table}
\renewcommand{\baselinestretch}{1.0}
\tiny
\settowidth{\digsp}{0}
\newcommand{\dig}{\hspace*{\digsp}}
\settowidth{\dotsp}{.}
\newcommand{\nodot}{\hspace*{\dotsp}}
\settowidth{\pmsp}{$\pm$00}
\newcommand{\nopm}{\hspace*{\pmsp}}
\begin{lrbox}{\thisbox}
\begin{tabular}{r  r@{\hspace*{0.05in}}r  r@{\hspace*{0.05in}}r  r@{\hspace*{0.05in}}r  r@{\hspace*{0.05in}}r  r@{\hspace*{0.05in}}r}
\hline
\hline
        & \multicolumn{2}{c}{[Fe II]}
        & \multicolumn{2}{c}{Pa$\beta$}
        & \multicolumn{2}{c}{H$_2$}
        & \multicolumn{2}{c}{Br$\gamma$}
        & \multicolumn{2}{c}{[S IX]}     \\
        & \multicolumn{2}{c}{\hrulefill}
        & \multicolumn{2}{c}{\hrulefill}
        & \multicolumn{2}{c}{\hrulefill}
        & \multicolumn{2}{c}{\hrulefill}
        & \multicolumn{2}{c}{\hrulefill}  \\
\cnc{\tiny Pos$^a$}
        & \cnc{\tiny Centroid$^b$}    & \cnc{\tiny FWHM$^c$}
        & \cnc{\tiny Centroid}        & \cnc{\tiny FWHM}
        & \cnc{\tiny Centroid}        & \cnc{\tiny FWHM}
        & \cnc{\tiny Centroid}        & \cnc{\tiny FWHM}
        & \cnc{\tiny Centroid}        & \cnc{\tiny FWHM} \\
\cnc{$''$}
        & \cnc{\tiny\kms}            & \cnc{\tiny\kms}
        & \cnc{\tiny\kms}            & \cnc{\tiny\kms}
        & \cnc{\tiny\kms}            & \cnc{\tiny\kms}
        & \cnc{\tiny\kms}            & \cnc{\tiny\kms}
        & \cnc{\tiny\kms}            & \cnc{\tiny\kms} \\
\hline \\
\multicolumn{11}{c}{\small NGC~4388, 30$^\circ$ slit} \\[4pt]
4\nodot\dig\dig
        &   18$\pm$20\dig                 &  353$\pm$60\dig
        &   60$\pm$20\dig                 &  192$\pm$75\dig
        & \ldots\dig\dig                  & \ldots\dig\dig 
        &  109$\pm$53\dig                 &  454$\pm$237
        & \ldots\dig\dig                  & \ldots\dig\dig  \\
3\nodot\dig\dig
        &   35$\pm$37\dig                 & $<150$\nopm
        &   45$\pm$16\dig                 &  334$\pm$64\dig
        & \ldots\dig\dig                  & \ldots\dig\dig 
        &   35$\pm$33\dig                 &  424$\pm$139
        & \ldots\dig\dig                  & \ldots\dig\dig  \\
2\nodot\dig\dig
        &   38$\pm$23\dig                 &  443$\pm$94\dig
        &    2$\pm$10\dig                 &  206$\pm$30\dig
        &   80$\pm$21\dig                 &  197$\pm$114
        &    1$\pm$18\dig                 &  284$\pm$71\dig
        & \ldots\dig\dig                  & \ldots\dig\dig  \\
1\nodot\dig\dig
        &   -2$\pm$13\dig                 &  249$\pm$42\dig
        &    4$\pm$11\dig                 &  168$\pm$40\dig
        &   48$\pm$14\dig                 &  338$\pm$53\dig
        &    7$\pm$14\dig                 &  286$\pm$52\dig
        &  -70$\pm$17\dig                 & $<150$\nopm \\
0\nodot\dig\dig
        &  -42$\pm$17\dig                 &  479$\pm$67\dig
        &  -12$\pm$12\dig                 &  249$\pm$43\dig
        &    8$\pm$10\dig                 &  306$\pm$34\dig
        &    6$\pm$15\dig                 & $<182$\nopm
        &   -5$\pm$28\dig                 & $<150$\nopm \\
-1\nodot\dig\dig
        &  -51$\pm$12\dig                 &  606$\pm$34\dig
        &  -21$\pm$11\dig                 &  366$\pm$48\dig
        &  -60$\pm$11\dig                 &  316$\pm$40\dig
        &   23$\pm$35\dig                 &  403$\pm$84\dig
        & -110$\pm$22\dig                 &  204$\pm$95\dig \\
-2\nodot\dig\dig
        &   52$\pm$12\dig                 &  624$\pm$31\dig
        &   12$\pm$28\dig                 &  321$\pm$62\dig
        &  -63$\pm$19\dig                 &  272$\pm$86\dig
        &  -38$\pm$15\dig                 &  392$\pm$53\dig
        & \ldots\dig\dig                  & \ldots\dig\dig  \\
-3\nodot\dig\dig
        &   76$\pm$17\dig                 &  195$\pm$64\dig
        &   16$\pm$18\dig                 &  383$\pm$70\dig
        &  -50$\pm$27\dig                 &  278$\pm$173
        &  158$\pm$29\dig                 &  524$\pm$98\dig
        & \ldots\dig\dig                  & \ldots\dig\dig  \\
-4\nodot\dig\dig
        &  138$\pm$22\dig                 &  357$\pm$91\dig
        &   24$\pm$27\dig                 & $<150$\nopm
        & \ldots\dig\dig                  & \ldots\dig\dig 
        & \ldots\dig\dig                  & \ldots\dig\dig 
        & \ldots\dig\dig                  & \ldots\dig\dig  \\[12pt]
\multicolumn{11}{c}{\small NGC~4388, 120$^\circ$ slit} \\[4pt]
2\nodot\dig\dig
        &   34$\pm$23\dig                 &  601$\pm$103
        &   49$\pm$19\dig                 &  218$\pm$81\dig
        &    3$\pm$16\dig                 &  253$\pm$80\dig
        & \ldots\dig\dig                  & \ldots\dig\dig 
        & \ldots\dig\dig                  & \ldots\dig\dig  \\
1\nodot\dig\dig
        &   10$\pm$21\dig                 &  624$\pm$75\dig
        &   91$\pm$20\dig                 &  291$\pm$54\dig
        &    2$\pm$10\dig                 &  221$\pm$34\dig
        &   56$\pm$17\dig                 &  404$\pm$41\dig
        & \ldots\dig\dig                  & \ldots\dig\dig  \\
0\nodot\dig\dig
        &   28$\pm$13\dig                 &  380$\pm$60\dig
        &   54$\pm$5\dig\dig              &  267$\pm$56\dig
        &   -4$\pm$11\dig                 &  211$\pm$40\dig
        &   54$\pm$16\dig                 &  286$\pm$62\dig
        &  -29$\pm$11\dig                 &  201$\pm$36\dig \\
-1\nodot\dig\dig
        &   14$\pm$17\dig                 &  332$\pm$100
        &   47$\pm$6\dig\dig              &  333$\pm$20\dig
        &  -37$\pm$9\dig\dig              &  284$\pm$30\dig
        &   71$\pm$15\dig                 &  331$\pm$58\dig
        & \ldots\dig\dig                  & \ldots\dig\dig  \\
-2\nodot\dig\dig
        &   13$\pm$21\dig                 &  252$\pm$140
        &    1$\pm$13\dig                 &  240$\pm$67\dig
        &  -42$\pm$12\dig                 &  249$\pm$442
        &   67$\pm$34\dig                 & $<194$\nopm
        & \ldots\dig\dig                  & \ldots\dig\dig  \\[12pt]
\hline \\
\end{tabular}
\end{lrbox}
\settowidth{\thiswid}{\usebox{\thisbox}}
\begin{center}
\begin{minipage}{\thiswid}
\caption{NGC~4388 Line Centroids and Widths}\label{ngc4388kinematics}
\usebox{\thisbox}
\tiny\begin{tabular}{l l}
$a$ & Positive position is towards the position angle of the slit as
quoted. \\
$b$ & Centroids are relative to 2525~\kms, the systemic velocity of NGC~4388. \\
$c$ & The resolution has been subtracted in quadrature. \\
\end{tabular}
\end{minipage}
\end{center}
\end{table}

\begin{table}
\renewcommand{\baselinestretch}{1.0}
\tiny
\settowidth{\digsp}{0}
\newcommand{\dig}{\hspace*{\digsp}}
\settowidth{\dotsp}{.}
\newcommand{\nodot}{\hspace*{\dotsp}}
\settowidth{\pmsp}{$\pm$00}
\newcommand{\nopm}{\hspace*{\pmsp}}
\begin{lrbox}{\thisbox}
\begin{tabular}{r  r@{\hspace*{0.05in}}r  r@{\hspace*{0.05in}}r  r@{\hspace*{0.05in}}r  r@{\hspace*{0.05in}}r  r@{\hspace*{0.05in}}r}
\hline
\hline
        & \multicolumn{2}{c}{[Fe II]}
        & \multicolumn{2}{c}{Pa$\beta$}
        & \multicolumn{2}{c}{H$_2$}
        & \multicolumn{2}{c}{Br$\gamma$} \\
        & \multicolumn{2}{c}{\hrulefill}
        & \multicolumn{2}{c}{\hrulefill}
        & \multicolumn{2}{c}{\hrulefill}
        & \multicolumn{2}{c}{\hrulefill}  \\
\cnc{\tiny Pos$^a$}
        & \cnc{\tiny Centroid$^b$}    & \cnc{\tiny FWHM$^c$}
        & \cnc{\tiny Centroid}        & \cnc{\tiny FWHM}
        & \cnc{\tiny Centroid}        & \cnc{\tiny FWHM}
        & \cnc{\tiny Centroid}        & \cnc{\tiny FWHM} \\
\cnc{$''$}
        & \cnc{\tiny\kms}            & \cnc{\tiny\kms}
        & \cnc{\tiny\kms}            & \cnc{\tiny\kms}
        & \cnc{\tiny\kms}            & \cnc{\tiny\kms}
        & \cnc{\tiny\kms}            & \cnc{\tiny\kms} \\
\hline \\
3\nodot\dig\dig
        &   19$\pm$26\dig                 &  382$\pm$105
        &   73$\pm$12\dig                 &  200$\pm$65\dig
        &   72$\pm$60\dig                 & $<191$\nopm
        &   24$\pm$28\dig                 &  265$\pm$145 \\
2\nodot\dig\dig
        &  -44$\pm$16\dig                 &  367$\pm$51\dig
        &  -58$\pm$36\dig                 &  255$\pm$125
        & \ldots\dig\dig                  & \ldots\dig\dig 
        &   74$\pm$23\dig                 & $<191$\nopm  \\
1\nodot\dig\dig
        & -363$\pm$19\dig                 & 1165$\pm$102
        & -306$\pm$14\dig                 & 1201$\pm$122
        &  -22$\pm$79\dig                 &  710$\pm$335
        & -153$\pm$46\dig                 &  899$\pm$210 \\
0\nodot\dig\dig
        & -190$\pm$13\dig                 & 1137$\pm$41\dig
        & -109$\pm$11\dig                 &  991$\pm$46\dig
        &   -2$\pm$49\dig                 &  639$\pm$99\dig
        &  -15$\pm$21\dig                 &  564$\pm$169 \\
-1\nodot\dig\dig
        &   54$\pm$13\dig                 &  501$\pm$56\dig
        &   -4$\pm$13\dig                 &  695$\pm$53\dig
        &   81$\pm$34\dig                 & $<191$\nopm
        &  127$\pm$19\dig                 &  446$\pm$85\dig \\
-2\nodot\dig\dig
        &  162$\pm$16\dig                 &  394$\pm$83\dig
        &  -84$\pm$61\dig                 &  532$\pm$220
        &  165$\pm$32\dig                 &  745$\pm$430
        &    3$\pm$18\dig                 & $<191$\nopm \\
-3\nodot\dig\dig
        &  -77$\pm$45\dig                 & 1117$\pm$248
        &  -75$\pm$21\dig                 & $<155$\nopm
        & \ldots\dig\dig                  & \ldots\dig\dig 
        & \ldots\dig\dig                  & \ldots\dig\dig  \\[5pt]
\hline \\
\end{tabular}
\end{lrbox}
\settowidth{\thiswid}{\usebox{\thisbox}}
\begin{center}
\begin{minipage}{\thiswid}
\caption{Mk~3 Line Centroids and Widths}\label{mk3kinematics}
\usebox{\thisbox}
\tiny\begin{tabular}{l l}
$a$ & Positive position is east towards the 113$^\circ$ position angle of the
slit. \\
$b$ & Centroids are relative to 4046~\kms, the systemic velocity of Mk~3. \\
$c$ & The resolution has been subtracted in quadrature. \\
\end{tabular}
\end{minipage}
\end{center}
\end{table}

}

\end{document}